\documentclass[twocolumn]{aastex631}

\usepackage{graphicx}
\usepackage{subfigure}
\usepackage{multirow}
\usepackage{comment}
\usepackage{hyperref}
\usepackage{mathtools}
\usepackage{longtable}
\usepackage{mathrsfs}
\usepackage{fontenc}
\usepackage{xcolor}
\usepackage{url}
\usepackage{gensymb}
\usepackage{pifont}
\usepackage[T1]{fontenc}

\newcommand{\angs}{$\mathring{\mathrm{A}}$}

\begin{document}
\title{A Two-zone Accretion Disk in the Changing-look Active Galactic Nucleus 1ES\,1927+654: Physical Implications for Tidal Disruption Events and Super-Eddington Accretion}

\author[0000-0001-8496-4162]{Ruancun Li}
\affil{Kavli Institute for Astronomy and Astrophysics, Peking University,
Beijing 100871, China}
\affil{Department of Astronomy, School of Physics, Peking University,
Beijing 100871, China}

\author[0000-0001-6947-5846]{Luis C. Ho}
\affil{Kavli Institute for Astronomy and Astrophysics, Peking University, Beijing 100871, China}
\affil{Department of Astronomy, School of Physics, Peking University, Beijing 100871, China}

\author[0000-0001-5231-2645]{Claudio Ricci}
\affil{Instituto de Estudios Astrof\'isicos, Facultad de Ingenier\'ia y Ciencias, Universidad Diego Portales, Av. Ej\'ercito Libertador 441, Santiago, Chile}
\affil{Kavli Institute for Astronomy and Astrophysics, Peking University, Beijing 100871, China}

\author[0000-0002-3683-7297]{Benny Trakhtenbrot}
\affil{School of Physics and Astronomy, Tel Aviv University, Tel Aviv 69978, Israel}

\begin{abstract}
The properties of slim accretion disks, while crucial for our understanding of black hole growth, have yet to be studied extensively observationally. We analyze the multi-epoch broad-band spectral energy distribution of the changing-look active galactic nucleus 1ES\,1927+654 to derive the properties of its complex, time-dependent accretion flow. The accretion rate decays as $\dot{M} \propto t^{-1.53}$, consistent with the tidal disruption of a $1.1\, M_\odot$ star. Three components contribute to the spectral energy distribution: a central overheated zone resembling a slim disk, an outer truncated thin disk, and a hot corona. Photon trapping in the slim disk triggered by the high initial $\dot{M}$ was characterized by a low radiation efficiency ($3\%$), which later more than doubled ($8\%$) after $\dot{M}$ dropped sufficiently low for the disk to transition to a geometrically thin state. The blackbody temperature profile $T \propto R^{-0.60}$ for the inner overheated zone matches the theoretical expectations of a slim disk, while the effective temperature profile of $T \propto R^{-0.69}$ for the outer zone is consistent with the predictions of a thin disk. Both profiles flatten toward the inner boundary of the disk as a result of Compton cooling in the corona. Our work presents compelling observational evidence for the existence of slim accretion disks and elucidates the key parameters governing their behavior, paving the way for further exploration in this area.
\end{abstract}


\section{Introduction}
\label{sec:sec1}

The gaseous accretion disk surrounding a supermassive black hole (BH) serves as the primary energy source in an active galactic nucleus (AGN; \citealt{Salpeter1964ApJ}). To elucidate the physical properties of accretion disks, \citet{Shakura1973AA} introduced the $\alpha$-prescription to represent viscosity. This framework led to the formulation of fundamental equations that have been supported since by numerous modern magnetohydrodynamic simulations (e.g., \citealp{Hirose2009ApJ,Fragile2018ApJ}). The resultant standard $\alpha$-disk model, later refined by \citet{Novikov1973blho} in a relativistic context, is commonly utilized in modeling accretion disk structures in most AGNs. Its popularity is largely due to the model's ability to predict temperatures that align fairly well with those inferred for the ``big blue bump'' observed in the optical-ultraviolet (UV) spectra of AGNs \citep{Shields1978,Pringle1981ARAA,Koratkar1999PASP}. The solution derived by \citet{Shakura1973AA} portrays an optically thick, geometrically thin accretion disk, a model that is applicable whenever the mass accretion rate ($\dot{M}$) falls slightly below the Eddington accretion rate, $\dot{M}_\mathrm{E} \equiv 16 L_\mathrm{E}/c^2$, where $L_{\rm E}\,=\,1.26 \times 10^{38}\, (M_{\rm BH}/M_{\odot})\,\rm erg\,s^{-1}$ is the Eddington luminosity of BH mass $M_{\rm BH}$. However, in scenarios where $\dot{M}$ drops substantially below $\dot{M}_\mathrm{E}$, the gas turns into an optically thin, advection-dominated accretion flow (ADAF: \citealp{Ichimaru1977ApJ,Narayan1994ApJ}). By contrast, when $\dot{M}$ exceeds $\dot{M}_\mathrm{E}$, the accreting gas becomes sufficiently optically thick to hinder the local radiation of all dissipated energy, and the trapped radiation gets advected inward with the accretion flow, which can be characterized aptly as a slim disk \citep{Katz1977ApJ,Abramowicz1988ApJ,Sadowski2009ApJS}. While some radiative magnetohydrodynamical simulations support the notion that photon trapping in slim disks reduces their radiative efficiency (e.g., \citealp{Ohsuga2005ApJ,Sadowski2014MNRAS}), \citet{Jiang2014ApJ} argue that magnetic buoyancy can sustain a high radiative efficiency. The presence of supermassive BHs in high-redshift ($z \gtrsim 6$) quasars (\citealp{Fan2006NewAR,Wu2015Natur,Banados2018Nature,Yang2021ApJ}) necessitates rapid early growth, underlining the importance of elucidating the nature of slim accretion disks.

\begin{deluxetable*}{ll}
\tablecaption{List of Symbols Used in This Work }\label{tab:symbols}
\tabletypesize{\scriptsize}
\tablehead{
      \colhead{Symbol}             &
      \colhead{Definition}   \\
}
\startdata
      \multicolumn{2}{l}{Observed Quantities} \\
      \hline
      $R_\mathrm{tr}$ & Inner truncation radius of the outer thin accretion disk \\
      $T_\mathrm{eff}$ & Effective temperature of the outer thin accretion disk at $R_\mathrm{tr}$ \\
      $L_\mathrm{bb}$ & Luminosity of the inner slim disk emitting blackbody radiation \\
      $T_\mathrm{bb}$ & Blackbody temperature of the slim disk radiation \\
      $\lambda_\mathrm{E}$ & Bolometric luminosity in Eddington units ($L_\mathrm{bol}/L_\mathrm{E}$) \\
      $\tau$ & Optical depth of the X-ray corona \\
      $\alpha_\mathrm{OX}$ & UV to X-ray spectral slope: $\alpha_\mathrm{OX} \equiv -\log{(L_\mathrm{X-ray}/L_\mathrm{UV})}/\log{(\nu_\mathrm{X-ray}/\nu_\mathrm{UV})}$ \\
      \hline
      \multicolumn{2}{l}{Derived Quantities}  \\
      \hline
      $\dot{M}$, $\Sigma$, $\Pi$, $H$, $T_c$, $v_R$ & Disk properties in physical units: mass accretion rate, disk surface density, vertically integrated \\
      & pressure, disk scale height, mid-plane disk temperature, and inward radial velocity, respectively. \\
      & These are derived from the observed quantities $R_\mathrm{tr}$ and $T_\mathrm{eff}$ using the equations in Section~\ref{sec:equs}. \\
      $\dot{m}$ & Mass accretion rate in Eddington units ($\dot{M}/\dot{M}_\mathrm{E}$) \\
      $\Sigma^\prime$, $T_c^\prime$ & Radially normalized disk surface density and mid-plane temperature, defined in Section~\ref{sec:Tsigma} \\
      $R_\mathrm{ph}$ & Photosphere size of the inner slim disk, derived from Equation~\ref{equ:radiusph} \\
      \hline
      \multicolumn{2}{l}{Other Quantities}  \\
      \hline
      $M_\mathrm{BH}$ & Mass of the central black hole \\
      $R_g$ & Schwarzschild radius, $R_g \equiv 2GM_\mathrm{BH}/c^2$ \\
      $R_\mathrm{in}$ & Innermost radius of the entire accretion disk \\
      $R_\mathrm{cor}$ & Size of the X-ray corona \\
      $r_\mathrm{tr}$, $r_\mathrm{cor}$ & Dimensionless truncation radius ($r_\mathrm{tr} \equiv R_\mathrm{tr}/R_g$) and corona size ($r_\mathrm{cor} \equiv R_\mathrm{cor}/R_g$) \\
\enddata
\tablecomments{Definition of symbols used in this work. The observed quantities are calculated directly from the broad-band SED analysis of \citet{Li2024paper2}. The derived quantities are calculated based on the observed quantities using the assumptions and equations listed in Sections~\ref{sec:equs} and \ref{sec:slimsize}.}
\end{deluxetable*}

Observations of systems undergoing rapid accretion offer a unique arena for testing theoretical models of slim accretion disks. Narrow-line Seyfert~1 galaxies such as I~Zwicky~1 \citep{Boroson1992ApJS, Ding2022} and RX~J0134.2$-$4258 \citep{Voges1999AA}, characterized by their distinctively narrow permitted emission lines and pronounced X-ray variability, have been suggested to host relatively low-mass BHs accreting at high rates \citep{Boroson2002,Shen2014}. The masses of their BHs, deduced through reverberation mapping of the broad-line region (BLR) clouds \citep{Blandford1982ApJ}, suggest that many are accreting at super-Eddington rates \citep{Du2014ApJ,Li2018ApJ}, thus providing excellent observational testbeds for slim disk models. Some ultraluminous X-ray sources (e.g., M82~X-1: \citealt{Pasham2014Natur}) represent an additional class of compelling super-Eddington systems. Characterized by X-ray luminosities that surpass the Eddington limit for stellar-mass BHs, ultraluminous X-ray sources have been hypothesized to be systems whose X-ray emission is beamed anisotropically, a possible outcome of the geometric effects intrinsic to slim accretion disks \citep{King2001ApJ, Roberts2007ApSS}. Lastly, dynamic, energetic sources such as tidal disruption events (TDEs; e.g., ASASSN-14li: \citealt{Holoien2016MNRAS}) and quasi-periodic eruptions (e.g., GSN~069: \citealt{Miniutti2023AA}) typically entail a dramatic surge in luminosity. In some instances, the accretion rate in these events may exceed the Eddington limit over a brief period of time, presenting a unique opportunity to investigate the transition from the standard to the slim disk regime, along with the consequent effects on the emitted spectrum \citep{Ulmer1999ApJ,Zauderer2011Natur}.

Continuum reverberation mapping, an effective tool for delineating the temperature profile of the accretion flow, has been leveraged for several AGNs presumed to be super-Eddington (e.g., Mrk~142: \citealt{Cackett2020ApJ}; Mrk~335: \citealt{Kara2023ApJ}; PG~1119+120: \citealt{Donnan2023MNRAS}). Intriguingly, the delay spectra do not support the $\tau \propto \lambda^2$ prediction from the slim accretion disk reprocessing scenario \citep{Starkey2017ApJ}. A shared trait among the three candidate super-Eddington systems is their soft X-ray spectra. Ultraluminous X-ray sources emit a distinct ``ultrasoft'' or ``broadened disk'' spectral component \citep{Gladstone2009MNRAS,Sutton2013MNRAS}. Some TDEs also exhibit extremely soft spectra that can be modeled by a hot thermal disk with a temperature of approximately $0.1\,$keV \citep{Miller2015Natur,Saxton2017AA,Hinkle2022ApJ}. As for narrow-line Seyfert~1 galaxies, they display a robust soft X-ray excess, the origin of which remains a topic of ongoing debate. Current explanations primarily include relativistically blurred ionized disk reflection \citep{Crummy2006MNRAS,Fabian2009Natur} and Compton emission from a warm corona in the inner disk \citep{Done2012MNRAS,Kawanaka2024}. Apart from these two models, \citet{Mineshige2000PASJ} propose that slim accretion disks can account for the soft X-ray excess in narrow-line Seyfert~1s, suggesting a peak temperature of $T_\mathrm{max} \approx 0.2\, (M_\mathrm{BH}/10^5 \,M_\odot)^{-1/4} \, \rm keV$. However, applying this model to real objects yields an extremely small emitting region, potentially smaller than the event horizon \citep{Mineshige2000PASJ}. To date, conclusive evidence for the existence of slim disks is still missing.

When considering the marked changes in both disk geometry and temperature profile that accompany different disk states, studying disk-state transitions—--evidenced as significant changes in the spectral energy distribution (SED)---provides a promising avenue for seeking evidence of slim disks \citep{Esin1997ApJ}. In the context of disk-state transition scenarios, changes in broad-band color have been employed to explain the ``changing-look'' phenomenon observed in several AGNs (e.g., \citealp{Noda2018MNRAS,Ruan2019ApJ}; see \citealp{Ricci2023NatAs} for a recent review). A potentially instructive example is the narrow-line Seyfert~1 galaxy RE~J1034+396, which exhibits a pronounced soft X-ray excess and a detectable X-ray quasi-periodic oscillation, along with a varying power law. \cite{Middleton2009MNRAS} posit that super-Eddington accretion rates may instigate changes in disk structure, and concurrently trigger the observed X-ray quasi-periodic oscillation. This hypothesis implicitly serves as evidence for the existence of slim disks in AGNs.

1ES\,1927+654, a previously classified ``true'' type~2 AGN (namely, an object that intrinsically lacks a BLR), underwent an outburst in March 2018 that manifested as an increase in $V$-band flux by at least 2 magnitudes \citep{Boller2003AA,Tran2011ApJ,Nicholls2018ATel}. Subsequent optical and X-ray spectroscopic follow-up observations over a three-year period revealed the emergence of broad Balmer line emission several weeks after the outburst, alongside dramatic spectral evolution in the X-rays \citep{Trakhtenbrot2019ApJ,Ricci2020ApJL,Ricci2021ApJS,Laha2022ApJ}. \citet{Li2022paper1} used the host galaxy's properties to estimate a BH mass of $M_\mathrm{BH} = 1.38_{-0.66}^{+1.25}\times 10^6 \,M_\odot$, thus establishing that the luminosity of the source after the outburst exceeded the Eddington limit (see also \citealt{Masterson2022} and \citealt{Li2024paper2}). \citet{Li2022paper1} performed detailed decomposition of the pre-outburst optical spectrum from \citet{Boller2003AA} along with 34 follow-up optical spectra from our monitoring program, revealing changes in both the luminosity and shape of the continuum and broad emission lines. Upon adjusting for the host galaxy's contribution, broad-band decomposition of the simultaneous SED \citep{Li2024paper2} reveals a $\Lambda$-shape on the plot of disk luminosity versus color, indicative of a potential disk-state transition during post-outburst evolution. This paper delves into the properties of the accretion flow, particularly seeking evidence for a state transition between a slim and a thin disk. We also scrutinize the attributes of the slim disk, demonstrating, to our knowledge for the first time, quantitative agreement with analytical models \citep{Abramowicz1988ApJ}.

The paper is structured as follows. Section~\ref{sec:sec2} elucidates the methodology applied to derive the fundamental physical quantities of the accretion flow and its mass accretion rate evolution. Section~\ref{sec:sec3} provides evidence of a transient slim accretion disk and characterizes its radiation properties, highlighting the potent soft X-ray emission from the inner overheated region. Section~\ref{sec:sec4} discusses the global SED anticipated from the accretion flow, based on a detailed analysis of the disk's temperature profile. Section~\ref{sec:sec5} explores the size of the slim disk and the corona region, and their implications for other disks in AGNs. The paper concludes in Section~\ref{sec:sec6}.

\begin{figure*}
\centering
\includegraphics[width=0.82\textwidth]{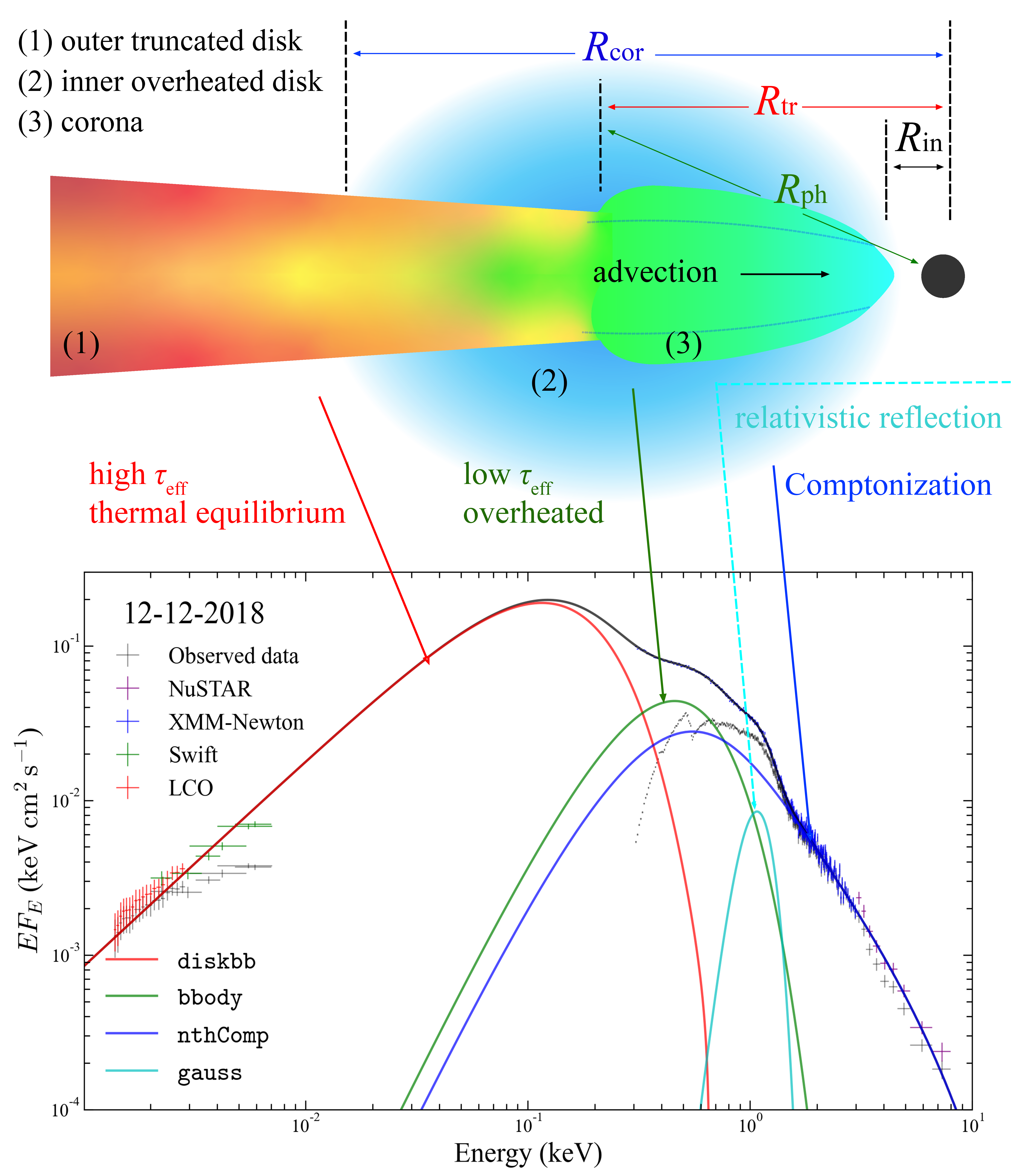}
\caption{Schematic illustration of the accretion flow following the optical outburst of 1ES\,1927+654, using the example SED from 12 December 2018. The upper panel gives our interpretation of the super-Eddington accretion flow, which consists of three distinct regimes: (1) an outer truncated disk in thermal equilibrium, depicted as a thin red disk with a vertical color gradient and truncated at radius $R_\mathrm{tr}$; (2) an inner overheated disk where advection becomes significant, represented in green without a vertical color gradient, whose photosphere size is denoted by $R_\mathrm{ph}$ that extends to the inner edge $R_\mathrm{in}$ from the BH; and (3) an X-ray corona, depicted as the blue component above the inner part of the disk, with radial size $R_\mathrm{cor}$. The three components contribute to the main observed SED, as indicated by the red, green, and blue arrows pointing to the spectral components in the lower panel. The outer thin disk is optically thick with a high effective optical depth $\tau_\mathrm{eff}$ and emits multicolor blackbody radiation, manifesting as the big blue bump. The inner disk, with a low $\tau_\mathrm{eff}$, is overheated due to strong advection and electron scattering effects, and it emits blackbody radiation at soft X-ray energies with a temperature determined by its photosphere. The corona upscatters the seed photons from the disk and contributes to the Comptonized emission in the hard X-ray band. In addition, relativistic reflection from the inner overheated disk can result in a broad 1\,keV feature, which can be phenomenologically modeled with a broad Gaussian component \citep{Masterson2022}. These four components effectively describe the observed SED, as demonstrated in the lower panel. The error bars display the simultaneous data as observed (grey) and as corrected for reddening and absorption from the Las Cumbres Observatory (red), Swift (green), XMM-Newton (blue), and NuSTAR (purple). The intrinsic SED model (black curve) comprises a multicolor thin disk (red), a blackbody (green), Comptonization emission (blue), and a phenomenological Gaussian feature (cyan), which are modeled with {\tt diskbb}, {\tt bbody}, {\tt nthComp}, and {\tt gauss} in \textsc{xspec}.}
\label{fig:disksed}
\end{figure*}

\section{Properties of the Accretion Flow}
\label{sec:sec2}

The thermal emission from the accretion disk contributes to the big blue bump (e.g., \citealp{Czerny1987ApJ}), peaking in the extreme-UV wavelength range. \citet{Li2024paper2} employ the {\tt diskbb} component in \textsc{xspec} to model the optical/UV emission in the broad-band SED of 1ES\,1927+654, interpolating between the UV photometric and soft X-ray observation. The model presumes an effective temperature profile consistent with a steady-state thin disk \citep{Mitsuda1984PASJ}, wherein the mass accretion rate remains constant with radius. Considering that the optical outburst of 1ES\,1927+654 highly likely was triggered by a TDE \citep{Trakhtenbrot2019ApJ,Ricci2020ApJL,Ricci2021ApJS,Li2022paper1}, its disk is not in a steady state. Given that the half-light radius of the accretion disk is usually estimated to be $R_e \lesssim 30\,R_g$ \citep{Zdziarski2022ApJ}, with $R_g \equiv 2GM_\mathrm{BH}/c^2$ the Schwarzschild radius, the viscous timescale within $R_e$ is $\sim 5$ hours. This estimate, based on the inward radial velocity calculated in Section~\ref{sec:mdot} (Table~\ref{tab:diskpro}), is significantly shorter than our observation duration of three years. Therefore, the assumption of a steady disk remains a sound approximation for the case study of 1ES\,1927+654. Additionally, the relatively low BH mass of $\sim 10^6\, M_\odot$ implies that the maximum temperature of the disk can reach $\sim 50\,$eV, which permits the disk to emit a flux comparable to the strong soft X-ray excess observed at 0.3\,keV (see Figure~\ref{fig:disksed}). Consequently, we can use SED modeling of the optical-UV and soft X-ray portions of the intrinsic SED, as described in \citet{Li2024paper2}, to constrain the inner truncation radius ($R_\mathrm{tr}$) and the effective temperature ($T_\mathrm{eff}$) at $R_\mathrm{tr}$. The symbols of the observed quantities derived from SED fitting are summarized in Table~\ref{tab:symbols}.

The BH mass, $M_\mathrm{BH} = 1.38_{-0.66}^{+1.25} \times 10^6\, M_\odot$ estimated from the stellar mass of the host galaxy bulge \citep{Li2022paper1}, implies $R_\mathrm{tr} \gtrsim 2.5\,R_g$ \citep{Li2024paper2}. This value will be used consistently throughout this work. In the early epochs, $R_\mathrm{tr} \approx 20\,R_g$, which implies that the thermalized disk has an inner edge that is quite distant from the BH (Figure~\ref{fig:radius}). We propose that the physical reason for the inner truncation is the low effective optical depth ($\tau_\mathrm{eff}$), which arises from the strong advection and electron scattering in the inner region of the disk (Section~\ref{sec:sec3}). Physically, the truncation radius is initially large not because the inner disk has disappeared; instead, the inner disk has overheated, causing it to emit in the soft X-ray ($0.3-1\,$keV) band. This corresponds to the blackbody component of the SED in \citet{Li2024paper2} and Figure~\ref{fig:disksed}. The blackbody soft X-ray excess, when combined with the hard X-ray Comptonized emission from the corona and a broad Gaussian component from relativistic reflection of the inner overheated disk, constitutes the total X-ray spectrum of 1ES\,1927+654 \citep{Ricci2020ApJL,Ricci2021ApJS,Masterson2022}.

Nevertheless, \citet{Li2024paper2} derive $R_\mathrm{tr}$ generally smaller than $20\,R_g$. The steady-state assumption still holds on account of the short viscous timescale at both $R_\mathrm{tr}$ and $R_e$, such that we can ascertain the disk properties locally at $R_\mathrm{tr}$ using the $\alpha$-prescription (e.g., \citealp{Shakura1973AA}). After first formulating the basic set of non-relativistic equations (Section~\ref{sec:equs}), we derive the local disk properties at $R_\mathrm{tr}$ and the evolution of the mass accretion rate (Section~\ref{sec:mdot}) and discuss the radiative efficiency (Section~\ref{sec:eta}). We find strong evidence for the existence of a slim disk at early epochs, when advection dominated the total cooling, but the accretion flow finally transformed back into a thin disk (Section~\ref{sec:scurve}).

\bigskip
\subsection{Basic Equations}
\label{sec:equs}

Under conditions of local thermal equilibrium \citep{Kato2008book}, we anticipate a radiative cooling rate

\begin{equation}
    Q_\mathrm{rad}^{-} = 2\sigma_{\rm SB} T_\mathrm{eff}^4,
\label{equ:tdobs}
\end{equation}

\noindent
where the factor of 2 represents the two sides of the disk and $\sigma_{\rm SB}$ is the Stefan-Boltzmann constant. Since the angular velocity of rotating matter is close to Keplerian ($\Delta \Omega/\Omega_{\rm K} \lesssim 0.04$) for supermassive BHs with $\dot{m} \lesssim 10$ \citep{Beloborodov1998MNRAS,Peng2021MNRAS}, where $\dot{m} \equiv \dot{M}/\dot{M}_\mathrm{E}$, the viscous heating rate can be expressed as

\begin{equation}
    Q_\mathrm{vis}^{+} = \frac{3GM_\mathrm{BH}\dot{M}}{4\pi R^3}\left(1-\sqrt{\frac{R_\mathrm{in}}{R}}\right),
\label{equ:qvisp}
\end{equation}

\noindent
with $G$ the gravitational constant, $M_\mathrm{BH}$ the mass of the central BH, and $R_\mathrm{in}$ the innermost radius of the entire accretion disk. At $R_\mathrm{tr}$, where the thin disk truncates, advection is significant for energy balance. Following \citet{Kato2008book},

\begin{equation}
    Q_\mathrm{adv}^{-} = \frac{\xi }{2\pi }\frac{\Pi \dot{M}}{\Sigma R^2},
\label{equ:qadv}
\end{equation}

\noindent
where $\Pi$ and $\Sigma$ are the vertically integrated pressure and density, respectively, as the disk may no longer be thin near $R_\mathrm{tr}$. Here radiation also dominates the total pressure. The parameter $\xi$, a quantity of $\mathcal{O}(1)$, is related with the local physical conditions \citep{Kato2008book}. The disk structure at $R_\mathrm{tr}$ is then given by energy conservation,

\begin{equation}
    Q_\mathrm{vis}^{+} = Q_\mathrm{rad}^{-} + Q_\mathrm{adv}^{-},
\label{equ:Econs}
\end{equation}

\noindent
mass conservation,

\begin{equation}
    -2\pi R \Sigma v_R = \dot{M},
\end{equation}

\noindent
and the equation of state,

\begin{equation}
\Pi = \frac{k_\mathrm{B}I_4}{\Bar{\mu}m_\mathrm{H}I_3}\Sigma T_c + \frac{2a}{3}I_4 T_c^4 H,
\label{equ:EoS}
\end{equation}

\noindent
where $I_3 = 0.457$ and $I_4 = 0.406$ are the vertical integration constants for an assumed $N = 3$ polytropic gas \citep{Kato2008book}, $k_\mathrm{B}$ is Boltzmann's constant, $\Bar{\mu} = 0.6$ is the mean particle weight, $m_\mathrm{H}$ is the mass of hydrogen, the radiation constant $a = 4\sigma_{\rm SB}/c$, $T_c$ is the temperature at the disk mid-plane, and $H$ is the disk scale height. Adopting the $\alpha$ prescription for the viscosity tensor $T_{R\phi}=-\alpha \Pi$ \citep{Shakura1973AA}, the equation of angular momentum conservation is given by

\begin{equation}
    \dot{M}(\sqrt{GMR}-\sqrt{GMR_\mathrm{in}}) = 2\pi \alpha \Pi R^2.
\label{equ:Lcons}
\end{equation}


\startlongtable
\begin{deluxetable*}{ccccccccc}
\tablecaption{Local Disk Properties at the Truncation Radius \label{tab:diskpro}}
\tablehead{
      \colhead{Number}   &
      \colhead{Date}   &
      \colhead{$\log{R_\mathrm{tr}}$} &
      \colhead{$\log{\dot{M}}$} &
      \colhead{$\log{\Sigma}$ }   &
      \colhead{$\log{\Pi}$ } &
      \colhead{$\log{H}$ } &
      \colhead{$\log{T_c}$ } &
      \colhead{$\log{v_R}$ } \\
        &
      (DD-MM-YYYY)  &
      \colhead{($\rm cm$) } &
      \colhead{($\rm g\,s^{-1}$)} &
      \colhead{($\rm g\,cm^{-2}$) }   &
      \colhead{($\rm g\,cm^{-1}\,s^{-1}$) } &
      \colhead{($\rm cm$) } &
      \colhead{($\rm K$) } &
      \colhead{($\rm km \, s^{-1}$) } \\
\colhead{(1)} &
\colhead{(2)} &
\colhead{(3)} &
\colhead{(4)} &
\colhead{(5)} &
\colhead{(6)} &
\colhead{(7)} &
\colhead{(8)} &
\colhead{(9)}
}
\startdata
  0 &  $20-05-2011$ & $11.99 \pm 0.04$ & $24.19 \pm 0.29$ & $21.88 \pm 0.13$ & $3.30 \pm 0.19$ & $11.62 \pm 0.26$ & $6.24 \pm 0.13$ & $3.10 \pm 0.29$  \\
  1 &  $17-05-2018$ & $12.89 \pm 0.03$ & $25.47 \pm 0.68$ & $22.32 \pm 0.66$ & $2.94 \pm 0.66$ & $13.38 \pm 0.26$ & $5.91 \pm 0.19$ & $3.84 \pm 0.68$  \\
  2 &  $31-05-2018$ & $12.89 \pm 0.03$ & $25.46 \pm 0.68$ & $22.32 \pm 0.66$ & $2.94 \pm 0.66$ & $13.36 \pm 0.26$ & $5.91 \pm 0.19$ & $3.83 \pm 0.68$  \\
  3 &  $05-06-2018$ & $12.88 \pm 0.03$ & $25.43 \pm 0.68$ & $22.31 \pm 0.65$ & $2.94 \pm 0.66$ & $13.34 \pm 0.26$ & $5.91 \pm 0.19$ & $3.81 \pm 0.68$  \\
  4 &  $14-06-2018$ & $12.90 \pm 0.03$ & $25.49 \pm 0.68$ & $22.34 \pm 0.66$ & $2.93 \pm 0.66$ & $13.40 \pm 0.26$ & $5.91 \pm 0.19$ & $3.87 \pm 0.68$  \\
  5 &  $10-07-2018$ & $12.83 \pm 0.03$ & $25.27 \pm 0.68$ & $22.21 \pm 0.66$ & $3.01 \pm 0.66$ & $13.19 \pm 0.26$ & $5.93 \pm 0.19$ & $3.64 \pm 0.68$  \\
  6 &  $24-07-2018$ & $12.81 \pm 0.03$ & $25.21 \pm 0.68$ & $22.17 \pm 0.66$ & $3.04 \pm 0.66$ & $13.12 \pm 0.26$ & $5.94 \pm 0.19$ & $3.56 \pm 0.68$  \\
  7 &  $07-08-2018$ & $12.80 \pm 0.03$ & $25.20 \pm 0.68$ & $22.17 \pm 0.66$ & $3.04 \pm 0.66$ & $13.12 \pm 0.26$ & $5.94 \pm 0.19$ & $3.56 \pm 0.68$  \\
  8 &  $23-08-2018$ & $12.71 \pm 0.08$ & $25.13 \pm 0.87$ & $22.22 \pm 0.66$ & $3.00 \pm 0.72$ & $13.02 \pm 0.60$ & $5.97 \pm 0.33$ & $3.63 \pm 0.87$  \\
  9 &  $03-10-2018$ & $12.65 \pm 0.06$ & $24.90 \pm 0.58$ & $22.05 \pm 0.42$ & $3.14 \pm 0.47$ & $12.78 \pm 0.41$ & $5.99 \pm 0.22$ & $3.31 \pm 0.58$  \\
 10 &  $19-10-2018$ & $12.65 \pm 0.09$ & $25.04 \pm 0.89$ & $22.20 \pm 0.64$ & $3.01 \pm 0.71$ & $12.92 \pm 0.63$ & $5.99 \pm 0.35$ & $3.59 \pm 0.89$  \\
 11 &  $23-10-2018$ & $12.66 \pm 0.08$ & $25.03 \pm 0.82$ & $22.18 \pm 0.60$ & $3.03 \pm 0.66$ & $12.91 \pm 0.58$ & $5.99 \pm 0.32$ & $3.54 \pm 0.82$  \\
 12 &  $21-11-2018$ & $12.61 \pm 0.10$ & $24.95 \pm 1.01$ & $22.16 \pm 0.74$ & $3.04 \pm 0.82$ & $12.83 \pm 0.70$ & $6.01 \pm 0.39$ & $3.50 \pm 1.01$  \\
 13 &  $06-12-2018$ & $12.61 \pm 0.10$ & $24.99 \pm 0.98$ & $22.20 \pm 0.71$ & $3.01 \pm 0.79$ & $12.85 \pm 0.70$ & $6.01 \pm 0.38$ & $3.57 \pm 0.98$  \\
 14 &  $12-12-2018$ & $12.43 \pm 0.02$ & $25.07 \pm 0.17$ & $22.48 \pm 0.11$ & $2.78 \pm 0.13$ & $12.84 \pm 0.12$ & $6.08 \pm 0.07$ & $4.06 \pm 0.17$  \\
 15 &  $28-03-2019$ & $12.37 \pm 0.02$ & $24.90 \pm 0.42$ & $22.38 \pm 0.40$ & $2.85 \pm 0.40$ & $12.67 \pm 0.16$ & $6.10 \pm 0.12$ & $3.89 \pm 0.42$  \\
 16 &  $07-05-2019$ & $12.22 \pm 0.04$ & $24.70 \pm 0.37$ & $22.31 \pm 0.26$ & $2.90 \pm 0.29$ & $12.39 \pm 0.27$ & $6.15 \pm 0.15$ & $3.78 \pm 0.37$  \\
 17 &  $02-11-2019$ & $12.03 \pm 0.03$ & $24.89 \pm 0.20$ & $22.59 \pm 0.11$ & $2.69 \pm 0.14$ & $12.33 \pm 0.18$ & $6.24 \pm 0.09$ & $4.38 \pm 0.20$  \\
 18 &  $03-05-2020$ & $12.06 \pm 0.03$ & $24.62 \pm 0.30$ & $22.32 \pm 0.19$ & $2.89 \pm 0.22$ & $12.15 \pm 0.23$ & $6.22 \pm 0.12$ & $3.88 \pm 0.30$  \\
 19 &  $16-09-2020$ & $12.08 \pm 0.03$ & $24.48 \pm 0.27$ & $22.17 \pm 0.18$ & $3.02 \pm 0.21$ & $12.04 \pm 0.21$ & $6.21 \pm 0.11$ & $3.58 \pm 0.27$  \\
 20 &  $12-01-2021$ & $12.04 \pm 0.03$ & $24.22 \pm 0.24$ & $21.92 \pm 0.11$ & $3.26 \pm 0.16$ & $11.74 \pm 0.22$ & $6.22 \pm 0.11$ & $3.13 \pm 0.24$  \\
     \enddata
      \tablecomments{Local disk properties at the truncation radius ($R_\mathrm{tr}$), derived by the equation system in Section~\ref{sec:equs}. Col. (1): Number of the broad-band datasets; see details in \citet{Li2024paper2}. Col. (2): Date of the observations. Col. (3): The inner truncation radius of the thin accretion disk derived from SED analysis in \citet{Li2024paper2}. Col. (4): Mass accretion rate of the BH. Col. (5): Disk gas surface density. Col. (6): Vertically integrated total pressure. Col. (7): Disk scale height. Col (8): The temperature at the disk mid-plane. Col (9): Inward radial velocity of the gas.
       }
       \end{deluxetable*}

\noindent
In the vertical direction, hydrodynamical equilibrium yields

\begin{equation}
    \Omega_{\rm K}^2 H^2 = 9\frac{\Pi}{\Sigma},
\label{equ:vertical}
\end{equation}

\noindent
where the numerical factor of 9 comes from vertical integration (e.g., \citealp{Sadowski2011}). Radiative cooling follows

\begin{equation}
    Q_\mathrm{rad}^{-} = \frac{8acT_c^4}{3\kappa_\mathrm{R} \Sigma},
\label{equ:flux}
\end{equation}

\noindent
where the Rosseland mean opacity coefficient $\kappa_\mathrm{R}$ can be expressed using Kramer’s formula,

\begin{equation}
\begin{aligned}
     \kappa_\mathrm{R} &= \kappa_\mathrm{es} + \kappa_\mathrm{ff} \\
     &= 0.34 + (0.64\times 10^{23})\frac{\Sigma}{2H}\left(\frac{2T_c}{3}\right)^{-7/2} \;\rm cm^2\,g^{-1},
\label{equ:opacity}
\end{aligned}
\end{equation}

\noindent
with $\kappa_\mathrm{es}$ and $\kappa_\mathrm{ff}$ the electron scattering and free-free opacity coefficients, respectively.

\subsection{Accretion Rate}
\label{sec:mdot}

Since $R_\mathrm{tr}$ and $T_\mathrm{eff}(R_\mathrm{tr})$ are directly derivable from our observations, $Q_\mathrm{rad}^{-}$ can be calculated using Equation~\ref{equ:tdobs}. In principle, the value of $\xi$ in Equation~\ref{equ:qadv} depends on both the pressure and surface density gradient, but this is challenging to estimate without resolving the global structure, especially because the disk is non-steady on large scales. Instead, throughout our analysis we fix it to a typical value of $\xi = 0.1$ \citep{Kato2008book}. We set the viscosity parameter to $\alpha = 0.1$, a fiducial value obtained from radiation magnetohydrodynamical simulations (e.g., \citealp{Hirose2009ApJ,Jiang2019aApJ}). The inner disk radius is fixed to $R_{\rm in} = 1.45\,R_g$, which corresponds to the size of the innermost stable circular orbit for a BH with spin parameter $a_\ast \approx 0.8$ \citep{Li2024paper2}. At each epoch, the quantities $\dot{M}$, $\Pi$, $\Sigma$, $H$, and $T_c$ can be solved analytically (Table~\ref{tab:diskpro}) using Equations \ref{equ:Econs}, \ref{equ:EoS}, \ref{equ:Lcons}, \ref{equ:vertical}, and \ref{equ:flux}\footnote{The choices of $\xi$, $\alpha$, and $R_\mathrm{in}$ impact the exact values of these five physical quantities, but slight changes in the normalization were tested and found not to affect any of the qualitative results presented in this paper (Appendix~\ref{app:test}).}. The parameter errors are then calculated through standard error propagation.

\begin{figure}
\centering
\includegraphics[width=0.45\textwidth]{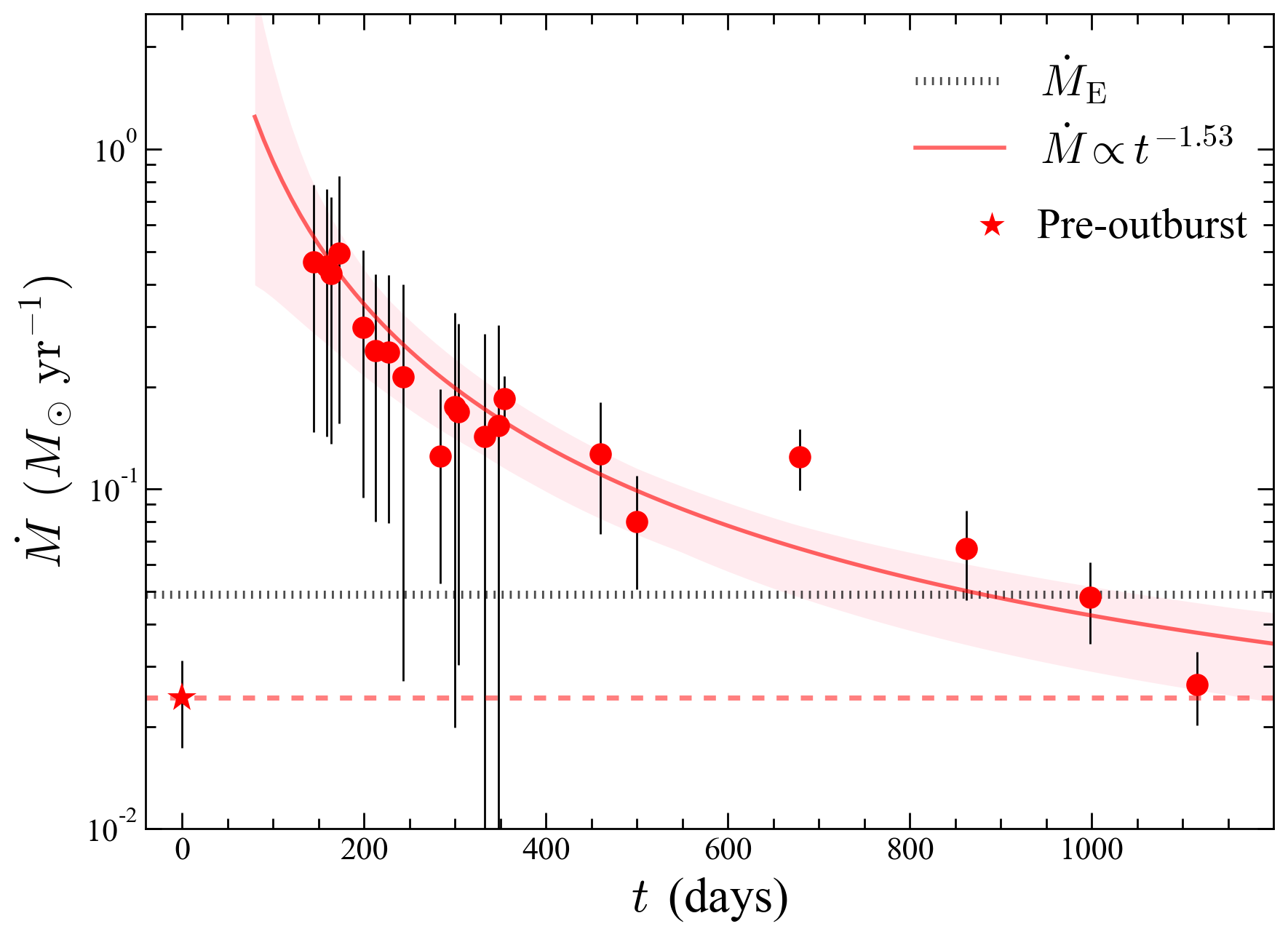}
\caption{Evolution of BH mass accretion rate as a function of time since the optical outburst on 23 December 2017 ($t$), calculated from the equation set in Section~\ref{sec:equs}. The trend is modeled with a power-law decline expressed as $\dot{M} \propto t^{-1.53}$ (Equation~\ref{equ:mdot} and Table~\ref{tab:mdot}), whose best-fit model and $1\,\sigma$ uncertainty are shown by the red curve and red shaded region. The red star and red dashed line mark the pre-outburst value calculated from the 2011 observation in \citet{Li2024paper2}. The dotted line marks the Eddington accretion rate for $M_\mathrm{BH} = 1.38_{-0.66}^{+1.25}\times 10^6 \,M_\odot$ \citep{Li2022paper1}.}
\label{fig:mdot}
\end{figure}

We parameterize the evolution of $\dot{M}$ (Figure~\ref{fig:mdot}) as

\begin{equation}
    \dot{M} = \dot{M}_{p}\left(\frac{t-t_0}{t_p-t_0}\right)^{-\gamma} + \dot{M}_0,
\label{equ:mdot}
\end{equation}

\noindent
where $t$ is defined as the time since the optical outburst on 23 December 2017, $\dot{M}_{p}$ is the peak mass accretion rate of the disk at time $t_p$, $t_0$ is the starting time of the optical outburst, $\gamma$ is the decline index, and $\dot{M}_0$ is the original mass accretion rate, since 1ES\,1927+654 was already an AGN before the optical outburst. To avoid potential parameter degeneracy between $t_p$ and $\dot{M}_{p}$, we fix $t_p = 80$ days since the time of the optical outburst on 23 December 2017, based on the $o$-band lightcurve of the Asteroid Terrestrial-impact Last Alert System \citep{Trakhtenbrot2019ApJ}. Using the nested-sampling method in the Python package {\tt dynesty} to sample the parameter posterior distribution, we obtain the best-fit values of $\dot{M}_{p}$, $t_0$, $\gamma$, and $\dot{M}_0$ (Table~\ref{tab:mdot}), whose uncertainties are calculated from the 16\% and 84\% values. The original mass accretion rate of $\dot{M}_0 = 1.20_{-0.13}^{+0.23} \times 10^{-2} \, M_\odot \rm  \, yr^{-1}$ is lower than that inferred from the pre-outburst observation in May 2011, $\dot{M} = (2.43\pm 0.70) \times 10^{-2}\, M_\odot \,\rm  yr^{-1}$ \citep{Li2024paper2}. The decline index $\gamma = 1.53_{-0.10}^{+0.10}$ is roughly consistent with the theoretical mass fallback rate of TDEs ($\gamma = 5/3$; \citealp{Rees1988Natur,Lodato2011MNRAS}), but steeper than that of the UV photometric light curve ($\gamma = 0.91\pm0.4$; \citealp{Laha2022ApJ}). The overall accreted mass budget of the transient event, calculated by integrating Equation~\ref{equ:mdot}, yields $\Delta M \simeq \dot{M}_{p}(t_p-t_0)/(\gamma -1) = 0.55_{-0.04}^{+0.05}\,M_\odot$. For a TDE scenario, we expect half of the disrupted star to have been consumed, yielding a mass of $M_\ast \simeq 2\Delta M = 1.1 \,M_\odot$ for the disrupted star. The semi-major axis of the most-bound orbit then can be expressed as \citep{Dai2015ApJ}

\begin{equation}
    a_\mathrm{mb} \simeq \left(\frac{M_\mathrm{BH}}{M_\ast}\right)^{2/3}R_\ast \simeq 1984\, \left(\frac{R_\ast}{R_\odot}\right)\,R_g.
\label{equ:rmb}
\end{equation}

\noindent
Interestingly, for a Sun-like star, $a_\mathrm{mb}$ is consistent to within a factor of 2 with the pericenter distance of the BLR clouds, which \cite{Li2022paper1} estimate as $a_\mathrm{BLR}(1-e) \approx 3400\, R_g$. This is in agreement with the proposal of \citet{Li2022paper1} that the BLR clouds may have originated from the outer part of the TDE disk.

\begin{deluxetable}{ccc}
\tablecaption{Parameters of $\dot{M}$ Evolution} \label{tab:mdot}
\tabletypesize{\scriptsize}
\tablehead{
      \colhead{Parameter}             &
      \colhead{Value}   &
      \colhead{Unit} \\
      \colhead{(1)} &
\colhead{(2)} &
\colhead{(3)}
}
\startdata
      $\dot{M}_p$ & $1.23_{-0.18}^{+0.21}$ &  $ M_\odot \,\rm yr^{-1}$ \\
      $t_0$ & $-10.7_{-13.0}^{+19.0}$ &  day \\
      $t_p$ & 80 &  day \\
      $\gamma$ & $1.53_{-0.10}^{+0.10}$& \\
      $\dot{M}_0$ & $1.20_{-0.13}^{+0.23}\times 10^{-2}$ &  $ M_\odot \,\rm yr^{-1}$ 
     \enddata
      \tablecomments{The best-fit parameters describing the evolution of $\dot{M}$, as formulated in Equation~\ref{equ:mdot}. Col. (1): Name of the parameter: $\dot{M}_\mathrm{p}$, peak mass accretion rate of the disk; $t_0$, time of the TDE event relative to the optical outburst on 23 December 2017; $t_p$, time when $\dot{M}$ peaked, relative to the optical outburst date; $\gamma$, decline index; and $\dot{M}_0$, steady mass accretion rate. Col (2): Best-fit parameter values and their lower and upper uncertainties. Col. (3): Units.
       }
      \end{deluxetable}

\subsection{Radiation Efficiency}
\label{sec:eta}

\begin{figure}
\centering
\includegraphics[width=0.48\textwidth]{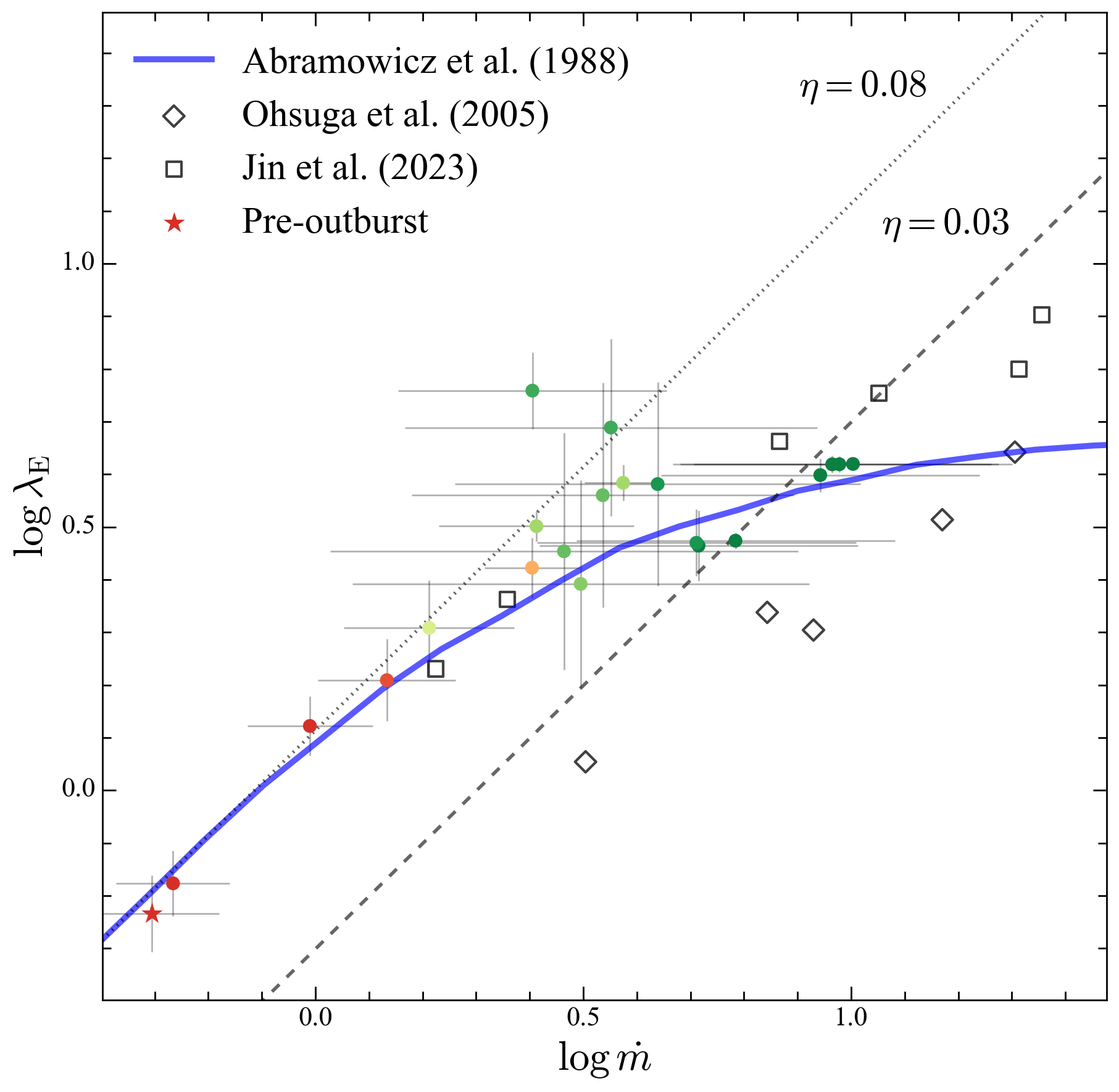}
\caption{The Eddington ratio ($\lambda_\mathrm{E}$) as a function of the dimensionless mass accretion rate ($\dot{m}$), color-coded by the time since the outburst (23 December 2017), from the beginning (green) to the end (red),  where the orange one denotes $t\approx 650$ days.. The red star shows the value in May 2011, before the outburst. The dotted line marks the $\lambda_\mathrm{E}-\dot{m}$ relation with constant radiation efficiency $\eta = 0.08$, while the dashed line denotes $\eta = 0.03$. The blue solid line shows the $\lambda_\mathrm{E}-\dot{m}$ relation for a typical slim accretion disk \citep{Abramowicz1988ApJ}. The open gray squares represent the six extreme narrow-line Seyfert~1 galaxies in \citet{Jin2023MNRAS}, and the open diamonds are derived from the super-Eddington accretion flows in the two-dimensional radiation-hydrodynamic simulations of \cite{Ohsuga2005ApJ}.
}
\label{fig:ptrap}
\end{figure}

The radiation efficiency, defined as $\eta \equiv L_\mathrm{bol}/\dot{M}c^2$, where $L_\mathrm{bol}$ is the bolometric luminosity, characterizes the efficiency with which gravitational energy is converted into radiation. By comparing the cumulative mass density of supermassive BHs with the integrated energy density released by AGNs, the median radiation efficiency is deduced to be $\eta \approx 0.1$ \citep{Soltan1982MNRAS,Yu2002MNRAS}. Apart from the last two epochs ($t \gtrsim 1000$\,days), the accretion in 1ES\,1927+654 appears to be super-Eddington (Figure~\ref{fig:mdot}). When the accretion rate approaches or exceeds the Eddington limit, $Q_\mathrm{adv}^-$ increases rapidly and eventually dominates the cooling process. Consequently, some photons are directly advected into the BH, resulting in a reduction of radiation efficiency, a phenomenon known as photon-trapping (e.g., \citealp{Abramowicz1988ApJ}). We calculate the dimensionless disk luminosity, $\lambda_\mathrm{E} \equiv L_\mathrm{bol}/L_\mathrm{E}$, where $L_\mathrm{bol}$ is obtained by summing over all three components of the SED described in \cite{Li2024paper2}. Plotting $\lambda_\mathrm{E}$ as a function of $\dot{m}$ (Figure~\ref{fig:ptrap}) reveals that initially the disk was radiatively inefficient, with $\eta \approx 0.03$, but as time transpired and $\dot{m}$ decreased, $\eta$ rose to $\sim 0.08$, exceeding the efficiency of a non-spinning standard accretion disk ($\eta = 0.057$). An efficiency of $\eta \approx 0.08$ corresponds to a relativistic thin disk with $a_\ast \approx 0.5$ \citep{Novikov1973blho}, less than the value of $a_\ast \approx 0.8$ estimated from \citet{Li2024paper2}, for which we anticipate $\eta \approx 0.12$. Given that the disk was not entirely uniformly thin at the end of our monitoring campaign, the slightly lower value of $\eta$ is consistent, within the uncertainties, with our expectations.

To gain deeper insight into the low radiation efficiency during the early epochs, we overplot as a blue curve in Figure~\ref{fig:ptrap} the $\lambda_\mathrm{E}-\dot{m}$ relationship for a slim disk \citep{Abramowicz1988ApJ}\footnote{The original $\lambda_\mathrm{E}-\dot{m}$ curve in \citet{Abramowicz1988ApJ} was derived for non-spinning BHs. We shifted the abscissa of their Figure~1 by $-0.15$ to align with the spin estimate of 1ES\,1927+654 ($a_\ast \approx 0.8$; \citealt{Li2024paper2}).}. Remarkably, the theoretical curve accurately describes most of our observations. Observations with $\dot{m} \gtrsim 2$, or those occurring before $t \approx 650$ days, deviate significantly from the constant $\eta = 0.08$ line, strongly favoring the logarithmic $\lambda_\mathrm{E}-\dot{m}$ dependence predicted for a slim disk. Interestingly, the radiation efficiency at $\dot{m} \approx 10$ falls between the value found in nearby narrow-line Seyfert~1 galaxies, for which $\dot{m}$ is derived from the optical-UV emission originating from the outer regions of the disk \citep{Jin2023MNRAS}, and the value from two-dimensional radiation-hydrodynamic simulations of super-Eddington accretion flows \citep{Ohsuga2005ApJ}. In the case of 1ES\,1927+654, whose $\dot{m}$ is estimated using the properties of its inner disk, the slightly lower $\eta$ compared to \citet{Jin2022ApJ} may suggest that $\dot{m}$ from the inner region better reflects the effects of advection cooling. Conversely, the higher $\eta$ in comparison to \citet{Ohsuga2005ApJ} might imply that outflows may not be as efficient in reducing the radiation efficiency as they appear in simulations.

\begin{figure}
\centering
\includegraphics[width=0.48\textwidth]{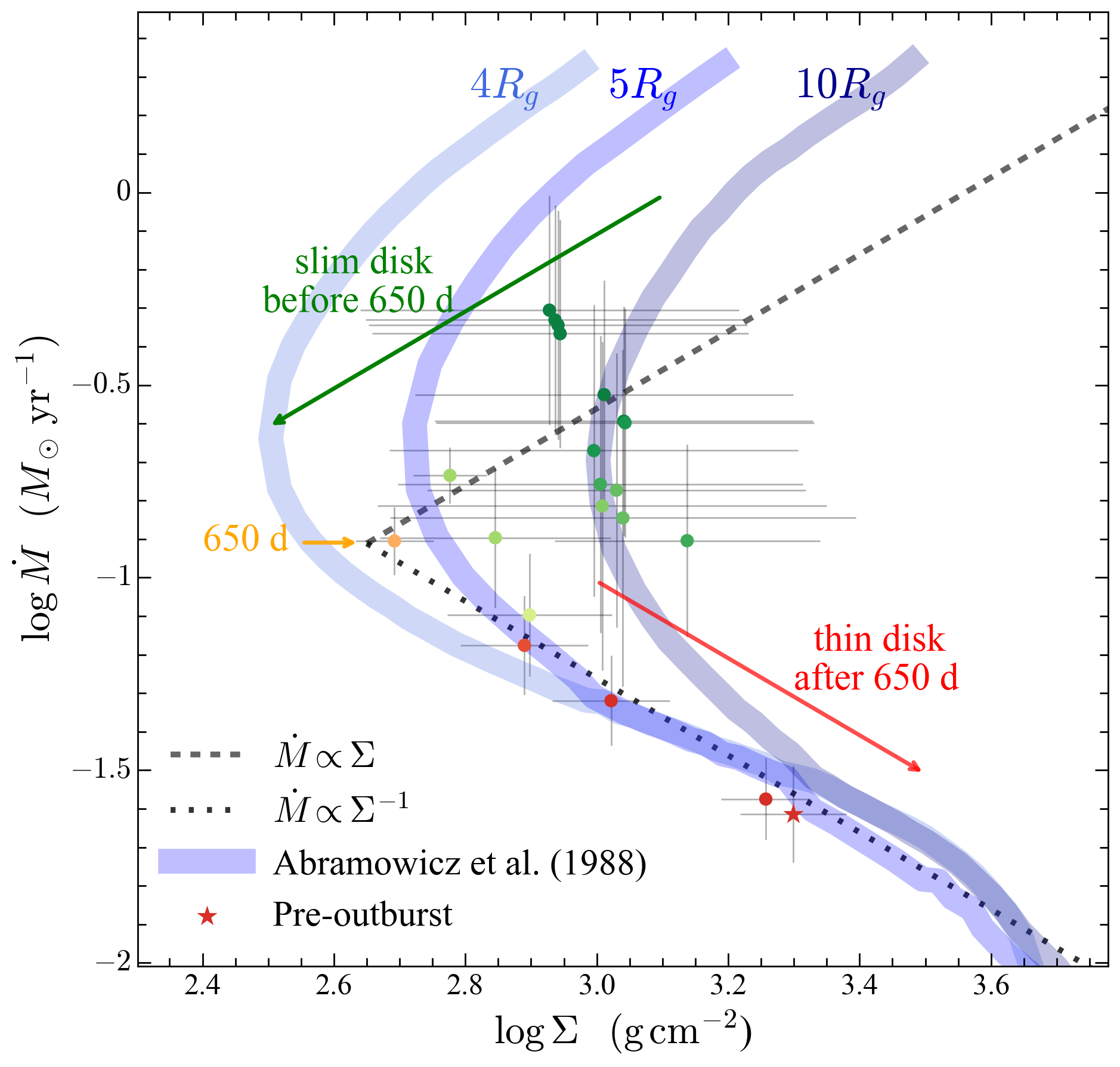}
\caption{The dimensionless mass accretion rate ($\dot{M}$) as a function of the disk mass surface density ($\Sigma$), designated as the $\dot{M}-\Sigma$ ``$S$-curve''. The color of the circles represents the time since the outburst (23 December 2017), from the beginning (green) to the end (red), where the orange one denotes $t\approx 650$ days. The red star shows the value in May 2011, before the outburst. The blue curves depict the $\dot{M}-\Sigma$ relationship at different radii predicted by the slim accretion disk solution of \citet{Abramowicz1988ApJ}. The dashed line indicates $\dot{M} \propto \Sigma$, as anticipated from the advection-dominated slim disk solution, while the dotted line represents $\dot{M} \propto \Sigma^{-1}$, as expected from a radiation pressure-dominated thin accretion disk. The two arrows indicate the direction of movement of 1ES\,1927+654 along the slim disk and thin disk branches of the $S$-curve before and after $t = 650$ days, respectively.}
\label{fig:scurve}
\end{figure}

\subsection{The Phase-transition Curve}
\label{sec:scurve}

The shift in radiation efficiency described in Section~\ref{sec:eta} (Figure~\ref{fig:ptrap}), together with the systematic change in both the optical bolometric correction ($\kappa_{5100}$) and the UV/X-ray spectral slope ($\alpha_\mathrm{OX} \equiv 0.3838\log{L_\mathrm{2500}/L_\mathrm{2\,keV}}$)\footnote{Following \citet{Li2024paper2}, we adopt a definition of $\alpha_\mathrm{OX}$ that has the opposite sign to that used by \cite{Avni1982}.} noted by \citet[][their Figures~6 and 7]{Li2024paper2} around the same, distinctive period of $t \approx 650$ days after the outburst, provides compelling evidence that the accretion disk experienced a physically distinct transition from an advection cooling-dominated (slim disk) phase to a radiative cooling-dominated (thin disk) phase. To better understand such a transition from a slim to a thin disk, it is helpful to invoke the concept of a phase-transition curve or $S$-curve, most commonly visualized through a plot of the accretion rate $\dot{M}$ versus the mass surface density $\Sigma$ (e.g., \citealp{Kato2008book}). In the $\dot{M}-\Sigma$ plot, the $S$-curve derives from the interplay between the heating and cooling mechanisms in the disk. Three branches characterize the optically thick regime: (1) the upper branch represents a stable, hot, geometrically thick, radiation pressure-dominated slim disk (Figure~\ref{fig:scurve}, dashed line; \citealp{Abramowicz1988ApJ}); (2) the middle branch traces an unstable, hot, geometrically thin, radiation pressure-dominated thin disk (Figure~\ref{fig:scurve}, dotted line; \citealp{Shakura1973AA}); and (3) the lower branch delineates a stable, cooler, geometrically thin disk wherein gas pressure dominates (not shown in Figure~\ref{fig:scurve}).

At constant radius $R$, linking $\Pi \sim \dot{M}$ in Equation \ref{equ:Lcons} with Equation~\ref{equ:qadv} leads to advection cooling $Q_\mathrm{adv}^- \propto \dot{M}^2 \Sigma^{-1}$. From Equation~\ref{equ:qvisp}, the viscous heating rate $Q_\mathrm{vis}^+ \propto \dot{M}$. For a slim accretion disk, the condition $Q_\mathrm{adv}^- \simeq Q_\mathrm{vis}^+$ produces the upper branch of the $S$-curve:

\begin{equation}
    \dot{M} \propto \Sigma.
    \label{equ:slimSc}
\end{equation}

\noindent
Meanwhile, if at fixed $R$ radiation pressure dominates the total pressure, we expect $\Pi \sim T_c^4 H$ (Equation~\ref{equ:EoS}). Since $H \sim (\Pi/\Sigma)^{1/2}$ (Equation~\ref{equ:vertical}), radiation cooling $Q_\mathrm{rad}^- \propto (\dot{M}/\Sigma)^{1/2}$. For a thin accretion disk, $Q_\mathrm{rad}^- \simeq Q_\mathrm{vis}^+$. The middle branch of the $S$-curve, which connects the upper and lower branches, is unstable because of the negative slope \citep{Kato2008book}

\begin{equation}
    \dot{M} \propto \Sigma^{-1}.
    \label{equ:thinSc}
\end{equation}

\noindent
When the disk temperature decreases, gas pressure dominates over radiation pressure, as indicated by the relationship $\Pi \sim \Sigma T_c$ (Equation~\ref{equ:EoS}). Consequently, the radiative cooling rate $Q_\mathrm{rad}^- $ is proportional to $\Pi^4 / \Sigma^5$. In the case of a thin accretion disk where $Q_\mathrm{rad}^- \simeq Q_\mathrm{vis}^+$, this leads to the formation of the lower branch of the $S$-curve, which is characterized by a positive slope,

\begin{equation}
    \dot{M} \propto \Sigma^{5/3}.
    \label{equ:thinSclower}
\end{equation}

\noindent
The $S$-curve, therefore, underlines the thermal and viscous instabilities inherent in accretion disk systems. Transitions between the branches of the $S$-curve can lead to dramatic changes in the disk state, often associated with outburst events in systems such as X-ray binaries or cataclysmic variables (e.g., \citealp{Orosz1997ApJ,Lasota2001NewAR}).

Within this backdrop, Figure~\ref{fig:scurve} plots the accretion rate $\dot{m}$ as a function of disk surface density $\Sigma$ at the transition radius $R_\mathrm{tr}$, as derived in Section~\ref{sec:mdot} for our multiepoch observations of 1ES\,1927+654. The distribution of points delineates an $S$-like sequence in time because the optical outburst qualitatively resembles the upper and middle branches of the $S$-curve. For reference, we overlay the theoretical $S$-curve from \citet{Abramowicz1988ApJ}, for three choices of radius. Considering that in 1ES\,1927+654 $R_\mathrm{tr}$ varies with time (see Figure~\ref{fig:radius}), we note that the points corresponding to the early epochs align more closely with the $10\,R_g$ curve, while the later epochs tend to gravitate toward the $4\,R_g$ curve. At $t \lesssim 650$ days, following marginal changes in $R_\mathrm{tr}$, the early epochs display a relationship $\dot{m} \propto \Sigma$, indicative of an accretion flow that is predominantly advective. We suggest that at early times 1ES\,1927+654 possessed a slim disk. By marked contrast, for the later epochs we observe $\dot{m} \propto \Sigma^{-1}$, in accord with the prediction for a standard, radiation pressure-dominated thin disk (the inner solution, in the terminology of \citealp{Shakura1973AA}). Such a solution implies thermal instability. Interestingly, the pre-outburst parameters derived from the 2011 observation also reside on this branch. 1ES\,1927+654 was classified as a true type~2 AGN, one that intrinsically lacks a BLR \citep{Tran2011ApJ}. Our present analysis suggests that the true type~2 characteristic of 1ES\,1927+654 may be linked with the thermal instability of its accretion disk. This instability may lead to fluctuations in the radiation output, potentially influencing the formation and persistence of the BLR. According to the radiation-hydrodynamic simulations of \cite{Proga2000ApJ}, thermal instabilities in a disk wind can give rise to BLR clouds.

Stellar BH binaries frequently undergo disk-state transitions on a timescale of $\sim 10$ days, as exemplified by GX\,339$-$4 \citep{Belloni2005AA}. However, the viscous timescale for a supermassive BH, $\sim 10^3-10^6$~yr for $M_{\rm BH} = 10^6-10^9\,M_\odot$, is too long to exhibit observable secular changes in disk accretion rate. Nevertheless, substantial variations in bolometric flux have been detected in several AGNs. For instance, Mrk\,1018 decreased its flux by more than a factor of 10 within a span of eight years, inducing a disk-state transition responsible for the ``changing-look'' event from a type~1 to a type~1.9 Seyfert \citep{Noda2018MNRAS}. Our observations of 1ES\,1927+654 suggest a transformation of its accretion flow from slim disk-dominated to thin disk-dominated at $t \approx 650$ days after the outburst. We propose that the timescale of disk-state transition largely depends on the rate of evolution of the mass accretion rate. In the case of 1ES\,1927+654, its change in accretion rate is highly likely to be externally induced, possibly due to a TDE, as discussed in Section~\ref{sec:mdot}. Future time-domain studies will provide valuable insights on the secular changes of mass accretion rate in AGNs and their possible manifestation as disk-state transitions.

\section{Radiation of the Transient Slim Disk}
\label{sec:sec3}

We have argued that 1ES\,1927+654 exhibits a transient slim disk for $t\lesssim 650$ days (Section~\ref{sec:scurve}). During the slim disk phase, the median soft X-ray emission of the blackbody component is roughly 10 times larger than during the thin disk phase \citep{Li2024paper2}. Meanwhile, during the slim disk phase $R_\mathrm{tr}$ is larger than $R_\mathrm{in}$, suggesting that the inner portion of the thermalized disk is truncated. The inner truncation of the thermalized emission from the accretion disk is not an unusual occurrence. It serves as a foundational framework for explaining state transitions in BH X-ray binaries. In this model, the accretion disk does not reach the innermost stable circular orbit of the BH, but is instead truncated at an inner radius, interior to which occupies a hot, optically thin, geometrically thick corona or ADAF (e.g., \citealp{Remillard2006ARAA,Done2007ARAA}). Some instances of inner disk truncation have been noted in supermassive BHs, but to date only in very low-luminosity, highly sub-Eddington AGNs \citep{Ho2008}. \cite{Quataert1999} interpreted the distinctive SEDs of the low-luminosity AGNs in M\,81 and NGC\,4579 \citep{Ho1999} as evidence of a truncated thin disk, interior to which lies a radiatively inefficient, optically thin ADAF. The high-luminosity nature of 1ES\,1927+654 disfavors the inner truncation due to an optically thin ADAF. Instead, we argue that the inner truncation is connected to the overheated nature of a slim or radiation pressure-dominated disk (Section~\ref{sec:overheat}).

The radiation from a slim disk has a higher color-temperature and manifests as the soft X-ray excess, which we represent as a blackbody component in the SED decomposition (Figure~\ref{fig:disksed}). However, a natural consequence of overheating in the inner disk is that Equation~\ref{equ:tdobs} is no longer applicable. To convert the observed temperature ($T_\mathrm{bb}$) and luminosity ($L_\mathrm{bb}$) into the physical properties of the overheated disk, two critical effects must be considered carefully. First, as $R$ decreases, the viscous timescale ($t_\mathrm{vis}$) becomes significantly shorter than the outward photon diffusion timescale ($t_\mathrm{dif}$), leading to a ``photon-trapping'' effect that limits the observed optical depth. Second, radiative transfer in the inner region is affected by strong scattering. When the opacity is dominated by electron scattering, the emerging flux decreases by roughly a factor of $(\kappa_\mathrm{ff}/\kappa_\mathrm{es})^{1/2}$ \citep{Wandel1988ApJ}. We provide detailed consideration of these effects in Sections~\ref{sec:ptrap} and \ref{sec:escatt}, and then estimate the size of the photospehere of the overheated disk in Section~\ref{sec:slimsize}.

\subsection{The Soft X-ray Excess from the Overheated Region}
\label{sec:overheat}

The inner truncation radius $R_\mathrm{tr}$ of the multicolor blackbody disk (fitted with {\tt diskbb}) decreased from $\sim 20\,R_g$ to $2.4\,R_g$ during the slim disk phase and remained stable in the thin disk phase (Figure~\ref{fig:radius}). Why does the multicolor blackbody emission truncate at a radius larger than $R_\mathrm{in}$? The disk between $R_\mathrm{tr}$ and $R_\mathrm{in}$ can be considered as a two-fluid, ``plasma + radiation'' flow. When radiation pressure dominates, we usually expect, following \citet{Shakura1973AA}, that the vertical mean gas density 

\begin{equation}
    \bar{\rho} \propto \alpha^{-1}\dot{m}^{-2}R^{3/2}.
\label{equ:rhoss}
\end{equation}

\noindent
As $R$ decreases, Maxwell stress increases, which in turn enhances the viscosity $\alpha$ via magnetorotational instabilities \citep{Jiang2019aApJ}. This suggests that the accreting plasma can have low density for BHs with high $\dot{m}$, leading to a reduced free-free emission capability,

\begin{equation}
\begin{aligned}
    \dot{w}_{\rm ff} &\simeq 1.6\times 10^{-27} n^2 \sqrt{T_c} \\
    &\simeq 1.6\times 10^{-27} \left(\frac{35\Sigma}{32H\mu m_\mathrm{H}}\right)^2 \sqrt{T_c}.
\label{equ:wff}
\end{aligned}
\end{equation}

\noindent
In the above equation, $n$ represents plasma number density, with the numerical factor derived from $\rho_c = \frac{35}{16}\bar{\rho}$ through integration. Meanwhile, the heating rate density

\begin{equation}
    \dot{w}^+ = Q_\mathrm{vis}^+/2H.
    \label{equ:wheat}
\end{equation}

\noindent
If $\dot{w}_{\rm ff} < \dot{w}^+$, the plasma cannot successfully reprocess the released energy into blackbody radiation. Consequently, a balance between viscous heating and radiative cooling requires a temperature $T_c \gg T_\mathrm{eff}$ \citep{Beloborodov1998MNRAS}. This regime is effectively optically thin, often referred to as an overheated disk (e.g., \citealp{Takahashi2016ApJ}). Observationally, researchers employ a color-temperature correction to fit the broad-band SED (e.g., \citealp{Done2012MNRAS}).

\begin{figure}
\centering
\includegraphics[width=0.48\textwidth]{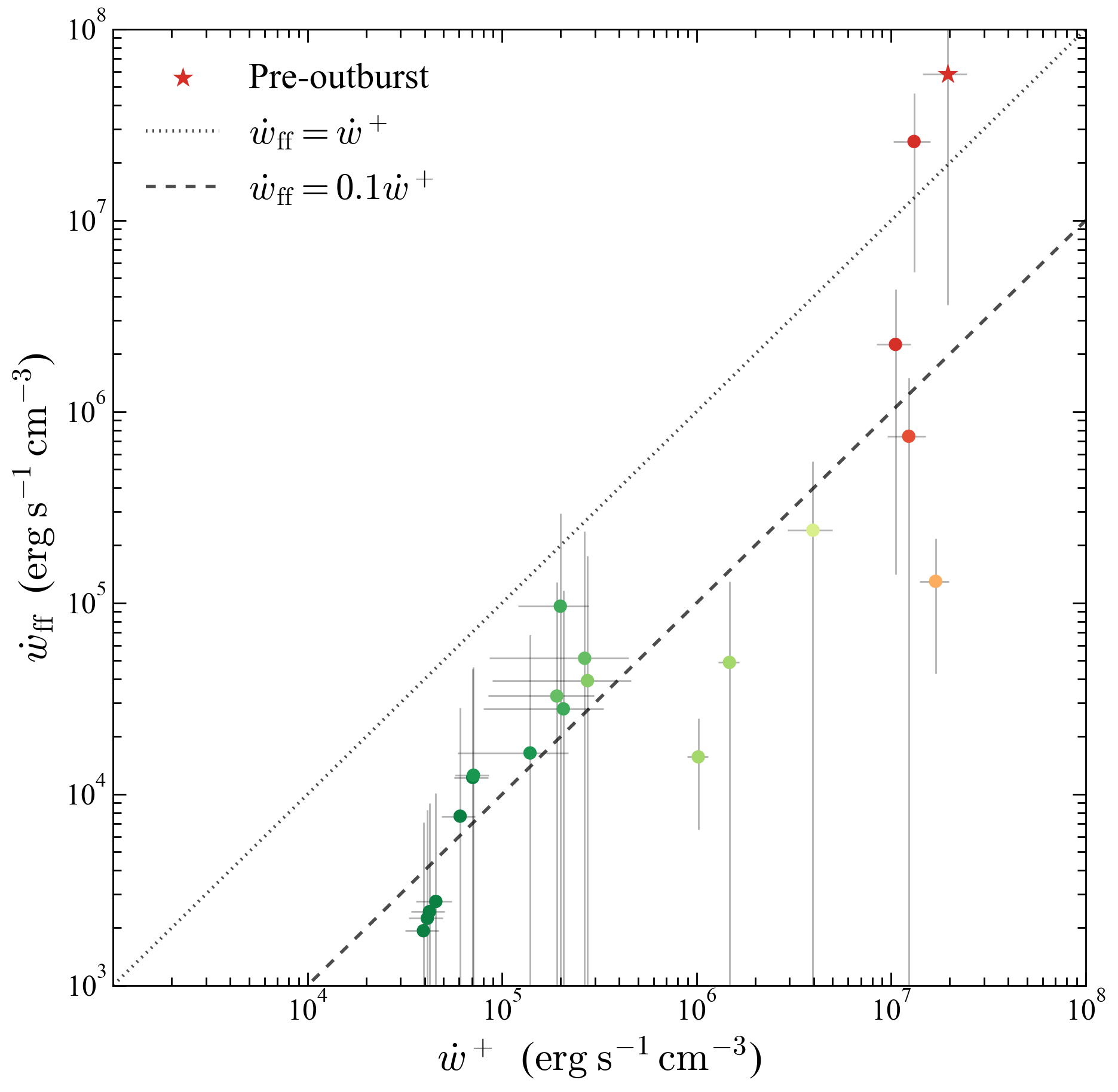}
\caption{The free-free emission capability ($\dot{w}_{\rm ff}$) versus the viscous heating rate density ($\dot{w}^+$) at $R=R_\mathrm{tr}$, color-coded by the time since the outburst (23 December 2017), from the beginning (green) to the end (red),  where the orange one denotes $t\approx 650$ days. The red star shows the value in May 2011, before the outburst. The black dashed line denotes $\dot{w}_{\rm ff} = 0.1\dot{w}^+$, while the dotted line gives for $\dot{w}_{\rm ff} = \dot{w}^+$.}
\label{fig:overh}
\end{figure}

In the case of 1ES\,1927+654, which has all the physical properties computed at $R_\mathrm{tr}$ (Table~\ref{tab:diskpro}), we can calculate $\dot{w}_{\rm ff}$ and $\dot{w}^+$ and their associated uncertainties to resolve the heating balance.  Except for the final epoch, 1ES\,1927+654 resides in the $\dot{w}_{\rm ff} < \dot{w}^+$ regime (Figure~\ref{fig:overh}), close to $\dot{w}_{\rm ff} = 0.1\dot{w}^+$, which suggests inefficient absorption of blackbody radiation at $R_\mathrm{tr}$. However, this does not imply that the truncation should be located at a larger radius, as we have not yet considered bound-free absorption, which would occur if heavy elements are present. Within $R_\mathrm{tr}$, the ratio $\dot{w}_{\rm ff}/\dot{w}^+$ can be further reduced due to an increase in viscosity \citep{Jiang2019aApJ}, resulting in a deviation from thermodynamic equilibrium in the inner overheated disk. As a result, the inner disk can overheat by roughly 5 times, reaching temperatures as high as $\sim 150\,$eV (\citealp{Li2024paper2}; see also Figure~\ref{fig:RT}), which manifests as the soft X-ray excess.

\subsection{Photon Trapping}
\label{sec:ptrap}

The inward radial velocity of the material increases at high $\dot{m}$, and can reach $\sim 0.1\,c$ at $R_\mathrm{tr}$ in the case of 1ES\,1927+654. Consequently, as shown in the schematic in Figure~\ref{fig:disksed}, photons may not have enough time to diffuse before being advected into the central BH. Only photons generated near the photosphere of the disk can escape into space. The optical depth that can be observed would be smaller than the total optical depth in the vertical direction. To quantify this effect, we first consider the viscous timescale

\begin{equation}
    t_\mathrm{vis} = \int_{R}^{R_\mathrm{in}} \frac{dr}{v_R(r)}.
\label{equ:tvis}
\end{equation}

\noindent
Photons would be trapped if their diffusion timescale $t_\mathrm{dif} \simeq H\tau/c$ is larger than $t_\mathrm{vis}$, where $\tau$ is the optical depth. The critical optical depth then can be calculated from

\begin{equation}
    \tau_\mathrm{c} \simeq \frac{ct_\mathrm{vis}}{H} \sim \frac{cR}{Hv_R},
\label{equ:tauc}
\end{equation}

\noindent
where photons produced at $\tau \gtrsim \tau_\mathrm{c}$ would be advected directly into the BH.

Calculating $t_\mathrm{vis}$ through Equation~\ref{equ:tvis} needs to assume a radial velocity profile $v_R(r)$. Following the solution of a radiation pressure-dominated thin disk \citep{Shakura1973AA}, we have $v_R \propto R^{-5/2}$ and $H \propto R^0$. Therefore, for a thin disk we expect $\tau_\mathrm{c}^\mathrm{td} \propto R^{7/2}$, which increases very quickly with radius. For a slim disk, combining the equations in Section~\ref{sec:mdot} and letting $Q_\mathrm{vis}^{+} \simeq Q_\mathrm{adv}^{-}$ give $v_R \propto R^{-1/2}$ and $H \propto R^{1/2}$, which yields a shallower radial dependence of optical depth, $\tau_\mathrm{c}^\mathrm{sd} \propto R$. As $R$ increases, $\tau_\mathrm{c}^\mathrm{sd}$ increase slower than $\tau_\mathrm{c}^\mathrm{td}$, such that the photon-trapping effect is stronger in a slim disk than in a thin disk. As both a slim and thin disk are simultaneously present in 1ES\,1927+654 (Figure~\ref{fig:scurve}), the actual value of $\tau_\mathrm{c}$ should lie in between $\tau_\mathrm{c}^\mathrm{td}$ and $\tau_\mathrm{c}^\mathrm{sd}$.

\subsection{Electron Scattering}
\label{sec:escatt}

As gas temperature increases at high mass accretion rate and when approaching the innermost disk, $\kappa_\mathrm{ff}$ decreases quickly and is eventually dominated by $\kappa_\mathrm{es}$ (Equation~\ref{equ:opacity}). The effective optical depth, $\tau_\mathrm{eff} \equiv \bar{\rho}H(\kappa_\mathrm{ff}\kappa_\mathrm{es})^{1/2}$, can be less than unity if the disk is dominated by electron scattering, resulting in a modified blackbody spectrum \citep{Wandel1988ApJ}. To calculate the emergent intensity, we first make the reasonable assumption \citep{Beloborodov1998MNRAS,Kato2008book} that the overheated disk is vertically isothermal. At each radius, scattering modifies the source function $S_\nu$ into \citep{Rybicki1986book}

\begin{equation}
    S_\nu = (1-\omega_\nu)B_\nu(T)+\omega_\nu J_\nu(\tau_\nu),
\end{equation}

\noindent
where $B_\nu$ is the Planck function, $J_\nu$ is the mean intensity, and $\omega_\nu \equiv \sigma_{\nu,s}/(\sigma_{\nu,a}+\sigma_{\nu,s})$ is the scattering albedo, with $\sigma_{\nu,a}$ and $\sigma_{\nu,s}$ the absorption and scattering cross section, respectively. For a plane slab, the solution to the radiation transfer Equation~8 of \citet{Zhu2019ApJ} yields a reduction of the emergent intensity $\chi_\nu \equiv I_\nu^\mathrm{out}/B_\nu < 1$, where $I_\nu^\mathrm{out}$ is the emergent intensity. For an isothermal slab, integrating Equation~11 of \citet{Zhu2019ApJ} over frequency gives

\begin{equation}
\begin{aligned}
    \chi &\equiv \frac{F^\mathrm{out}}{\sigma T^4} = (1-e^{-\tau_d/\mu}) \\
   &\times \left(1-\bar{\omega} \frac{e^{-\sqrt{3(1-\bar{\omega})
   }\tau}+ e^{\sqrt{3(1-\bar{\omega})
   }(\tau-\tau_d)}}{e^{-\sqrt{3(1-\bar{\omega})
   }\tau_d}(1-\sqrt{1-\bar{\omega}}) + (1-\sqrt{1-\bar{\omega}})} \right) \\[4pt]
   &\mathrm{with}\; \tau = 2\mu \tau_d/(3\tau_d+1),\label{equ:chiI}
\end{aligned}
\end{equation}

\noindent
where $\tau_d$ and $\tau$ denote the total and variable optical depth in the vertical direction, respectively. The mean albedo is denoted by $\bar{\omega} \simeq \kappa_\mathrm{es}/(\kappa_\mathrm{abs}+\kappa_\mathrm{es})$, while $\mu \equiv \cos{i}$, with $i$ the disk inclination angle. The absorption opacity $\kappa_\mathrm{abs}$ typically comprises both bound-free and free-free absorption. In a plasma containing heavy elements, which is particularly plausible in an AGN disk (e.g., \citealt{Wang2023}, and references therein), the absorption coefficient can significantly increase, with $\kappa_\mathrm{abs} \approx 30\, \kappa_\mathrm{ff}$ for solar abundances \citep{Wandel1988ApJ}. Qualitatively, Equation~\ref{equ:chiI} implies that the emergent flux decreases monotonically with decreasing optical depth and increasing scattering albedo.

\begin{figure}
\centering
\includegraphics[width=0.48\textwidth]{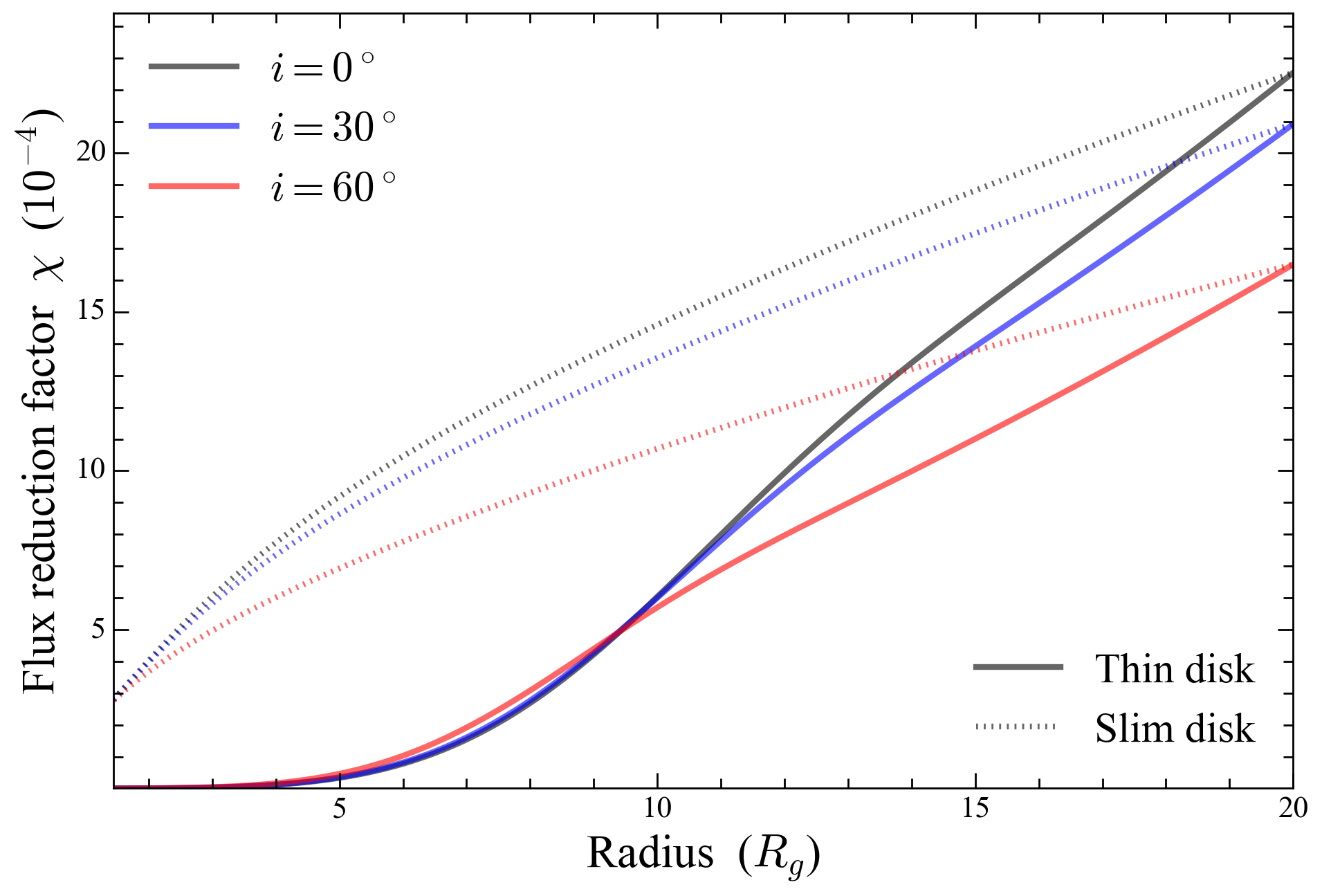}
\caption{Flux reduction factor ($\chi \equiv F^\mathrm{out}/\sigma T^4$) as a function of radius of the inner disk overheated because of photon trapping (Section~\ref{sec:ptrap}) and electron scattering (Section~\ref{sec:escatt}), as deduced for the 17 May 2018 observations of 1ES\,1927+654. Solid lines assume the $\bar{\rho}$ and $T$ profile for a radiatively cooled accretion flow (thin disk), while dotted lines pertain to an optically thick, advection-dominated accretion flow (slim disk). Different colors represent different inclination angles of the disk.}
\label{fig:fluxredu}
\end{figure}

The mean albedo $\bar{\omega}$ also varies with $R$, as both gas density and temperature fluctuate. For a radiation pressure-dominated thin disk, $\rho \propto R^{3/2}$ and $T_c \propto R^{-3/8}$ \citep{Shakura1973AA}, while in an advection-dominated slim disk $\rho \propto R^{-1}$ and $T_c \propto R^{-1/2}$ \citep{Abramowicz1988ApJ}, such that $\kappa_\mathrm{ff}^\mathrm{td} \propto R^{45/16}$ and $\kappa_\mathrm{ff}^\mathrm{sd} \propto R^{5/4}$. As such, the closer one gets to the inner region, the more significant both photon trapping (Section~\ref{sec:ptrap}) and electron scattering become in reducing the emergent flux, regardless of the specific details of the accretion flow. Figure~\ref{fig:fluxredu} illustrates the flux reduction effect of our 17 May 2018 observation. For the overheated disk in this epoch, during which $R_\mathrm{tr}\approx 20\,R_g$, after considering bound-free absorption we calculate $\kappa_\mathrm{abs}/\kappa_\mathrm{es} \approx 30\,\kappa_\mathrm{ff}/\kappa_\mathrm{es} = 8.7\times10^{-7}$ and $\tau_c = 14$ from Equations~\ref{equ:opacity} and \ref{equ:tauc}. This demonstrates that the electron scattering effect is quite significant.

\subsection{Photosphere Size}
\label{sec:slimsize}

As discussed above, both the photon trapping and electron scattering effects need to be taken into account to investigate the emission properties of the inner overheated disk. Within $R_\mathrm{tr}$, the overall luminosity reduction factor can be integrated through

\begin{equation}
    \chi_L = \frac{\int_{R_\mathrm{in}}^{R_\mathrm{tr}}2\pi R \chi(R) \sigma T(R)^4 dR}{\int_{R_\mathrm{in}}^{R_\mathrm{tr}} 2\pi R \sigma T(R)^4 dR}.
\label{equ:chiL}
\end{equation}

\noindent
The above integration needs to assume a radial profile for both $\tau_c$ and $\kappa_\mathrm{ff}$. For our 20 epochs of observations and setting $i = 60^\circ$ \citep{Li2024paper2}, we use Equation~\ref{equ:chiL} (see also Sections~\ref{sec:ptrap} and \ref{sec:escatt}) to calculate $\chi_L^\mathrm{td} = 7.81_{-2.56}^{+3.78} \times 10^{-4}$ for the thin disk profile and $\chi_L^\mathrm{sd} = 11.71_{-3.42}^{+3.60} \times 10^{-4}$ for the slim disk profile, where the upper and lower error stand for the 16\% and 84\% value of $\chi_L$, respectively. The scatter is rather small from epoch to epoch. Without prior knowledge of inclination, Equation~\ref{equ:chiI} reduces to $\chi \sim \sqrt{1-\bar{\omega}}$ \citep{Zhu2019ApJ}, and we expect $\chi_L^\mathrm{td} \approx 7.0 \times 10^{-4}$ and $\chi_L^\mathrm{sd} \approx 10.5 \times 10^{-4}$. The difference of $\chi_L$ between different epochs or types of accretion flow is much less than our measurement uncertainty ($\sim 0.4\,$dex). Therefore, we simply adopt $\chi_L = 1 \times 10^{-3}$ for the ensuing calculations.

A radiation-driven outflow from a super-Eddington accretion disk is likely to be optically thick \citep{Ohsuga2011ApJ,Jiang2014ApJ}. Therefore, the disk photosphere often lies above the usual effective absorption opacity photosphere \citep{Jiang2014ApJ}; see schematic in Figure~\ref{fig:disksed}. We simply define the photosphere size of the inner overheated disk as

\begin{equation}
    R_\mathrm{ph} \equiv \sqrt{\frac{L_\mathrm{bb}}{\pi \chi_L \sigma T_\mathrm{bb}^4} + R_\mathrm{in}^2}.
\label{equ:radiusph}
\end{equation}

\begin{figure}
\centering
\includegraphics[width=0.48\textwidth]{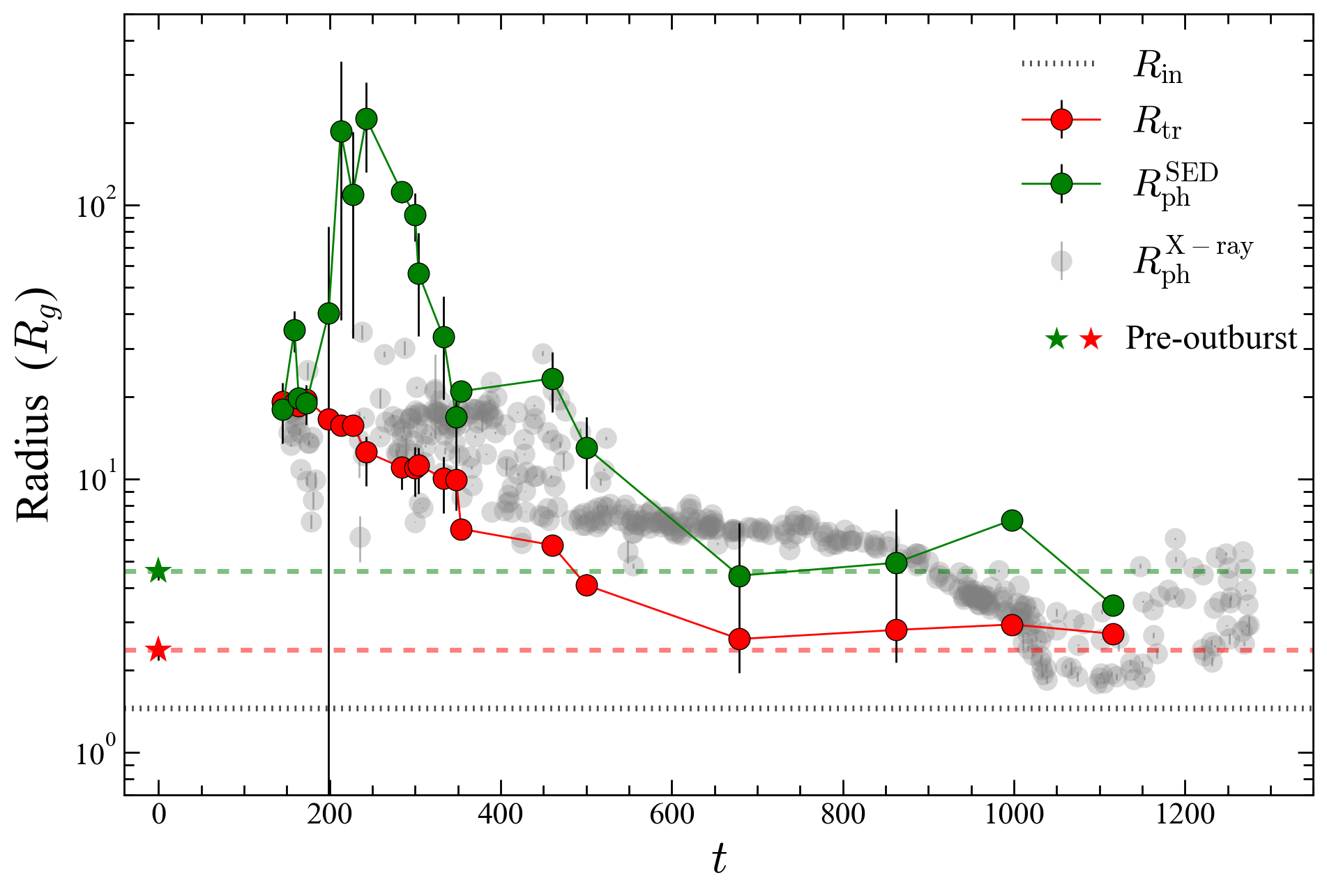}
\caption{Characteristic radius of the accretion flow as a function of time since the optical outburst. Red points give the inner truncation radius ($R_\mathrm{tr}$) of the outer thin disk, which was derived directly from broad-band SED fitting \citep{Li2024paper2}. Green points show the photosphere size ($R_\mathrm{ph}^\mathrm{SED}$) calculated through Equation~\ref{equ:radiusph}, with $T_\mathrm{bb}$ and $L_\mathrm{bb}$ derived from broad-band SED fitting \citep{Li2024paper2}. The grey points correspond to the photosphere size ($R_\mathrm{ph}^{\rm X-ray}$) obtained from X-ray observations alone \citep{Masterson2022}. The stars and the horizontal dashed lines mark the pre-outburst values of $R_\mathrm{tr}$ (red) and $R_\mathrm{ph}^\mathrm{SED}$ (green) calculated from the 2011 observation \citep{Li2024paper2}. The black dotted line shows the inner disk radius ($R_\mathrm{in} = 1.45\,R_g$) adopted in this paper.}
\label{fig:radius}
\end{figure}

\noindent
The results are displayed in Figure~\ref{fig:radius}, where $R_\mathrm{ph}^\mathrm{SED}$ (green points) uses $T_\mathrm{bb}$ and $L_\mathrm{bb}$ from our 20-epoch broad-band observations \citep{Li2024paper2} and $R_\mathrm{ph}^\mathrm{X-ray}$ (grey points) stems from the 435 epochs of NICER observations by \cite{Masterson2022}. The two results generally align with each other, even though \citet{Masterson2022} did not include any emission from the outer thin disk. On the whole, $R_\mathrm{ph}$ (grey points) lies above $R_\mathrm{tr}$ (red points), which corroborates our interpretation of the photosphere size. Between $t = 200$ to 300 days, $R_\mathrm{ph}$ is $\sim 10$ times larger than $R_\mathrm{tr}$, which may be related to the strong outflow responsible for the rapid X-ray spectral evolution \citep{Li2024paper2}. Its reflection also can explain the broad 1\,keV feature \citep{Masterson2022}.

\section{Global Spectral Energy Distribution}
\label{sec:sec4}

The effects of both photon trapping (Section~\ref{sec:ptrap}) and electron scattering (Section~\ref{sec:escatt}) rapidly decrease as the radius expands, causing the disk to become optically thick and geometrically thin beyond $R_\mathrm{tr}$. Consequently, the global SED emerges as an interplay between the inner overheated disk, the outer disk in thermal equilibrium, and the X-ray corona (Figure~\ref{fig:disksed}), as a result of the systematic variation in both temperature and radius (\citealp{Li2024paper2}; also Figure~\ref{fig:radius}). We exploit these circumstances to examine the disk temperature profile of 1ES\,1927+654, both for the inner and outer disk (Section~\ref{sec:tprofile}). Following this, we discuss Compton cooling within the corona region and its effects on the disk effective temperature (Section~\ref{sec:Ccooling}). Finally, we explore the implications for the UV spectral index of AGNs (Section~\ref{sec:UVindex}).

\subsection{Temperature Profile}
\label{sec:tprofile}

Similar to the blackbody temperature evolution discussed by \cite{Masterson2022} and \cite{Li2024paper2}, both the photosphere radius of the overheated disk and the inner truncation radius of the thin disk show systematic changes over time (Figure~\ref{fig:radius}). This is significant for two reasons. First, the slope of the equatorial temperature profile reflects the accretion disk state. For a slim disk, which lies on the upper branch of the accretion disk $S$-curve, the temperature profile often is solved numerically to satisfy the transonic condition near the inner region \citep{Abramowicz1988ApJ,Sadowski2011}. In the case of a thin disk \citep{Shakura1973AA}, if it is radiation pressure-dominated and situated on the middle branch of the $S$-curve, the temperature follows $T_c \propto R^{-3/8}$, while if gas pressure dominates and the disk lies on the bottom branch of the $S$-curve, $T_c \propto R^{-3/4}$. The effective temperature profile of the outer disk analyzed here greatly influences the overall shape of the global SED: for a disk in thermal equilibrium, $T_\mathrm{eff} \propto R^{-1/2}$ for a slim disk and $T_\mathrm{eff} \propto R^{-3/4}$ for a thin disk.

\subsubsection{Inner Disk}
\label{sec:Tprofile_i}

\begin{figure}
\centering
\includegraphics[width=0.48\textwidth]{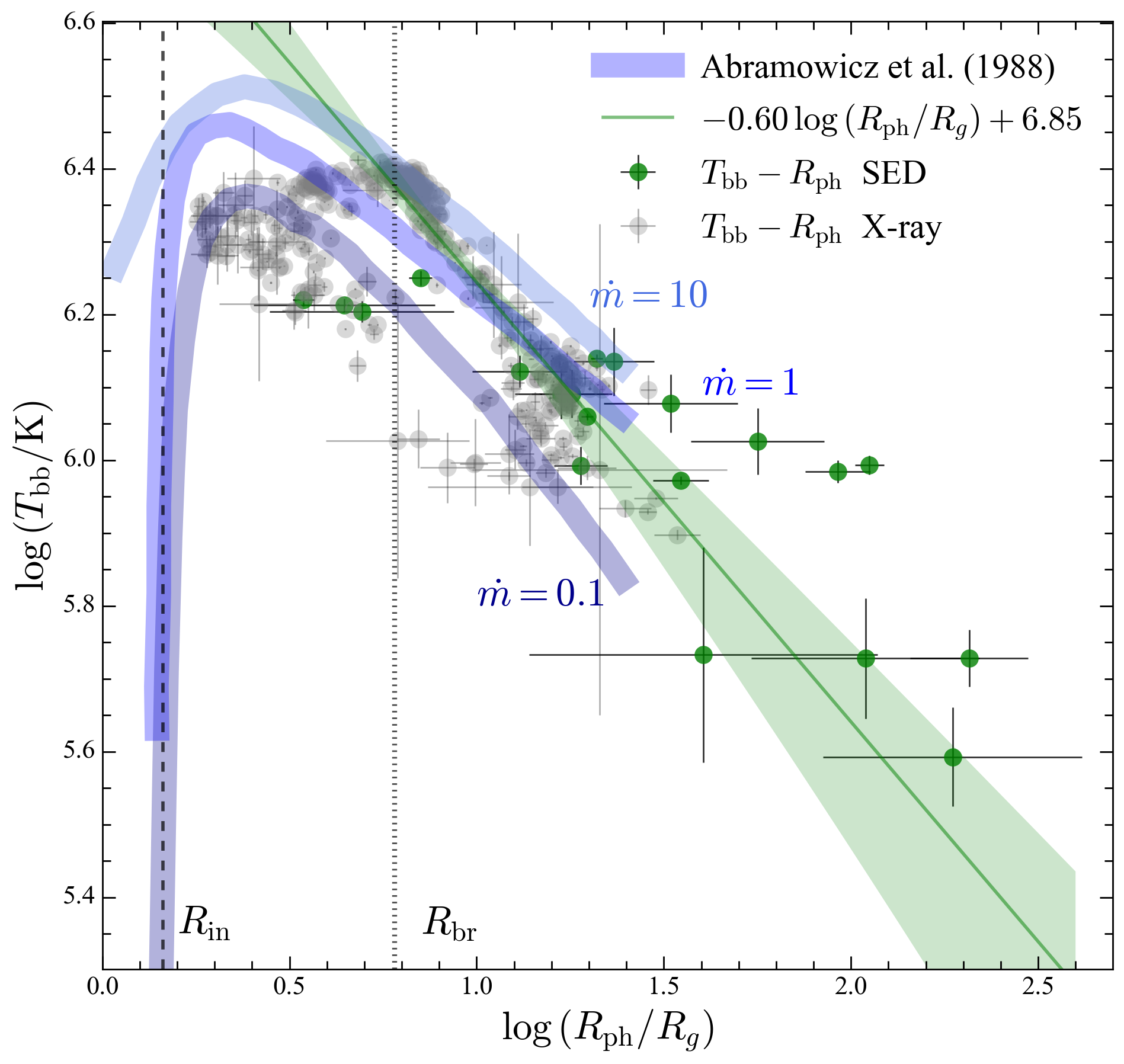}
\caption{The blackbody temperature ($T_\mathrm{bb}$) as a function of photosphere size ($R_\mathrm{ph}$, Equation~\ref{equ:radiusph}) of the inner overheated disk from this work (green points) and from the NICER X-ray observations (gray points) of \cite{Masterson2022}. The green solid line gives the linear fit for the points with $R_\mathrm{tr} \gtrsim R_\mathrm{br}$, $\log{T_\mathrm{bb}} = -0.60 \log{(R_\mathrm{ph}/R_g)} +6.85$, with the $1\,\sigma$ uncertainty shown as the shaded region. The vertical lines mark the inner disk radius ($R_\mathrm{in}$; dashed) and the break radius ($R_\mathrm{br}$; dotted) in Equation~\ref{equ:Teff_broken}. The three blue shaded curves depict the temperature-size relation of the slim accretion disk solution of \citet{Abramowicz1988ApJ} for $\dot{m} = 10$, 1, and 0.1.}
\label{fig:RT}
\end{figure}

In the study by \citet{Masterson2022}, the evolution of the soft X-ray excess is categorized into three distinct phases. Chronologically, the blackbody temperature ($T_\mathrm{bb}$) initially increases (Phase~1), then remains constant (Phase~2), and finally decreases (Phase~3). Utilizing the photospheric radius ($R_\mathrm{ph}$) calculated in Equation~\ref{equ:radiusph}, the temperature profile of the inner overheated disk is illustrated in Figure~\ref{fig:RT}\footnote{This temperature profile is dynamic, not static, as the accretion rate also fluctuates over time, as seen in Figure~\ref{fig:mdot}.}. Notably, since $R_\mathrm{ph}$ consistently decreases over time, the three phases identified by \citet{Masterson2022} translate into three regions in Figure~\ref{fig:RT}. In the outer region where $R_\mathrm{ph} \gtrsim 6\,R_g$, $T_\mathrm{bb}$ shows an inverse correlation with $R_\mathrm{ph}$, maintaining a similar slope. The constant-$T_\mathrm{bb}$ phase now resides in the region where $4\,R_g \lesssim R_\mathrm{ph} \lesssim 6\,R_g$, with $T_\mathrm{bb}$ exceeding $2\times 10^{6}$ K. Where $R_\mathrm{ph} \lesssim 4\,R_g$, $T_\mathrm{bb}$ decreases as it approaches the inner radius of the accretion disk, corresponding to the steep $L_\mathrm{bb} \propto T_\mathrm{bb}^{11}$ slope of Phase~3 in \citet{Masterson2022}. Theoretically, this temperature decrease near the inner edge signifies that the solution for the accretion flow must be transonic, where the local sound speed equals $v_R$ at $R_\mathrm{son} \sim R_\mathrm{in}$ (e.g., \citealp{Abramowicz1988ApJ,Sadowski2011}).

In Figure~\ref{fig:RT}, we overlay the theoretical slim accretion disk solutions for different values of $\dot{m}$. Since the solution in \citet{Abramowicz1988ApJ} considers a non-spinning $10\,M_\odot$ Schwarzschild BH, we adjusted the $\log{R}$ values by $0.32$~dex to match the inner radius and the $\log{T}$ values by $1.35$~dex based on the scaling relation $T \propto M_\mathrm{BH}^{-1/4}$. The observationally inferred temperature profile of the inner overheated disk at
$R_\mathrm{ph} \gtrsim 6\,R_g$ matches the slim disk solution between $\dot{m} = 1$ to $\dot{m} = 10$, in excellent agreement with the mass accretion rate of 1ES\,1927+654 (Figure~\ref{fig:mdot}). For the region $R_\mathrm{ph} \gtrsim 6\,R_g$, fitting the $T_\mathrm{bb}-R_\mathrm{ph}$ relation using both the results from our SED analysis \citep{Li2024paper2} and those derived solely from X-rays \citep{Masterson2022} yields

\begin{equation}
    \log{(T_\mathrm{bb}/{\rm K})} = (-0.60\pm0.05) \log{(R_\mathrm{ph}/R_g)} + (6.85\pm0.04),
\label{equ:Tbb_profile}
\end{equation}

\noindent
which has an intrinsic scatter of $0.08$~dex. Interestingly, the $-0.6$ slope of the temperature profile of the inner overheated disk closely matches with that inferred from a self-similar solution of a slim disk ($T \propto R^{-5/8}$; \citealp{Wang1999ApJ}), as opposed to the $-3/8$ slope anticipated from the inner region of a thin disk \citep{Shakura1973AA}. These findings lend further credence to the notion that the inner overheated disk of 1ES\,1927+654 is a slim disk, where the cooling is dominated by advection.

\subsubsection{Outer Disk}
\label{sec:Tprofile_o}

\begin{figure}
\centering
\includegraphics[width=0.48\textwidth]{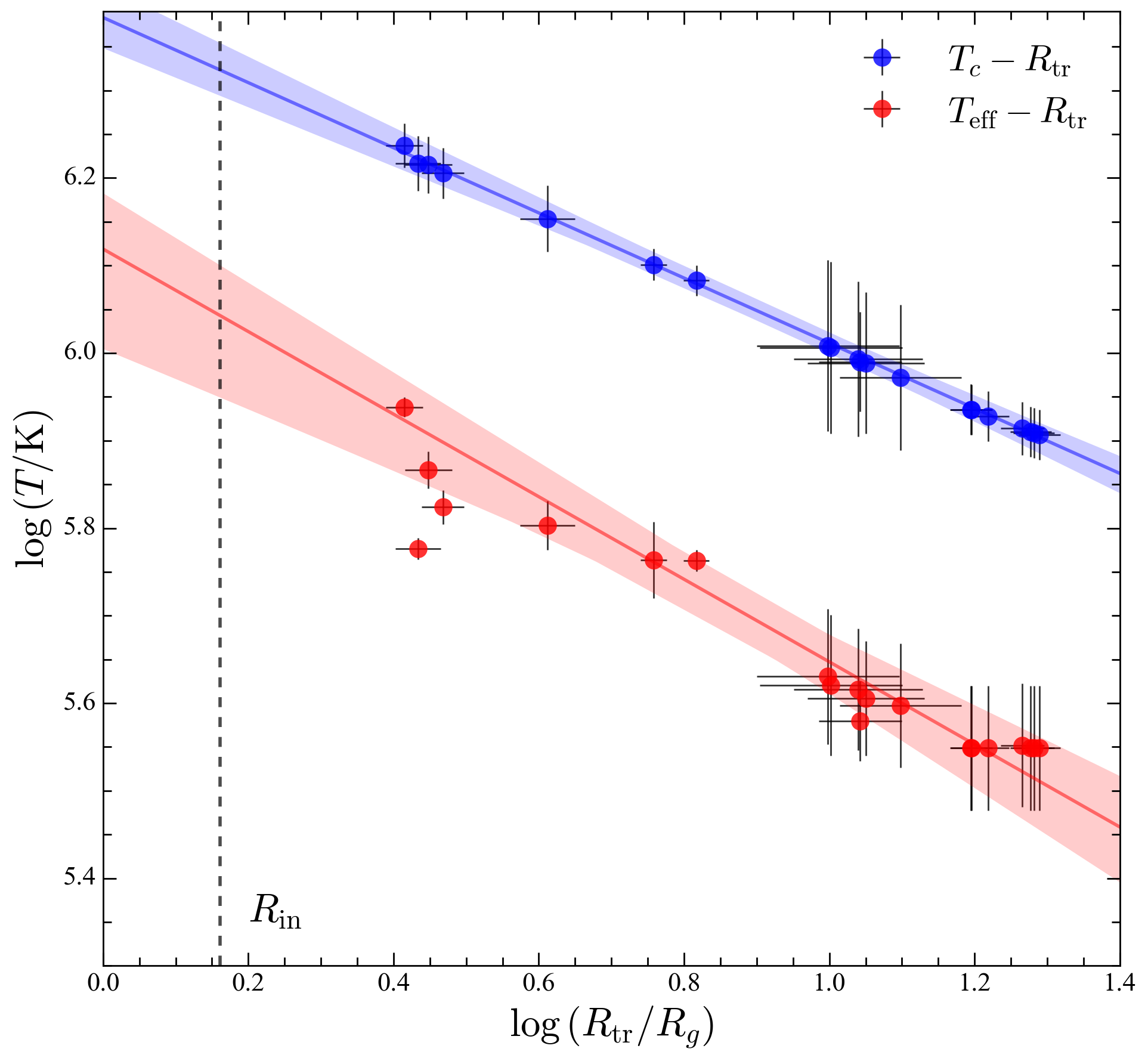}
\caption{The variation with $R_\mathrm{tr}$ of the disk middle-plane temperature ($T_c$, blue points) and effective temperature ($T_\mathrm{eff}$, red points) at $R_\mathrm{tr}$. The corresponding linear fits and $1\,\sigma$ uncertainty are $\log{(T_c/{\rm K})} = (-0.37\pm0.02) \log{(R_\mathrm{tr}/R_g)} + (6.38\pm0.02)$ (blue line) and $\log{T_\mathrm{eff}} = (-0.47\pm0.04) \log{(R_\mathrm{tr}/R_g)} + (6.11\pm0.04)$ (red line). The vertical black dashed line marks the inner disk radius ($R_\mathrm{in}$).}
\label{fig:Rtr_T}
\end{figure}

Separated by the truncation radius, the outer accretion disk can reach local thermal equilibrium because of its higher gas density, and thus its emitted spectrum can be calculated by integration of Equation~\ref{equ:tdobs}. We first examine the equatorial temperature ($T_c$) profile, which at the $R_\mathrm{tr}$ for each epoch of observation is derived by the set of equations in Section~\ref{sec:mdot}. A linear regression to the blue points in Figure~\ref{fig:Rtr_T} yields

\begin{equation}
    \log{(T_c/{\rm K})} = (-0.37\pm0.02) \log{(R_\mathrm{tr}/R_g)} + (6.38\pm0.02),
\label{equ:Tc_profile}
\end{equation}

\noindent
which has an intrinsic scatter of only 0.004~dex, despite $\dot{m}$ being quite different among the different epochs. The observed slope of $-0.37\pm0.02$, along with the weak dependence on accretion rate, agrees with expectations from the temperature profile of a radiation pressure-dominated thin accretion disk, which follows $T_c \propto \dot{m}^0 R^{-3/8}$ \citep{Shakura1973AA}. Physically, as $R_\mathrm{tr}$ shrinks from $\sim 20\,R_g$ to $\sim 2\,R_g$, Equation~\ref{equ:Tc_profile} traces the temperature profile exterior to the $R_\mathrm{tr}$ of the previous epoch. These results suggest that the outer disk, at least within $\sim 20\,R_g$, is a radiation pressure-dominated thin disk. In combination with the results in Section~\ref{sec:Tprofile_i}, at each epoch $R_\mathrm{tr}$ not only divides the accretion flow into two regions---an inner overheated disk and an outer disk in thermal equilibrium---but also distinguishes an inner slim disk within $R_\mathrm{tr}$ from a outer thin disk beyond $R_\mathrm{tr}$.

\begin{figure}
\centering
\includegraphics[width=0.48\textwidth]{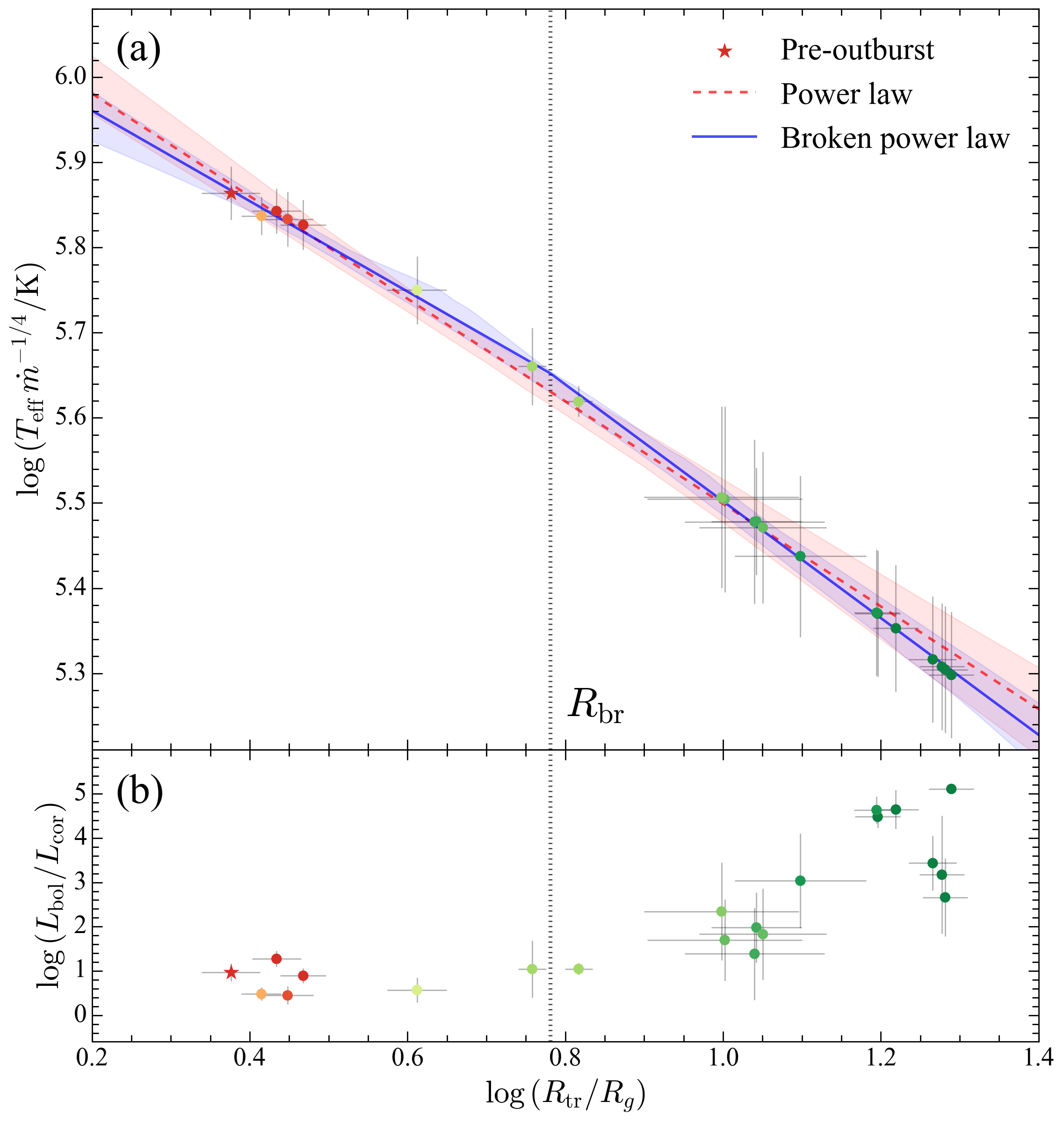}
\caption{The variation with truncation radius ($R_\mathrm{tr}$) of (a) the normalized effective temperature ($T_\mathrm{eff} \, \dot{m}^{- 1/4}$) and (b) corona bolometric correction ($L_\mathrm{bol}/L_\mathrm{cor}$). The color of the circles represents the time since the outburst (23 December 2017), from the beginning (green) to the end (red), where the orange one denotes $t\approx 650$ days. The red star shows the value in May 2011, before the outburst. The red dashed line shows the power-law fit of Equation~26 and its $1\,\sigma$ uncertainty (shaded region), while the corresponding broken power-law fit of Equation~27 is depicted as the blue solid line and shaded region. The vertical black dotted line marks the break radius ($R_\mathrm{br}$).}
\label{fig:RtrTeff}
\end{figure}

For a thin disk in thermal equilibrium, we expect the effective temperature $T_\mathrm{eff} \propto R^{-3/4}$. However, the radial dependence of the effective temperature of 1ES\,1927+654 follows a different slope ($-0.47\pm0.04$; Figure~\ref{fig:Rtr_T}), and the correlation has larger intrinsic scatter ($0.043$~dex) than the $T_c - R_\mathrm{tr}$ relation. This is primarily due to the fact that the observations were taken at different epochs, over the time span of which the mass accretion rate underwent significant changes (Figure~\ref{fig:mdot}). Unlike the equatorial temperature, for which $T_c \propto \dot{m}^0$, the effective temperature varies with accretion rate as $T_\mathrm{eff} \propto \dot{m}^{1/4}$. To account for the non-simultaneity issue, we normalize the effective temperature by the dimensionless mass accretion rate (Figure~\ref{fig:RtrTeff}a),

\begin{equation}
\begin{aligned}
    \log{(T_\mathrm{eff}\,\dot{m}^{-1/4}/{\rm K})} &= \\
    (-0.60 \pm 0.03) &\log{(R_\mathrm{tr}/R_g)} + (6.10 \pm0.02).
\end{aligned}
\label{equ:Teff_profile}
\end{equation}

\noindent
The intrinsic scatter is now significantly reduced to $0.014$~dex. However, the revised slope of $-0.6\pm0.03$ fits the expectation of neither the slim disk nor the thin disk model. Closer inspection of the residuals of the single power-law fit reveals a systematic trend with radius (shallower for small $R_\mathrm{tr}$ and steeper for large $R_\mathrm{tr}$) that motivates us to consider replacing the fit with a broken power-law model

\begin{equation}
\begin{aligned}
    \log{(T_\mathrm{eff}\,\dot{m}^{-1/4}/{\rm K})} &=  \\
    (-0.53 \pm 0.08) &\log{(R_\mathrm{tr}/R_g)} + (6.07\pm0.04) \;\;(R<R_\mathrm{br}) \\
    (-0.69 \pm 0.09) &\log{(R_\mathrm{tr}/R_g)} + (6.19\pm0.12) \;\;(R>R_\mathrm{br}),
\end{aligned}
\label{equ:Teff_broken}
\end{equation}

\noindent
where $R_\mathrm{br} = 6.04 \pm 1.60 R_g$ is the best-fit value of the break radius. The intrinsic scatter now reduces to merely $0.005$~dex.

\subsection{Compton Cooling}
\label{sec:Ccooling}

The corona in an AGN is commonly thought to be a layer of high-temperature, magnetically active plasma situated above and below the accretion disk \citep{Haardt1991ApJ,Haardt1994ApJ}. This geometry is especially apparent in 1ES\,1927+654, where the corona's optical depth correlates with the disk surface density \citep{Li2024paper2}. The high-energy electrons in the corona Compton upscatter the thermal photons from the disk, enhancing their energy to produce the hard X-ray emission depicted in Figure~\ref{fig:disksed}. This process of Compton cooling transfers energy to the corona \citep{Svensson1994ApJ} and can reduce the energy dissipated thermally in the accretion disk \citep{Petrucci2020AA}.

The effective temperature of the thermalized disk in 1ES\,1927+654 exhibits a shallower slope within $R_\mathrm{br} \approx 6\, R_g$, wherein $T_\mathrm{eff}$ attains values lower than predicted by the thin disk. Intriguingly, within the same radial range the temperature of the overheated disk also shows a marginal increase, as depicted in Figure~\ref{fig:RT}. Both of these observations hint at some additional cooling process at play in the inner disk region. A plausible explanation could be Compton cooling from the X-ray corona, as the corona's size is often reported to be $\sim 10\,R_g$ based on X-ray reverberation mapping experiments (see review by \citealp{Cackett2021iSci}). Assuming that the optically thin corona has a slab-like geometry above the accretion disk, some of the upscattered photons from the corona may re-enter the disk, contribute to its thermal energy, and thus impact its $T_\mathrm{eff}$. However, this heating effect may be counteracted or even surpassed by the cooling effect caused by the loss of lower energy photons from the disk to the corona. In addition, the disk could lose energy to the corona through Comptonization, which would lower its $T_\mathrm{eff}$. The above considerations lead us to deduce that the corona size is $R_\mathrm{cor} \simeq R_\mathrm{br} \approx 6\,R_g$.

To gain additional insights into the above picture, we calculate the corona bolometric correction, defined as the ratio of the bolometric luminosity to the corona luminosity ($L_\mathrm{cor}$), deduced from SED fitting \citep{Li2024paper2}. Plotting $\kappa_\mathrm{cor} \equiv L_\mathrm{bol}/L_\mathrm{cor}$ as a function of $R_\mathrm{tr}$ reveals a positive correlation for the early epochs with $R_\mathrm{tr} \gtrsim 6\,R_g$. The physical explanation is that at $R_\mathrm{tr} > R_\mathrm{cor}$ the corona resides within the inner overheated disk where efficient free-free cooling dominates over Compton cooling \citep{Li2024paper2}. This results in the high observed $\kappa_\mathrm{cor}$. As $R_\mathrm{tr}$ decreases, the corona becomes gradually exposed, such that once $R_\mathrm{tr} < R_\mathrm{cor}$ the corona is fully exposed and $\kappa_\mathrm{cor}$ becomes independent of $R_\mathrm{tr}$ (Figure~\ref{fig:RtrTeff}b).

More interestingly, for the later epochs, the overheated disk resides within the corona ($R_\mathrm{tr} \lesssim R_\mathrm{cor}$). The roughly constant temperature observed over $4\,R_g \lesssim R_\mathrm{ph} \lesssim 6\,R_g$ (Figure~\ref{fig:RT}) can be attributed to additional Compton cooling. If the corona geometry is layered, both heating and Compton cooling rates are proportional to the dissipation rate of accretion energy \citep{Kawanaka2024}. Thus, the soft X-rays can maintain a steady temperature of $\sim 2\times 10^{6}\,$K within $R_\mathrm{cor}$.

\subsection{Continuum Spectral Index}
\label{sec:UVindex}

The effective temperature profile of the outer thin disk is vital in determining the spectral index of the AGN optical-UV continuum. Given that $T_\mathrm{eff} \propto R^{-p}$, we can derive $\alpha_\nu = 3-2/p$ by integrating the Planck function from $R_\mathrm{in}$ to infinity. Fitting a single power-law to $T_\mathrm{eff}$ versus $R$ using Equation~\ref{equ:Teff_profile}, we obtain $\alpha_\nu \approx -0.32$. However, the steeper slope of the outer disk broken power-law fit (Equation~\ref{equ:Teff_broken}) produces a bluer slope of $\alpha_\nu \approx 0.08$. Regardless, both temperature profiles yield a redder SED compared to the expectations of a standard thin disk, for which $\alpha_\nu = 1/3$. The observed slope of the optical to UV continuum of quasars is even redder ($\alpha_\nu \approx -0.4$; e.g., \citealp{Francis1991ApJ,VandenBerk2001AJ}).

We offer a possible resolution of this apparent discrepancy between observation and theory. For a thin accretion disk in which radiative cooling roughly balances viscous heating, $T_\mathrm{eff} \propto M_\mathrm{BH}^{-1/4}\dot{m}^{1/4} r^{-3/4}$ (Equations~\ref{equ:tdobs} and \ref{equ:qvisp}), with $r \equiv R/R_g$ the normalized radius. For the temperature profile defined in Equation \ref{equ:Teff_broken}, the median BH mass of $M_{\rm BH} = 4\times 10^8\,M_\odot$ \citep{Shen2011ApJS} accreting at $\dot{m} = 0.03$ would have $T_{\rm eff} = 6\times 10^4\,$K at $3\,R_g$. If we suppose a break radius $R_\mathrm{br} \approx 10\, R_g$ based on the typical sizes of coronae (\citealp{Cackett2021iSci}) and integrate the disk spectrum from $3\,R_g$ to $10^4 \,R_g$, roughly the self-gravity radius \citep{Laor1989MNRAS}, fitting a power law to the resulting spectrum over the wavelength range 1300--5500\,\angs\, yields a spectral index $\alpha_\nu \approx -0.37$. This value is notably close to the continuum slope for radio-quiet quasars, which show $\alpha_\nu \approx -0.32$ over the same wavelength range \citep{Francis1991ApJ}. In conclusion, we speculate that the lower $T_\mathrm{eff}$ in the inner region of the disk of 1ES\,1927+654 may be a consequence of additional cooling from the corona. This concept potentially can be generalized to typical quasars, thereby resolving the long-standing conflict between the observed and predicted optical-UV spectral indices.

\section{Discussion}
\label{sec:sec5}

\subsection{Size of the Overheated Disk}
\label{sec:diskscale}

From the framework introduced in Section~\ref{sec:overheat} (Equations~\ref{equ:rhoss} and \ref{equ:wff}), we deduce that in the inner disk the free-free emission capability strongly correlates with the dimensionless mass accretion rate and radius:

\begin{equation}
    \dot{w}_\mathrm{ff} \propto \dot{m}^{-4} R^3.
    \label{equ:wffapprox}
\end{equation}

\noindent
We also expect the inner disk to be geometrically thick, whereupon $H/R \simeq 1$ \citep{Kato2008book}. Combining Equations~\ref{equ:qvisp} and \ref{equ:wheat}, the heating rate density can be given approximately by

\begin{equation}
    \dot{w}^+ \propto \dot{m} R^{-4}.
    \label{equ:wplusapprox}
\end{equation}

\noindent
The above relations imply that as the inner radius of the disk is approached, $\dot{w}^+$ rapidly overwhelms $\dot{w}_\mathrm{ff}$, and at sufficiently small $R$ the disk is always overheated, especially for high $\dot{m}$ \citep{Beloborodov1998MNRAS}. More importantly, based on the temperature profile in Figure~\ref{fig:RT}, we deduce that the overheated region takes the form of a slim disk.

\begin{figure}
\centering
\includegraphics[width=0.48\textwidth]{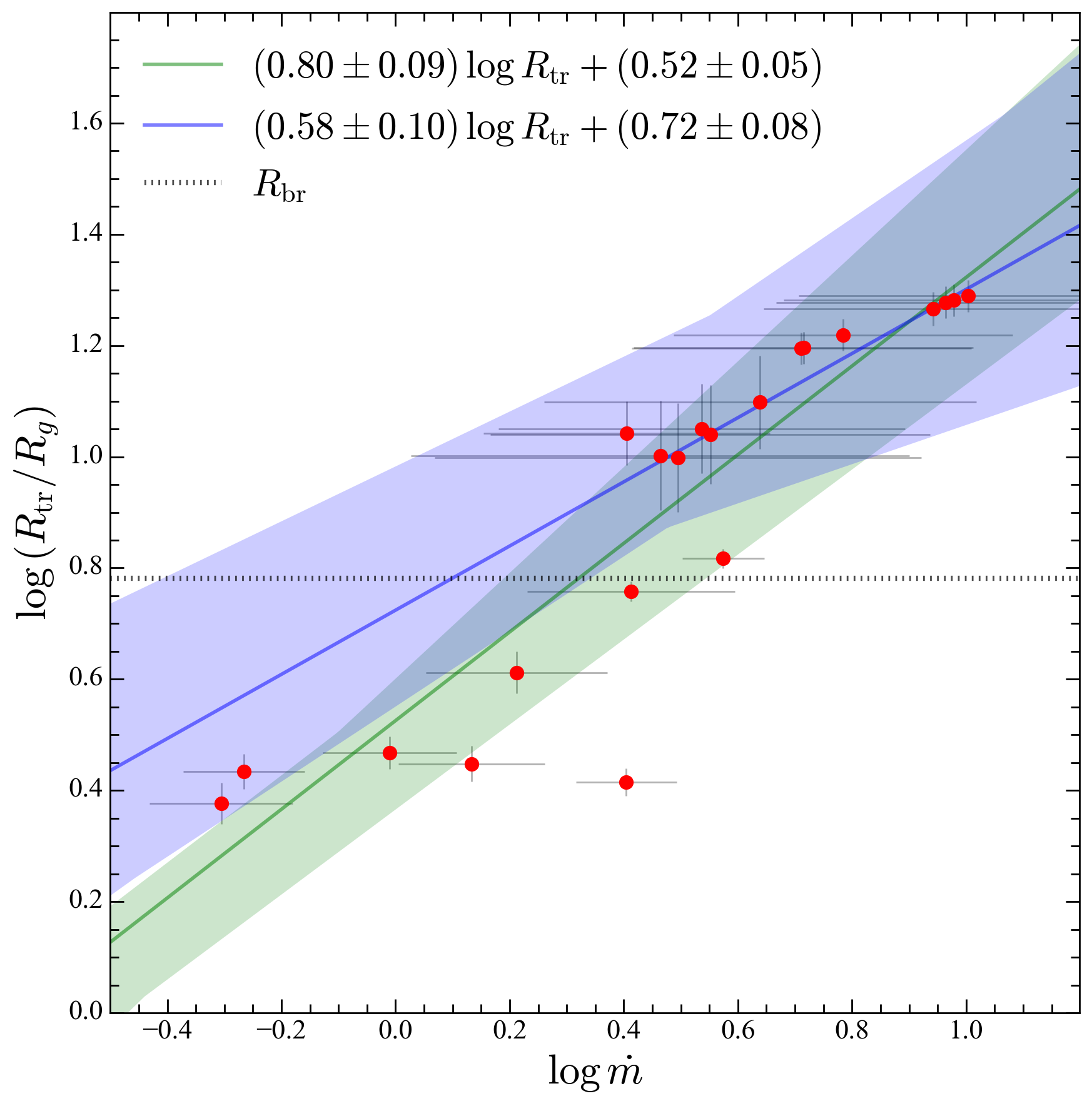}
\caption{The dependence of the truncation radius ($R_\mathrm{tr}$) on the dimensionless mass accretion rate ($\dot{m}$). The horizontal dotted line marks the break radius defined in Equation~\ref{equ:Teff_broken}. The green solid line shows a linear fit for all the points, while the blue solid line only fits the points with $R_\mathrm{tr} > R_\mathrm{br}$; the shaded regions denote the $1\,\sigma$ uncertainties of the fits.}
\label{fig:mdot_r}
\end{figure}

The truncation radius ($R_\mathrm{tr}$) defines the inner boundary of the outer thermalized disk. Within $R_\mathrm{tr}$, the disk overheats, primarily contributing to soft X-ray emission (Section~\ref{sec:sec3}). Assuming that at $R_\mathrm{tr}$ the condition $\dot{w}_\mathrm{ff} \simeq \dot{w}^+$ applies, $R_\mathrm{tr} \propto \dot{m}^{0.71}$ (Equations~\ref{equ:wffapprox} and \ref{equ:wplusapprox}). Figure~\ref{fig:mdot_r} explores this correlation for 1ES\,1927+654. The 21 observational epochs define a relation $\log{r_\mathrm{tr}} = (0.80\pm0.09) \log{\dot{m}} + (0.52\pm0.05$) with $0.14$~dex intrinsic scatter, where $r_\mathrm{tr} \equiv R_\mathrm{tr}/R_g$ denotes the dimensionless truncation radius. Within the error budget, the slope is consistent with the na\"ive expectation. If we focus only on the points where $R_\mathrm{tr} > R_\mathrm{br}$, with the break radius defined in Equation~\ref{equ:Teff_broken}, the intrinsic scatter ($0.07$~dex) reduces by a factor of 2, and the correlation becomes $\log{r_\mathrm{tr}} = (0.58\pm0.10) \log{\dot{m}} + (0.72\pm0.08$). This is to be expected, if, as discussed in Section~\ref{sec:Ccooling}, we associate $R_\mathrm{br}$ with the size of the corona ($R_\mathrm{cor}$). Compton cooling becomes significant within $R_\mathrm{cor}$ and mitigates overheating, thereby thinning the disk \citep{Svensson1994ApJ}. Consequently, we expect an even smaller $R_\mathrm{tr}$ for the region of the disk that is smaller than $R_\mathrm{cor}$. Crucially, however, $R_\mathrm{tr} \simeq R_\mathrm{cor}$ occurs at $\dot{m} \approx 3$ (Figure~\ref{fig:mdot_r}), an accretion rate much higher than typically achieved by most type~1 AGNs. This suggests that most AGNs are characterized by $R_\mathrm{tr} < R_\mathrm{cor}$, and that Compton cooling plays a significant role in determining the precise value of $r_\mathrm{tr}$.

We mention, as an aside, the implications of our results for interpreting a long-standing puzzling property of the soft X-ray excess in AGNs. The observation that the temperature of the soft X-ray excess in 1ES\,1927+654 remains nearly constant between $4\,R_g$ to $6\,R_g$ (Figure~\ref{fig:RT}) is informative. A similar phenomenon has been studied extensively using large samples of AGNs (e.g., \citealt{Crummy2006MNRAS}). Within the context of our framework for the disk structure of 1ES\,1927+654, Compton cooling plays a significant role when $R_\mathrm{tr} < R_\mathrm{br}$, perhaps to a degree that it can sustain a constant temperature (e.g., \citealp{Gierlinski2004MNRAS}). Taking into account the temperature profile of the inner overheated disk (Equation~\ref{equ:Tbb_profile}) and assuming $R_\mathrm{tr} \sim R_\mathrm{ph}$, we obtain $T_\mathrm{bb} \propto M_\mathrm{BH}^{-1/4} r_\mathrm{tr}^{-0.6}$. If $T_\mathrm{bb}$ remains constant for all BHs, in view of the $\dot{m}$ dependence in Figure~\ref{fig:mdot_r}, we can then formulate

\begin{equation}
r_\mathrm{tr} \propto \dot{m}^{0.60} M_\mathrm{BH}^{-0.42}.
\label{equ:littlertr}
\end{equation}

\noindent
From this expression, we can anticipate that the relative strength of the soft X-ray excess, $L_\mathrm{SX}/L_\mathrm{E}$, would be proportional to $R_\mathrm{tr}^2/L_\mathrm{E} \propto \lambda_\mathrm{E}^{1.20} M_\mathrm{BH}^{0.16}$, assuming $\dot{m} \sim \lambda_\mathrm{E}$. This implies a strong dependence of the strength of the soft X-ray excess on $\lambda_\mathrm{E}$ but a weak dependence on $M_\mathrm{BH}$, in excellent agreement with the observations of \citet{Gliozzi2020MNRAS}.

\subsection{Size of the X-ray Corona}
\label{sec:corsize}

Another key scale to consider is the size of the corona. At $R \lesssim R_\mathrm{cor}$, Compton cooling reduces both the temperature of the overheated disk and the effective temperature of the outer thermalized disk, leading to a bending of both profiles (Figures~\ref{fig:RT} and \ref{fig:RtrTeff}). The corona physically can exist only in a location where the disk has a sufficiently low surface density, or else thermal radiation dominates over Compton cooling \citep{Jiang2019bApJ}. The boundary of the corona should be located at the radius that has a certain specific value of disk surface density.  For 1ES\,1927+654, we can infer $R_\mathrm{tr} \simeq R_\mathrm{cor}$ at $t \approx 400$ days (Figure~\ref{fig:radius}), at which time $\Sigma\, |_{R_\mathrm{cor}} \simeq \Sigma\, |_{R_\mathrm{tr}} \approx 700 \, \rm g\, cm^{-2}$ (Table~\ref{tab:diskpro}). Hereinafter, we denote $\Sigma_0 = 700 \, \rm g\, cm^{-2}$.

In the inner region of a thin accretion disk, the surface density increases in the radial direction as $\Sigma \propto R^{3/2}$ \citep{Kato2008book}. Therefore, at some fixed radius, for instance $R = 10\,R_g$, we expect $\Sigma\, |_{10R_g}/\Sigma_0 = (10/r_\mathrm{cor})^{3/2}$, where $r_\mathrm{cor} \equiv R_\mathrm{cor}/R_g$ denotes the dimensionless corona radius. This indicates that the size of the corona is inversely correlated with the surface density of the inner disk, or $r_\mathrm{cor} \sim {\Sigma \, |_{10 R_g}}^{-2/3}$. Assuming, for simplicity, that the luminosity of the corona $L_\mathrm{cor} \propto r_\mathrm{cor}^2$, we expect for the UV/X-ray flux ratio $\alpha_\mathrm{OX} \propto r_\mathrm{cor}^{-2} \sim {\Sigma \, |_{10 R_g}}^{4/3}$. Making use of the correlation between corona plasma optical depth $\tau$ and inner disk surface density reported in \citet{Li2024paper2}, $\Sigma \propto \tau ^{2/5}$, we arrive at $\alpha_\mathrm{OX} \propto \tau^{0.53}$. This agrees reasonably well with the slope $0.43\pm 0.07$ of the $\alpha_\mathrm{OX}-\tau$ relation \citep{Li2024paper2}. Since $\alpha_\mathrm{OX}$ correlates with $\Sigma$, the relation between $\alpha_\mathrm{OX}$ and $\dot{m}$ should follow an $S$-curve qualitatively similar to that of the $\Sigma - \dot{m}$ relation (Section~\ref{sec:scurve}). The disk of 1ES\,1927+654 is initially situated on the branch of a stable slim disk (Figure~\ref{fig:scurve}a), where $\Sigma \propto \dot{m}$, such that $r_\mathrm{cor}$ increases as $\dot{m}$ decreases. Later, $r_\mathrm{cor}$ begins to decrease when the disk transitions onto the $\Sigma \propto \dot{m}^{-1}$ branch, which results in the change of slope in the $\alpha_\mathrm{OX}-L_\mathrm{2500}$ relation \citep{Li2024paper2}. The disk in most type~1 AGNs is expected to be located on the stable lower branch of the $S$-curve, where $\Sigma \propto \dot{m}^{3/5}$ (Equation~\ref{equ:thinSclower}), which predicts an inverse correlation between $r_\mathrm{cor}$ and $\dot{m}$.

Reverberation mapping overcomes the limitations of spatial resolution, using light echoes to ascertain the geometry and kinematics of the central regions of AGNs (see review by \citealp{Cackett2021iSci}). X-ray reverberation mapping, in particular, is a valuable tool for studying the geometry of the X-ray corona (e.g., \citealp{Fabian2009Natur,Uttley2014ARAA}). X-ray reverberation mapping leverages X-ray variability, generally characterized as a red-noise process. Due to this characteristic, the lag-frequency spectrum furnishes the time lag between correlated variability of the light curves in the soft and hard X-ray band. The amplitude of the soft lag, which indicates that the soft band trails the hard band, typically measures the distance between the corona and the disk surfaces that produce the relativistic blurred reflection spectrum in the soft X-ray band. The size of the corona scales with $M_\mathrm{BH}$ \citep{Kara2016MNRAS}, or $R_\mathrm{cor} = r_\mathrm{cor}R_g \propto M_\mathrm{BH}$. The corona size also correlates positively with the intrinsic hard X-ray luminosity in some highly variable AGNs \citep{Kara2013MNRAS,Alston2020NatAs}. Assuming that AGNs maintain a nearly constant $\dot{m}$ because of their long viscous times, thermal instability in the inner disk perhaps induces surface density fluctuations. A decrease in $\Sigma$ would initially increase the hard X-ray flux, based on the $\Sigma-\alpha_\mathrm{OX}$ correlation in \citet{Li2024paper2}, and subsequently would lead to an increase in $r_\mathrm{cor}$ (as discussed in Section~\ref{sec:diskscale}). The same argument may apply to the BH transient MAXI\,J1820+070, whose corona size correlates positively with X-ray hardness while the count rate remains nearly constant \citep{Kara2019Natur}.

\subsection{The $T_c-\Sigma$ Phase-transition Curve}
\label{sec:Tsigma}

The evolution of 1ES\,1927+654 on the $\dot{M}-\Sigma$ phase-transition curve (Figure~\ref{fig:scurve}) traces an evolutionary path in which prior to $t = 650$ days the source lies on the slim disk branch ($\dot{M} \propto \Sigma$), while thereafter it transitions to the radiation pressure-dominated, thin disk branch ($\dot{M} \propto \Sigma^{-1}$). At fixed radius, the equatorial temperature $T_c$ also depends differently on $\Sigma$ for the two states. Since radiation pressure dominates in both disk states, from Equations~\ref{equ:EoS} and \ref{equ:vertical} we infer that the total pressure $\Pi \propto T_c^4 H \propto T_c^4(\Pi/\Sigma)^{1/2}$, or $\Pi \propto T_c^8/\Sigma$. In conjunction with $\dot{M} \propto T_c^8 \Sigma^{-1}$ (Equation~\ref{equ:Lcons}), we can recast the slim disk branch ($\dot{M} \propto \Sigma$) into

\begin{equation}
    T_c \propto \Sigma^{1/4}
    \label{equ:tsigslim}
\end{equation}

\noindent
and the radiation pressure-dominated thin disk branch ($\dot{M} \propto \Sigma^{-1}$) into

\begin{equation}
    T_c \propto \Sigma^{0}.
    \label{equ:tsigthin}
\end{equation}

\noindent
Note, however, that we cannot directly place 1ES\,1927+654 on the $T_c - \Sigma$ plot using parameters derived at $R_\mathrm{tr}$ (Table~\ref{tab:diskpro}), for the evolution of $T_c$ at different epochs comes not from variation of accretion rate but from changes in $R_\mathrm{tr}$ (Equation~\ref{equ:Tc_profile}). Unlike $\dot{M}$, which is constant with radius for a steady disk, $T_c$ should be normalized to the same radius before we can use it for the $T_c-\Sigma$ phase-transition curve. Recalling that the accretion flow within $R_\mathrm{tr}$ is an overheated slim disk described by the temperature profile in Equation~\ref{equ:Tbb_profile}, we extrapolate the temperature profile inward to normalize the equatorial temperature as

\begin{equation}
    {T_c}^\prime \equiv T_c \, |_{R_\mathrm{tr}} \times (R_\mathrm{tr}/R_g)^{0.60}.
    \label{equ:tcprime}
\end{equation}

\noindent
In the same spirit, we normalize the surface density, assuming a radial profile of $\Sigma \propto R^{-1/2}$ for a slim disk \citep{Kato2008book}:

\begin{equation}
    \Sigma ^\prime \equiv \Sigma \, |_{R_\mathrm{tr}} \times (R_\mathrm{tr}/R_g)^{0.50}.
    \label{equ:sigprime}
\end{equation}

\begin{figure}
\centering
\includegraphics[width=0.48\textwidth]{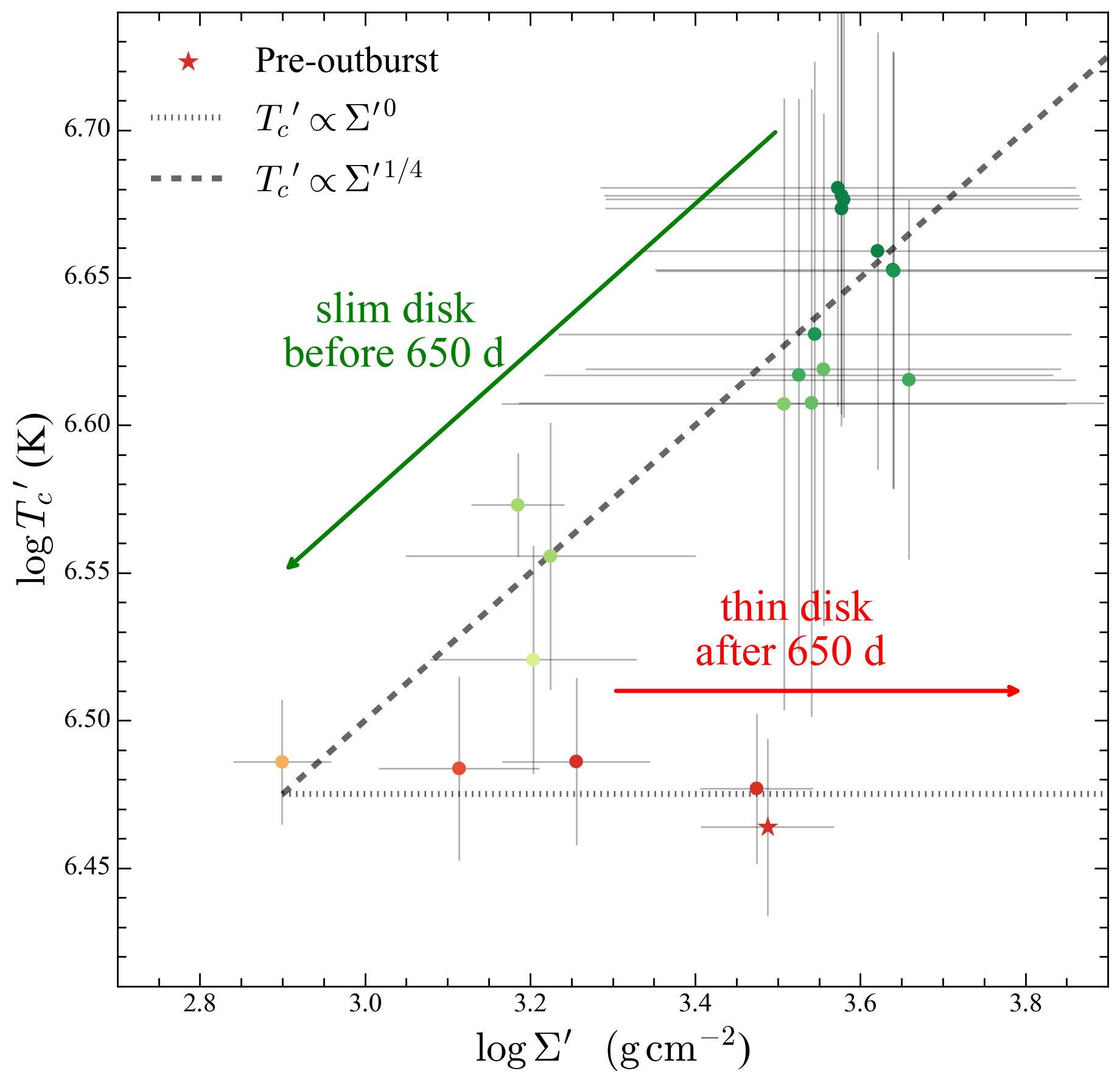}
\caption{The normalized disk equatorial temperature (${T_c}^\prime$, Equation~\ref{equ:tcprime}) as a function of normalized disk mass surface density ($\Sigma^\prime$, Equation~\ref{equ:sigprime}). The color of the circles represents the time since the outburst (23 December 2017), from the beginning (green) to the end (red), where the orange one denotes $t\approx 650$ days. The red star shows the value in May 2011, before the outburst. The dashed line shows $T_c^\prime \propto \Sigma^{\prime 1/4}$, which is expected from an advection-dominated slim disk, while the dotted line represents $T_c^\prime \propto \Sigma^{\prime  0}$ as expected from a radiation pressure-dominated thin disk. The two arrows indicate the direction of movement of 1ES\,1927+654 along the slim disk and thin disk branches of the $S$-curve before and after $t = 650$ days, respectively.}
\label{fig:Tsigma}
\end{figure}

\noindent
In the normalized ${T_c}^\prime-\Sigma^\prime$ plane (Figure~\ref{fig:Tsigma}), prior to $t = 650$ days 1ES\,1927+654 evolves tightly along the slim disk branch, where $T_c \propto \Sigma^{1/4}$; after $t = 650$ days, 1ES\,1927+654 follows the $T_c \propto \Sigma^{0}$ track of a radiation pressure-dominated thin disk. These results strongly reinforce the slim-to-thin disk transition discussed in Section~\ref{sec:scurve}.

It is worth mentioning that when normalizing $T_c$ through Equation~\ref{equ:tcprime}, we chose to extrapolate inward, using the slim disk profile for the inner overheated disk (Equation~\ref{equ:Tbb_profile}), normalizing both $T_c$ and $\Sigma$ to values at small radii. We could have chosen to normalize $T_c$ by adopting the temperature profile of the outer thin disk ($T \propto R^{-3/8}$). Such an outward normalization would result in ${T_c}^\prime \simeq \text{constant}$, because Equation~\ref{equ:Tc_profile} is extremely close to $T \propto R^{-3/8}$. This is naturally expected. Once we normalize $T_c$ outwardly, at $R \gtrsim R_\mathrm{tr}$ the disk is always a thin disk, and only the $T_c \propto \Sigma^{0}$ branch  (${T_c}^\prime \simeq \text{constant}$) remains on the $T_c-\Sigma$ $S$-curve.

\subsection{Estimating the Disk Mass Surface Density}
\label{sec:disksigma}

The companion study of 1ES\,1927+654 by  \citet{Li2024paper2} reports a two-branch correlation between the corona optical depth ($\tau$) and the dimensionless mass accretion rate. From comparing the slope of the $\tau - \dot{m}$ correlation with the $\dot{M}-\Sigma$ relation ($S$-curve), \citet{Li2024paper2} suggested that there exists a strong disk-corona connection, whereby disk surface density and coronal optical depth are related as $\Sigma \propto \tau^{2/5}$. This proposal links $\Sigma$, a quantity that cannot be measured directly, with $\tau$, which can be inferred from X-ray spectra.

\begin{figure}
\centering
\includegraphics[width=0.48\textwidth]{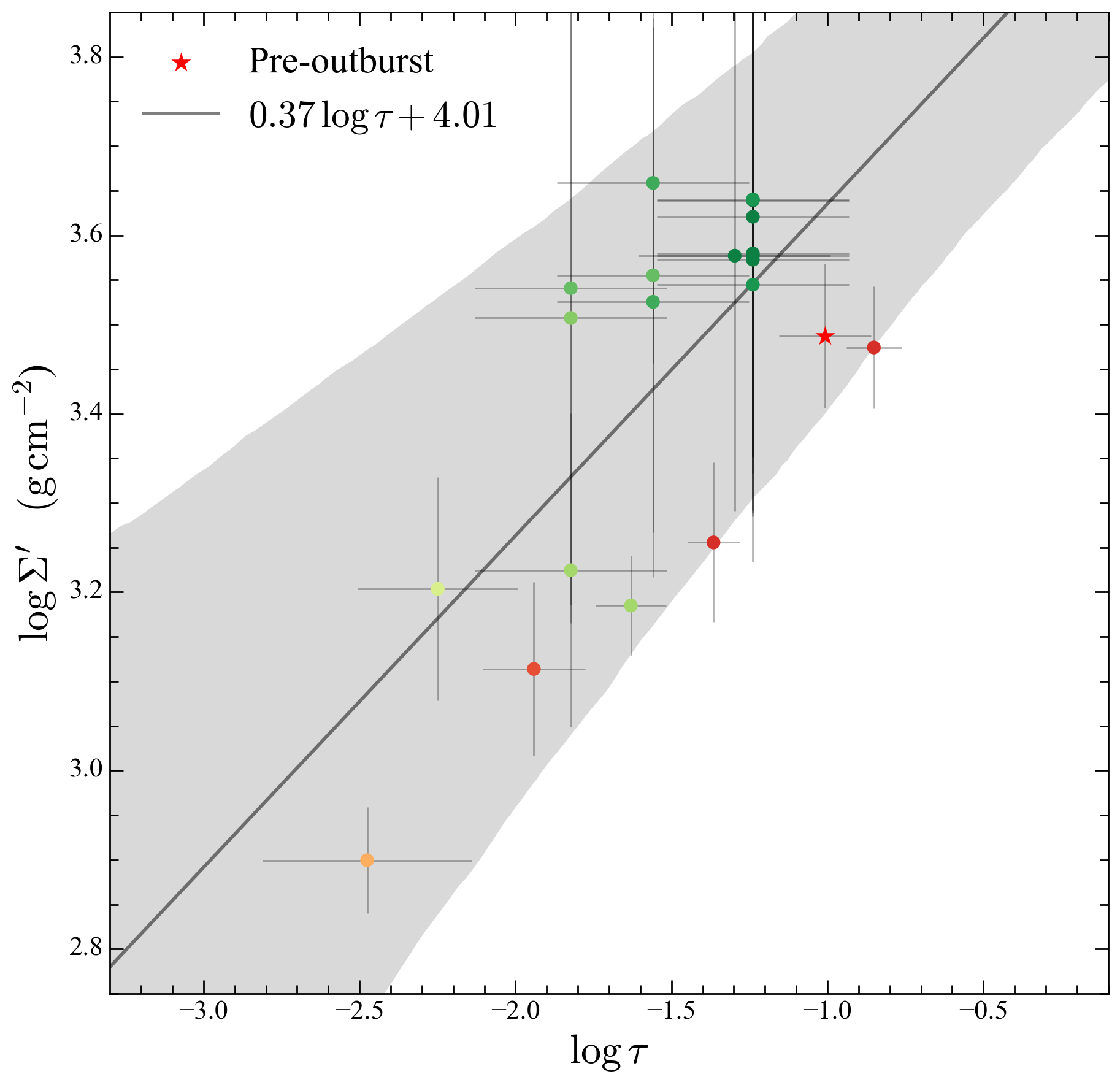}
\caption{The normalized disk mass surface density ($\Sigma^\prime$, Equation~\ref{equ:sigprime}) as a function of the corona optical depth ($\tau$). The color of the circles represents the time since the outburst (23 December 2017), from the beginning (green) to the end (red),  where the orange one denotes $t\approx 650$ days. The red star shows the value in May 2011, before the outburst. The solid line gives the linear fit $\log{\Sigma^\prime} = (0.37 \pm 0.08) \log{\tau} + (4.01 \pm 0.13)$, with the gray shaded region representing the $1\,\sigma$ uncertainty of the fit.}
\label{fig:sigmatau}
\end{figure}

As mentioned in Section~\ref{sec:corsize}, $\Sigma$ here should represent the value at a fixed radius near the BH. Comparison of the corona optical depth derived in \citet{Li2024paper2} with the normalized disk surface density (Equation~\ref{equ:sigprime}) produces a strong correlation of the form

\begin{equation}
    \log{(\Sigma^\prime/\mathrm{g\,cm^{-2}})} = (0.37 \pm 0.08) \log{\tau} + (4.01 \pm 0.13),
    \label{equ:sigprimefit}
\end{equation}

\noindent
which has an intrinsic scatter of 0.14~dex (Figure~\ref{fig:sigmatau}). The slope of $0.37 \pm 0.08$ is consistent with the slope of 0.4 proposed in \citet{Li2024paper2}. This analysis warrants further refinement using a statistical sample, to test the viability of using Equation~\ref{equ:sigprimefit} as a tool to estimate the inner properties of accretion disks.

\subsection{The Unstable Disk Solution}
\label{sec:unstable}

The transition from a slim to a thin disk in 1ES\,1927+654 is smooth, gradually traversing through the unstable ($\dot{M} \propto \Sigma$) region dominated by radiation pressure (Figure~\ref{fig:scurve}). Theoretically, thermal instability in regions dominated by radiation pressure provokes a swift transition between the slim disk and gas pressure-dominated thin disk  solutions \citep{Kato2008book}, a concept encapsulated in the ``limit-cycle oscillation'' model. One-dimensional hydrodynamic calculations of thermal instability (e.g., \citealp{Watarai2003ApJ}) elegantly explain the periodic transition between the super-Eddington and near-Eddington states observed in the BH X-ray binary star system GRS\,1915+105 \citep{Belloni2000AA}. \citet{Ohsuga2006ApJ} also successfully used two-dimensional simulations to reproduce periodic variations stemming from disk instabilities under radiation pressure-dominated conditions. In this content, the unstable trajectory of 1ES\,1927+654 during the late epochs of its evolution becomes a fascinating phenomenon.

The ``limit-cycle oscillation'' phenomenon, if directly applicable to AGNs, would occur on a much longer timescale in supermassive BHs. Given that the thermal timescale is comparable to the dynamical timescale $\sqrt{r^3}M_\mathrm{BH}$, the transition timescale of approximately 100 seconds seen in GRS\,1915+105 would translate to $\sim 100$ days for 1ES\,1927+654. This duration is much shorter than the accretion timescale of this changing-look AGN (Figure~\ref{fig:mdot}), suggesting that the change in the accretion state primarily stems from variations in the fallback rate of matter from the TDE. Alternatively, thermal instability could be suppressed by magnetic fields \citep{Svensson1994ApJ, Hirose2009ApJ}.

\subsection{Implications for AGN Disks}
\label{sec:impli}

In response to its rapidly changing mass accretion following a TDE, the post-outburst SED of 1ES\,1927+654 underwent dramatic and distinct variations \citep{Li2024paper2} that we ascribe to three physical components: a lower temperature multicolor blackbody, a higher temperature blackbody, and optically thin Comptonised emission. While the lower temperature blackbody has an equatorial temperature profile consistent with the theoretical predictions of a thin disk (Section~\ref{sec:Tprofile_o}), the hotter blackbody exhibits a photosphere size evolution akin to that of the inner boundary of the outer thin disk (Figure~\ref{fig:radius}). Upon consideration of the effects of photon trapping and electron scattering within the inner disk, we conclude that the blackbody component peaking in the soft X-rays emanates from an inner overheated disk, positioned between the innermost stable circular orbit and the inner boundary of the outer thin disk. The temperature profile of the overheated disk resembles that of a slim accretion disk (Figure~\ref{fig:RT}; \citealp{Abramowicz1988ApJ,Wang1999ApJ}). For the optically thin Comptonised component or corona, \citet{Li2024paper2} underscore the correlation between the corona's optical depth and the inner disk's surface density, leading us to favor a slab geometry for the corona, positioned over the inner accretion disk. Figure~\ref{fig:disksed} gives a  schematic picture of the two-zone accretion disk in 1ES\,1927+654.

\begin{figure*}
\centering
\includegraphics[width=0.98\textwidth]{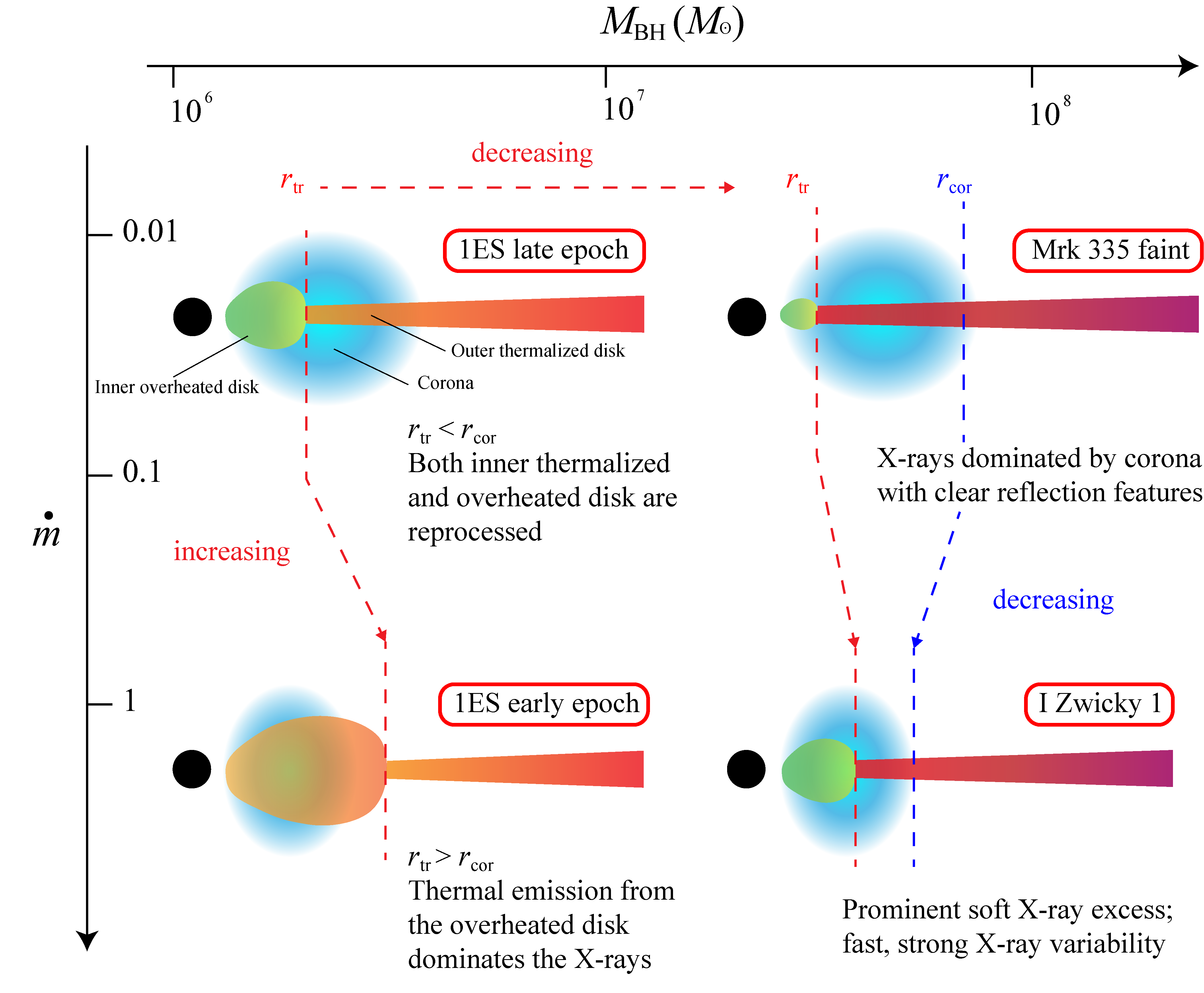}
\caption{Cartoon to illustrate the characteristic scales of the accretion disk discussed in Section~\ref{sec:impli} and their relation with BH mass ($M_\mathrm{BH}$) and dimensionless mass accretion rate ($\dot{m}$). The black circle depicts the BH, exterior to which is an outer thin truncated disk (red), an inner puffed-up overheated disk (green when Compton cooling is dominant, but otherwise orange for low $M_\mathrm{BH}$ and high $\dot{m}$), and an X-ray corona (blue). As $M_\mathrm{BH}$ increases, the truncation radius ($R_\mathrm{tr}$) decreases but the radius of the corona ($R_\mathrm{cor}$) remains constant (Equation~\ref{equ:littlertr}). By contrast, for 1ES\,1927+654 and normal type~1 AGNs such as Mrk~335 and I~Zwicky~1, $R_\mathrm{tr}$ increases while $R_\mathrm{cor}$ generally decreases with increasing $\dot{m}$.}
\label{fig:cartoon2}
\end{figure*}

Are the generic properties of the two-zone accretion disk of 1ES\,1927+654 applicable to other AGNs? The structure of the two-zone accretion disk is largely determined by $\dot{m}$ and $M_\mathrm{BH}$ (Equation~\ref{equ:littlertr}), as encapsulated in the schematic in Figure~\ref{fig:cartoon2}. During the early epochs of 1ES\,1927+654's evolution, when $\dot{m} \gtrsim 3$, the truncation radius exceeds the radius of the corona, and the inner overheated disk emits thermal radiation. However, when the accretion rate subsides to $\dot{m} \lesssim 1$, $r_\mathrm{tr}$ retracts significantly.

Within the corona, some of the seed photons from both the overheated disk and the thermalized disk within $r_\mathrm{cor}$ upscatter, which lowers the effective temperature of the temperature profile (as demonstrated in Figures~\ref{fig:RT} and \ref{fig:RtrTeff}). Our basic model for 1ES\,1927+654 can be applied to AGNs with more massive BHs ($M_\mathrm{BH} \gtrsim 10^7 \,M_\odot$). As $M_\mathrm{BH}$ increases, $r_\mathrm{tr}$ decreases, according to Equation~\ref{equ:littlertr}. Since $\Sigma$ depends only weakly on $M_\mathrm{BH}$ for all three disk states on the $S$-curve, the size of the corona should remain largely invariant with $M_\mathrm{BH}$ because of its sensitivity to $\Sigma$ (Section~5.2). Thus, for $\dot{m}$ of merely a few percent, a value common for most type~1 AGNs, the X-ray spectrum is mainly dominated by a cut-off power law. Lacking the soft X-ray excess, the disk is less ionized and hence more reflection-dominated \citep{Nandra2007MNRAS}. A typical example in this regime might be the faint state of Mrk\,335, during which \cite{Mastroserio2020MNRAS} detect a strong K$\alpha$ reflection feature. As $\dot{m}$ rises to or above the Eddington limit, and consequently $r_\mathrm{tr}$ expands, the soft X-ray excess emitted by the inner overheated disk also gains prominence. In view of the fact that in most cases $r_\mathrm{tr} < r_\mathrm{cor}$, Compton cooling in the overheated disk maintains a near-constant temperature, as observed in narrow-line Seyfert~1s \citep{Gierlinski2004MNRAS}. It is also noteworthy that, within the context of our picture, the corona partially overlays the inner overheated disk, while its remainder extends above the more extensive outer disk. This depiction broadly describes the geometry of the corona in I~Zwicky~1 inferred from X-ray reverberation mapping \citep{Wilkins2017MNRAS}.

\subsection{Do Slim Disks Exist in other AGNs?}
\label{sec:otherAGN}

The inner portion of an accretion disk can become overheated by reducing the free-free emission capability (Equation~\ref{equ:wff}) or elevating the heating rate density (Equation~\ref{equ:wheat}), conditions that might arise when the mass accretion rate is elevated or the BH mass is low (Equation~\ref{equ:rhoss}). Interestingly, the truncation radius depends on both $\dot{m}$ and $M_\mathrm{BH}$ (Equation~\ref{equ:littlertr}). From a theoretical standpoint, there is no fixed critical accretion rate above which the disk would transform into a slim state \citep{Sadowski2011}. The transition point on the $S$-curve is always a function of radius. For instance, in the case of 1ES\,1927+654 the critical value of $\dot{m} \approx 3$ at $\sim 5\,R_g$ (Figure~\ref{fig:scurve}). Nevertheless, as $r_\mathrm{tr}$ increases the slim disk component can dominate the SED, as evidenced by the spectral transition observed in \citet{Li2024paper2}.

Can we find evidence for slim disks in other galactic nuclei? Figure~\ref{fig:scurve} delineates the transition branch from a thin disk to a slim disk, situated at $\dot{m}$ values ranging from $\sim 0.5$ to 2. Within this branch, $\Sigma$ decreases as it approaches the slim disk state. For AGNs with high values of $\lambda_\mathrm{E}$ ($\gtrsim 1$)---an observational proxy for $\dot{m} \gtrsim 1$---less massive BHs are expected to exhibit a slim disk-dominated SED because they possess larger $r_\mathrm{tr}$ (Equation~\ref{equ:littlertr}). In contrast, more massive BHs would present a thin disk-dominated SED. Consequently, as $\dot{m}$ increases the gas within $r_\mathrm{cor}$ is more prone to settle onto a slim disk when $M_{\rm BH}$ is low. Because the disk resides in the middle branch of the $S$-curve, for the same $\dot{m} \approx 1$, $\Sigma$ would be smaller for less massive BHs. Drawing from the $\alpha_\mathrm{OX}-\Sigma$ correlation identified by \citet{Li2024paper2}, we anticipate less massive BHs to exhibit a higher $\alpha_\mathrm{OX}$ than their more massive counterparts at $\lambda_\mathrm{E} \gtrsim 1$, with this disparity reducing as $\lambda_\mathrm{E}$ decreases. This hypothesis is corroborated by the findings of \citet[][their Figure~10a]{Liu2021ApJ}, which illustrate that AGNs with smaller $M_\mathrm{BH}$ consistently possess higher $\alpha_\mathrm{OX}$ values at elevated values of $\lambda_\mathrm{E}$. In light of these observations and the theoretical framework articulated in this paper, it is reasonable to infer that slim disks are prevalent in highly accreting AGNs. Furthermore, $\alpha_\mathrm{OX}$ emerges as an effective indicator for identifying such systems.

\section{Summary}
\label{sec:sec6}

The changing-look AGN 1ES\,1927+654 underwent dramatic changes in both flux and spectral shape in the aftermath of an optical outburst on 23 December 2017 \citep{Trakhtenbrot2019ApJ,Ricci2020ApJL}. Based on a BH mass of $M_\mathrm{BH} = 1.38_{-0.66}^{+1.25}\times 10^6\, M_\odot$ derived from the properties of the host galaxy, the accretion rate following the optical outburst was sufficiently high to qualify as super-Eddington \citep{Li2022paper1}. We explore the physical properties of the accretion flow by comparing basic predictions from accretion disk theory with physical parameters derived from in-depth decomposition of the SEDs obtained from 21 epochs of simultaneous broad-band observations. Our key results and conclusions include:

\begin{enumerate}

\item The long-term observing campaign tracks a time-dependent accretion rate that declines as a power law with index $\gamma = 1.53_{-0.10}^{+0.10}$, consistent with the theoretical predictions for a tidal disruption event. The total accreted mass adds up to $0.55_{-0.04}^{+0.05}\, M_\odot$, suggesting that the disrupted star had an original mass of $\sim 1.1\,M_\odot$.

\item During the initial phase when the accretion rate was high, the radiation efficiency was relatively low with $\eta \approx 0.03$, systematically increasing to $\eta \approx 0.08$ at late times as the accretion rate declined. This trend agrees well with the theoretical predictions of a slim accretion disk, wherein photon trapping at high accretion rates leads to a suppression of radiation efficiency.

\item The mass accretion rate and disk surface density trace a systematic evolutionary trend that transitions from a slim disk at $t \lesssim 650$ days after the outburst to a truncated outer thin disk thereafter.

\item The slim disk-dominated accretion phase was distinguished by an overheated spectrum that radiated predominantly as a soft X-ray excess. The blackbody temperature profile $T \propto R^{-0.60}$ of the inner overheated disk matches the theoretical expectations of a slim disk obtained from numerical simulations and analytical self-similar solutions.

\item The temperature profile of the outer disk follows the predictions of a thin disk beyond a break radius of $\sim 6\,R_g$ but flattens interior to it as a consequence of Compton cooling with the overlying X-ray corona.

\end{enumerate}

The inner truncation of the thermalized disk in 1ES\,1927+654 is due to low gas density. The inner disk is overheated as a result of the free-free emission capability being less than the heating rate density. A noticeable correlation between the disk truncation radius and the dimensionless mass accretion rate suggests that the disk becomes increasingly susceptible to overheating when the accretion rate is high. The constancy of the soft X-ray excess temperature, along with its relationship to the disk truncation radius, endorses the role of Compton cooling in maintaining temperature stability and thinning the disk. The size of the corona, which directly influences the cooling and temperature of the disk, appears to correlate distinctly with the disk surface density. A reduction in this density would trigger a growth in the corona size and a corresponding rise in hard X-ray flux. We anticipate that disk surface density correlates inversely with the size of the corona's boundary.

In terms of wider implications, the structure of a two-zone accretion disk akin to that of 1ES\,1927+654 is predominantly determined by the mass accretion rate and the BH mass. Variations in these parameters lead to variations in the relative contribution to the total emission of the overheated disk and thermalized disk, impacting the observed X-ray spectrum. Our model may be broadly applicable to the physical understanding of other AGNs.

\begin{acknowledgments}
We are grateful to referee for many insightful and constructuve comments. We thank Erin Kara and Ken Ohsuga for insightful discussions that greatly aided in the interpretation of our results. This work was supported by the National Science Foundation of China (11721303, 11991052, 12011540375, 12233001), the National Key R\&D Program of China (2022YFF0503401), and the China Manned Space Project (CMS-CSST-2021-A04, CMS-CSST-2021-A06). CR acknowledges support from Fondecyt Regular grant 1230345 and ANID BASAL project FB210003.
\end{acknowledgments}

\vspace{5mm}
\facilities{Swift(XRT and UVOT), XMM-Newton (EPIC and OM), Las Cumbres Observatory}

\software{George \citep{Ambikasaran2015ITPAM}, SAS (v19.0.0; \citealp{Gabriel2004ASPC}), xspec (12.11.1; \citealp{Arnaud1996ASPC})}

\appendix
\section{Dependence of results on different values of $\alpha$}
\label{app:test}

\begin{figure*}
\figurenum{A1}
\centering
\includegraphics[width=0.48\textwidth]{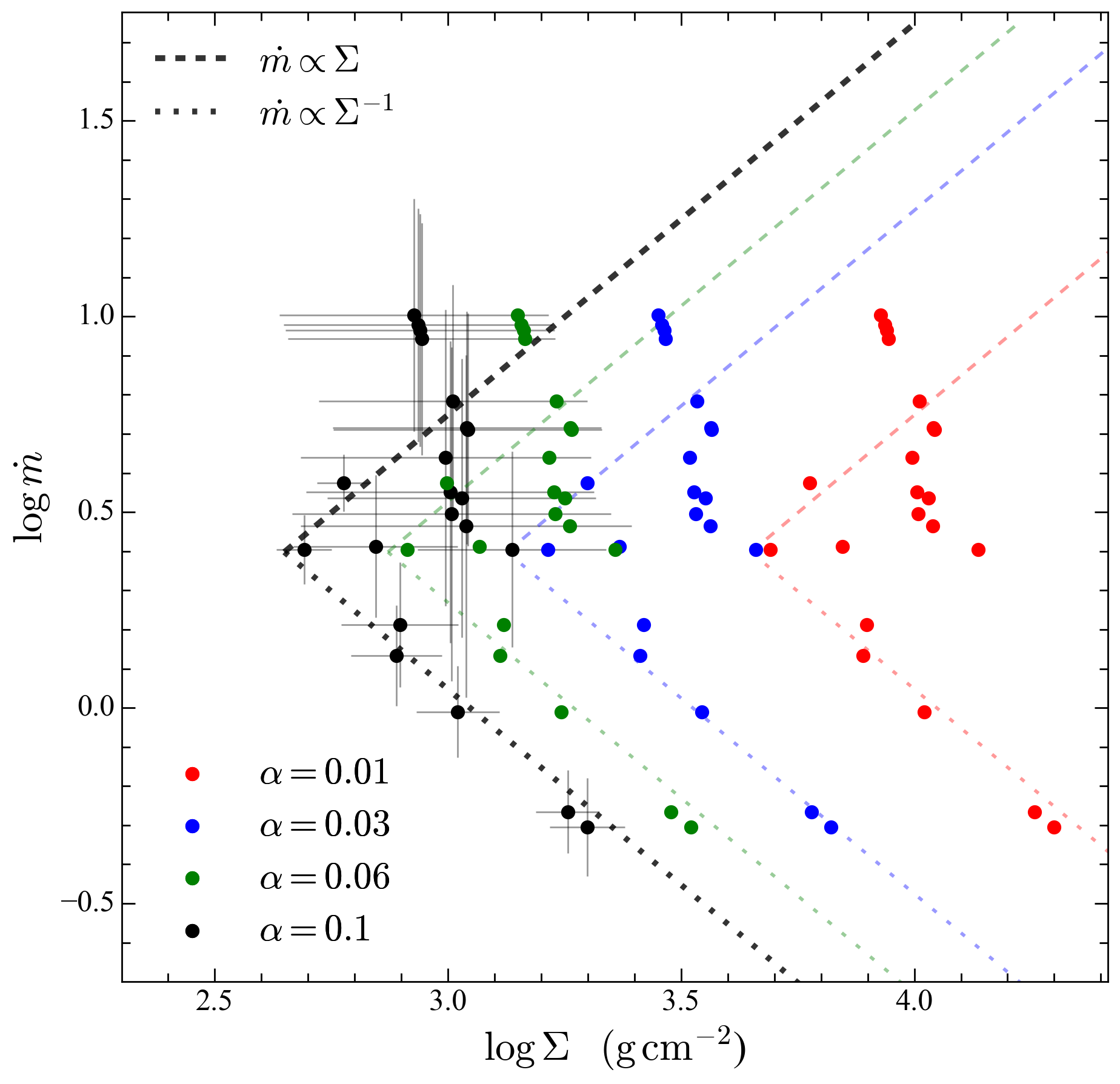}
\includegraphics[width=0.48\textwidth]{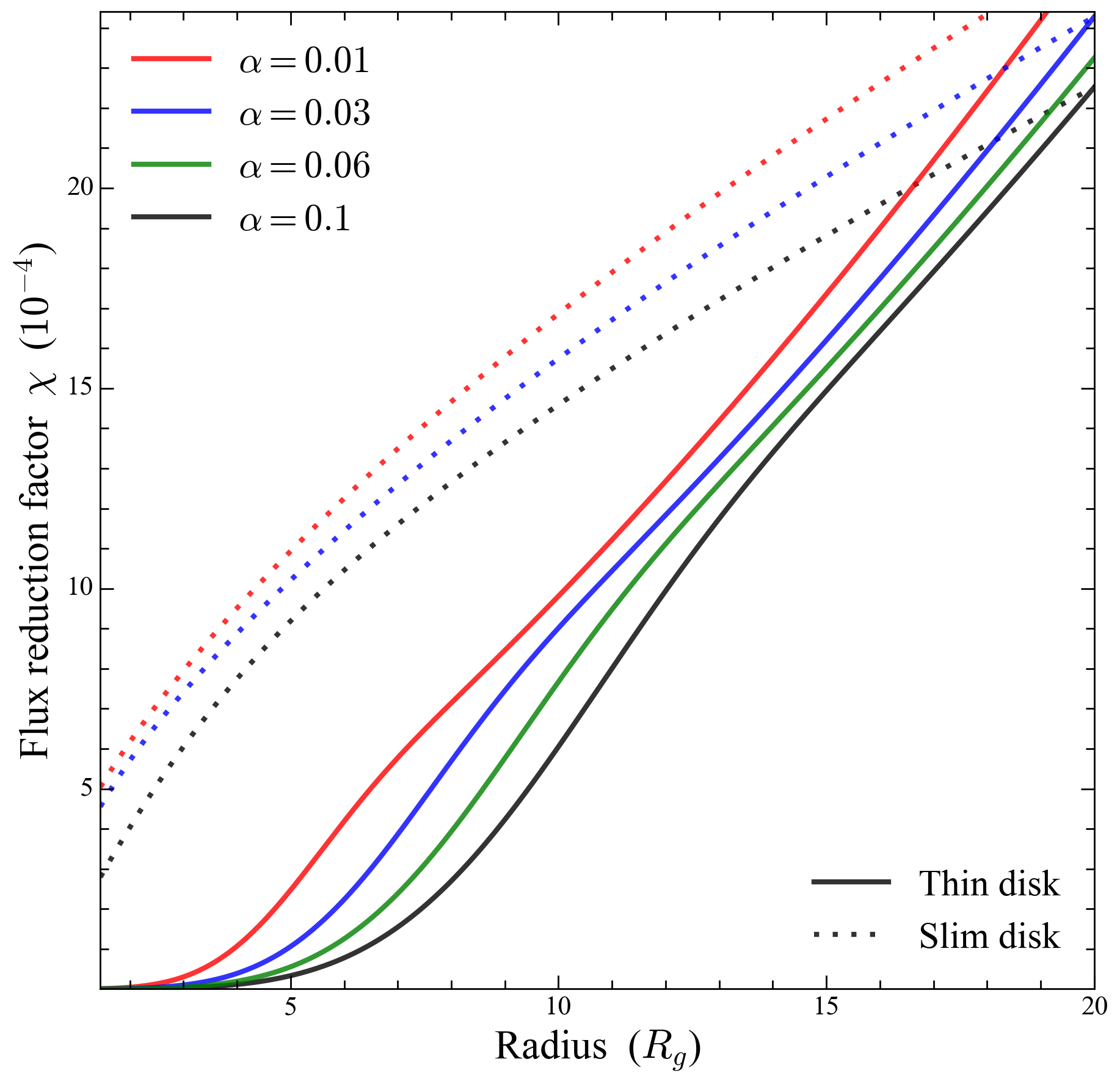}
\caption{ The $\dot{m}-\Sigma$ ``$S$-curve'' (left) and flux reduction factor (right) with different values of the viscosity parameter ($\alpha$). The black error bars indicate the $\dot{m}$ and $\Sigma$ derived from our fiducial choice $\alpha=0.1$ (Table~\ref{tab:diskpro}), and the black dashed line represents the slim disk branch $\dot{m} \propto \Sigma$, while the black dotted line represents the radiation pressure dominated thin disk branch $\dot{m} \propto \Sigma^{-1}$. The colored points of $\dot{m}$ and $\Sigma$ are calculated from the same equation set in Section~\ref{sec:equs} but with different values of $\alpha$: 0.01 (red), 0.03 (blue), and 0.06 (green). The two branches in the left panel are also shifted accordingly. The right panel shows the flux reduction factor ($\chi$), defined in Section~\ref{sec:escatt}, as a function of radius, the same as in Figure~\ref{fig:fluxredu} with $i=0^\circ$. Solid lines assume the $\bar{\rho}$ and $T$ profile for a radiatively cooled accretion flow (thin disk), while dotted lines pertain to an optically thick, advection-dominated accretion flow (slim disk). Colors represent different $\alpha$: 0.01 (red), 0.03 (blue), 0.06 (green), and 0.1 (black).} 
\label{fig:alphatest}
\end{figure*}

In Section~\ref{sec:sec2}, when deriving local disk properties, such as disk surface density ($\Sigma$) and mid-plane temperature ($T_c$) at the truncation radius, we set the viscosity parameter to $\alpha = 0.1$. In radiation magnetohydrodynamical simulations, the value of $\alpha$ is mainly determined by Reynolds and Maxwell stresses \citep{Penna2013MNRAS,Jiang2019aApJ}. Because in these simulations the magnetic pressure usually increases as the BH is approached, $\alpha$ typically increases with decreasing radius, for instance from $\alpha = 0.01$ at $\sim 20 R_g$ to $\alpha = 0.1$ at $\lesssim 4 R_g$ \citep{Jiang2019aApJ}. Without knowing the details of the magnetic field strength in the disk of 1ES\,1927+654, in this study we fix $\alpha = 0.1$ for all epochs. Here, we reevaluate the results presented in this paper by exploring different values of $\alpha$. In $\alpha$-disk theory, given $R_\mathrm{tr}$ and $T_\mathrm{eff}$, $\dot{M}$ does not depend on $\alpha$, but we expect $\Sigma \propto 1/\alpha$. This is exactly what we see in the left panel of Figure~\ref{fig:alphatest}. Even though the value of $\Sigma$ changes, if we shift the $\dot{m} \propto \Sigma$ and $\dot{m} \propto \Sigma^{-1}$ black lines accordingly by $\log{10}$, $\log{\frac{10}{3}}$, and $\log{\frac{10}{6}}$ dex for $\alpha$ values of 0.01, 0.03, and 0.06, respectively, the points still move very well along the phase-transition curve. Therefore, our result that 1ES\,1927+654 went through a slim-to-thin disk transition along the theoretical $\dot{m}-\Sigma$ ``$S$-curve'' still holds (Figure~\ref{fig:scurve}). For the same reason, the qualitative results in the $T_c-\Sigma$ phase-transition curve (Figure~\ref{fig:Tsigma}) also are not affected by different values of $\alpha$. Because the exact value of local disk properties changes with $\alpha$, the flux reduction factor due to electron scattering should also change, as shown in the right panel of Figure~\ref{fig:alphatest}. Using the expression $\chi \sim \sqrt{1-\bar{\omega}}$ \citep{Zhu2019ApJ}, the radially integrated luminosity reduction factor (Equation~\ref{equ:chiL}) is $\chi_L^\mathrm{td} = 8.6_{-2.5}^{+3.7} \times 10^{-4}$ and $\chi_L^\mathrm{sd} = 12.2_{-3.5}^{+3.7} \times 10^{-4}$ for $\alpha = 0.01$, $\chi_L^\mathrm{td} = 7.9_{-2.4}^{+3.6} \times 10^{-4}$ and $\chi_L^\mathrm{sd} = 11.4_{-3.3}^{+3.5} \times 10^{-4}$ for $\alpha = 0.03$, and $\chi_L^\mathrm{td} = 7.4_{-2.4}^{+3.5} \times 10^{-4}$ and $\chi_L^\mathrm{sd} = 10.9_{-3.2}^{+3.3} \times 10^{-4}$ for $\alpha = 0.06$. The $\chi_L$ values derived from different $\alpha$ values are consistent with the case of $\alpha = 0.1$: $\chi_L^\mathrm{td} \approx 7.0 \times 10^{-4}$ and $\chi_L^\mathrm{sd} \approx 10.5 \times 10^{-4}$ (Section~\ref{sec:slimsize}). Therefore, adopting different values of $\alpha$ does not affect our calculation of $R_\mathrm{ph}$ through Equation~\ref{equ:radiusph}. In summary: although the exact disk properties in Table~\ref{tab:diskpro} depend on $\alpha$, as predicted by $\alpha$-disk theory, the main results and conclusions of this study are not affected by the assumed value of $\alpha$.

\begin{thebibliography}{}
\expandafter\ifx\csname natexlab\endcsname\relax\def\natexlab#1{#1}\fi
\providecommand{\url}[1]{\href{#1}{#1}}
\providecommand{\dodoi}[1]{doi:~\href{http://doi.org/#1}{\nolinkurl{#1}}}
\providecommand{\doeprint}[1]{\href{http://ascl.net/#1}{\nolinkurl{http://ascl.net/#1}}}
\providecommand{\doarXiv}[1]{\href{https://arxiv.org/abs/#1}{\nolinkurl{https://arxiv.org/abs/#1}}}

\bibitem[{{Abramowicz} {et~al.}(1988){Abramowicz}, {Czerny}, {Lasota}, \& {Szuszkiewicz}}]{Abramowicz1988ApJ} {Abramowicz}, M.~A., {Czerny}, B., {Lasota}, J.~P., \& {Szuszkiewicz}, E. 1988, \apj, 332, 646

\bibitem[{{Alston} {et~al.}(2020){Alston}, {Fabian}, {Kara}, {Parker}, {Dovciak}, {Pinto}, {Jiang}, {Middleton}, {Miniutti}, {Walton}, {Wilkins}, {Buisson}, {Caballero-Garcia}, {Cackett}, {De Marco}, {Gallo}, {Lohfink}, {Reynolds}, {Uttley}, {Young}, \& {Zogbhi}}]{Alston2020NatAs} {Alston}, W.~N., {Fabian}, A.~C., {Kara}, E., {et~al.} 2020, Nature Astronomy, 4, 597

\bibitem[{{Ambikasaran} {et~al.}(2015){Ambikasaran}, {Foreman-Mackey}, {Greengard}, {Hogg}, \& {O'Neil}}]{Ambikasaran2015ITPAM} {Ambikasaran}, S., {Foreman-Mackey}, D., {Greengard}, L., {Hogg}, D.~W., \& {O'Neil}, M. 2015, IEEE Transactions on Pattern Analysis and Machine Intelligence, 38, 252

\bibitem[{{Arnaud}(1996)}]{Arnaud1996ASPC} {Arnaud}, K.~A. 1996, in Astronomical Society of the Pacific Conference Series, Vol. 101, Astronomical Data Analysis Software and Systems V, ed. G.~H.  {Jacoby} \& J.~{Barnes} (San Francisco: ASP), 17

\bibitem[Avni \& Tananbaum(1982)]{Avni1982} Avni, Y., \& Tananbaum, H. 1982, \apj, 262, L17

\bibitem[{{Ba{\~n}ados} {et~al.}(2018){Ba{\~n}ados}, {Venemans},
  {Mazzucchelli}, {Farina}, {Walter}, {Wang}, {Decarli}, {Stern}, {Fan},
  {Davies}, {Hennawi}, {Simcoe}, {Turner}, {Rix}, {Yang}, {Kelson}, {Rudie}, \&
  {Winters}}]{Banados2018Nature}
{Ba{\~n}ados}, E., {Venemans}, B.~P., {Mazzucchelli}, C., {et~al.} 2018, \nat,
  553, 473

\bibitem[{{Belloni} {et~al.}(2005){Belloni}, {Homan}, {Casella}, {van der Klis}, {Nespoli}, {Lewin}, {Miller}, \& {M{\'e}ndez}}]{Belloni2005AA} {Belloni}, T., {Homan}, J., {Casella}, P., {et~al.} 2005, \aap, 440, 207

\bibitem[{{Belloni} {et~al.}(2000){Belloni}, {Klein-Wolt}, {M{\'e}ndez}, {van der Klis}, \& {van Paradijs}}]{Belloni2000AA} {Belloni}, T., {Klein-Wolt}, M., {M{\'e}ndez}, M., {van der Klis}, M., \& {van Paradijs}, J. 2000, \aap, 355, 271

\bibitem[{{Beloborodov}(1998)}]{Beloborodov1998MNRAS} {Beloborodov}, A.~M. 1998, \mnras, 297, 739

\bibitem[{{Blandford} \& {McKee}(1982)}]{Blandford1982ApJ} {Blandford}, R.~D., \& {McKee}, C.~F. 1982, \apj, 255, 419

\bibitem[{{Boller} {et~al.}(2003){Boller}, {Voges}, {Dennefeld}, {Lehmann}, {Predehl}, {Burwitz}, {Perlman}, {Gallo}, {Papadakis}, \& {Anderson}}]{Boller2003AA} {Boller}, Th., {Voges}, W., {Dennefeld}, M., {et~al.} 2003, \aap, 397, 557

\bibitem[{{Boroson} (2002)}]{Boroson2002} Boroson, T. A. 2002, \apj, 565, 78

\bibitem[{{Boroson} \& {Green}(1992)}]{Boroson1992ApJS} {Boroson}, T.~A., \& {Green}, R.~F. 1992, \apjs, 80, 109

\bibitem[{{Cackett} {et~al.}(2021){Cackett}, {Bentz}, \& {Kara}}]{Cackett2021iSci} {Cackett}, E.~M., {Bentz}, M.~C., \& {Kara}, E. 2021, iScience, 24, 102557

\bibitem[{{Cackett} {et~al.}(2020){Cackett}, {Gelbord}, {Li}, {Horne}, {Wang}, {Barth}, {Bai}, {Bian}, {Carroll}, {Du}, {Edelson}, {Goad}, {Ho}, {Hu}, {Khatu}, {Luo}, {Miller}, \& {Yuan}}]{Cackett2020ApJ} {Cackett}, E.~M., {Gelbord}, J., {Li}, Y.-R., {et~al.} 2020, \apj, 896, 1

\bibitem[{{Crummy} {et~al.}(2006){Crummy}, {Fabian}, {Gallo}, \& {Ross}}]{Crummy2006MNRAS} {Crummy}, J., {Fabian}, A.~C., {Gallo}, L., \& {Ross}, R.~R. 2006, \mnras, 365, 1067

\bibitem[{{Czerny} \& {Elvis}(1987)}]{Czerny1987ApJ} {Czerny}, B., \& {Elvis}, M. 1987, \apj, 321, 305

\bibitem[{{Dai} {et~al.}(2015){Dai}, {McKinney}, \& {Miller}}]{Dai2015ApJ} {Dai}, L., {McKinney}, J.~C., \& {Miller}, M.~C. 2015, \apjl, 812, L39


\bibitem[Ding et al.(2022)]{Ding2022} Ding, Y., Li, R., Ho, L. C., \& Ricci, C. 2022, \apj, 931, 77

\bibitem[{{Done} {et~al.}(2012){Done}, {Davis}, {Jin}, {Blaes}, \& {Ward}}]{Done2012MNRAS} {Done}, C., {Davis}, S.~W., {Jin}, C., {Blaes}, O., \& {Ward}, M. 2012, \mnras, 420, 1848

\bibitem[{{Done} {et~al.}(2007){Done}, {Gierli{\'n}ski}, \& {Kubota}}]{Done2007ARAA} {Done}, C., {Gierli{\'n}ski}, M., \& {Kubota}, A. 2007, \aapr, 15, 1

\bibitem[{{Donnan} {et~al.}(2023){Donnan}, {Hern{\'a}ndez Santisteban}, {Horne}, {Hu}, {Du}, {Li}, {Xiao}, {Ho}, {Aceituno}, {Wang}, {Guo}, {Yang}, {Jiang}, \& {Yao}}]{Donnan2023MNRAS} {Donnan}, F.~R., {Hern{\'a}ndez Santisteban}, J.~V., {Horne}, K., {et~al.} 2023, \mnras, 523, 545

\bibitem[{{Du} {et~al.}(2014){Du}, {Hu}, {Lu}, {Wang}, {Qiu}, {Li}, {Bai}, {Kaspi}, {Netzer}, {Wang}, \& {SEAMBH Collaboration}}]{Du2014ApJ} {Du}, P., {Hu}, C., {Lu}, K.-X., {et~al.} 2014, \apj, 782, 45


\bibitem[{{Esin} {et~al.}(1997){Esin}, {McClintock}, \& {Narayan}}]{Esin1997ApJ} {Esin}, A.~A., {McClintock}, J.~E., \& {Narayan}, R. 1997, \apj, 489, 865

\bibitem[{{Fabian} {et~al.}(2009){Fabian}, {Zoghbi}, {Ross}, {Uttley}, {Gallo}, {Brandt}, {Blustin}, {Boller}, {Caballero-Garcia}, {Larsson}, {Miller}, {Miniutti}, {Ponti}, {Reis}, {Reynolds}, {Tanaka}, \& {Young}}]{Fabian2009Natur} {Fabian}, A.~C., {Zoghbi}, A., {Ross}, R.~R., {et~al.} 2009, \nat, 459, 540

\bibitem[{{Fan}(2006)}]{Fan2006NewAR} {Fan}, X. 2006, \nar, 50, 665

\bibitem[{{Fragile} {et~al.}(2018){Fragile}, {Etheridge}, {Anninos}, {Mishra},
  \& {Klu{\'z}niak}}]{Fragile2018ApJ}
{Fragile}, P.~C., {Etheridge}, S.~M., {Anninos}, P., {Mishra}, B., \&
  {Klu{\'z}niak}, W. 2018, \apj, 857, 1

\bibitem[{{Francis} {et~al.}(1991){Francis}, {Hewett}, {Foltz}, {Chaffee}, {Weymann}, \& {Morris}}]{Francis1991ApJ} {Francis}, P.~J., {Hewett}, P.~C., {Foltz}, C.~B., {et~al.} 1991, \apj, 373, 465

\bibitem[{{Gabriel} {et~al.}(2004){Gabriel}, {Denby}, {Fyfe}, {Hoar}, {Ibarra}, {Ojero}, {Osborne}, {Saxton}, {Lammers}, \& {Vacanti}}]{Gabriel2004ASPC} {Gabriel}, C., {Denby}, M., {Fyfe}, D.~J., {et~al.} 2004, in Astronomical Society of the Pacific Conference Series, Vol. 314, Astronomical Data Analysis Software and Systems (ADASS) XIII, ed. F.~{Ochsenbein}, M.~G.  {Allen}, \& D.~{Egret} (San Francisco: ASP), 759

\bibitem[{{Gierli\'nski} \& {Done}(2004)}]{Gierlinski2004MNRAS} {Gierli\'nski}, M., \& {Done}, C. 2004, \mnras, 349, L7

\bibitem[{{Gladstone} {et~al.}(2009){Gladstone}, {Roberts}, \& {Done}}]{Gladstone2009MNRAS} {Gladstone}, J.~C., {Roberts}, T.~P., \& {Done}, C. 2009, \mnras, 397, 1836

\bibitem[{{Gliozzi} \& {Williams}(2020)}]{Gliozzi2020MNRAS} {Gliozzi}, M., \& {Williams}, J.~K. 2020, \mnras, 491, 532

\bibitem[{{Haardt} \& {Maraschi}(1991)}]{Haardt1991ApJ} {Haardt}, F., \& {Maraschi}, L. 1991, \apjl, 380, L51

\bibitem[{{Haardt} {et~al.}(1994){Haardt}, {Maraschi}, \& {Ghisellini}}]{Haardt1994ApJ} {Haardt}, F., {Maraschi}, L., \& {Ghisellini}, G. 1994, \apjl, 432, L95


\bibitem[{{Hinkle} {et~al.}(2022){Hinkle}, {Holoien}, {Shappee}, {Neustadt}, {Auchettl}, {Vallely}, {Shahbandeh}, {Kluge}, {Kochanek}, {Stanek}, {Huber}, {Post}, {Bersier}, {Ashall}, {Tucker}, {Williams}, {de Jaeger}, {Do}, {Fausnaugh}, {Gruen}, {Hopp}, {Myles}, {Obermeier}, {Payne}, \& {Thompson}}]{Hinkle2022ApJ} {Hinkle}, J.~T., {Holoien}, T. W.~S., {Shappee}, B.~J., {et~al.} 2022, \apj, 930, 12

\bibitem[{{Hirose} {et~al.}(2009){Hirose}, {Krolik}, \& {Blaes}}]{Hirose2009ApJ} {Hirose}, S., {Krolik}, J.~H., \& {Blaes}, O. 2009, \apj, 691, 16

\bibitem[Ho(1999)]{Ho1999} Ho, L. C. 1999,  \apj, 516, 672

\bibitem[Ho(2008)]{Ho2008} Ho, L. C. 2008, ARA\&A, 46, 475

\bibitem[{{Holoien} {et~al.}(2016){Holoien}, {Kochanek}, {Prieto}, {Stanek}, {Dong}, {Shappee}, {Grupe}, {Brown}, {Basu}, {Beacom}, {Bersier}, {Brimacombe}, {Danilet}, {Falco}, {Guo}, {Jose}, {Herczeg}, {Long}, {Pojmanski}, {Simonian}, {Szczygie{\l}}, {Thompson}, {Thorstensen}, {Wagner}, \& {Wo{\'z}niak}}]{Holoien2016MNRAS} {Holoien}, T.~W.~S., {Kochanek}, C.~S., {Prieto}, J.~L., {et~al.} 2016, \mnras, 455, 2918

\bibitem[{{Ichimaru}(1977)}]{Ichimaru1977ApJ} {Ichimaru}, S. 1977, \apj, 214, 840

\bibitem[{{Jiang} {et~al.}(2019){Jiang}, {Blaes}, {Stone}, \& {Davis}}]{Jiang2019bApJ} {Jiang}, Y.-F., {Blaes}, O., {Stone}, J.~M., \& {Davis}, S.~W. 2019, \apj, 885, 144

\bibitem[{{Jiang} {et~al.}(2014){Jiang}, {Stone}, \& {Davis}}]{Jiang2014ApJ} {Jiang}, Y.-F., {Stone}, J.~M., \& {Davis}, S.~W. 2014, \apj, 796, 106

\bibitem[{{Jiang} {et~al.}(2019){Jiang}, {Stone}, \& {Davis}}]{Jiang2019aApJ} {Jiang}, Y.-F., {Stone}, J.~M., \& {Davis}, S.~W. 2019, \apj, 880, 67

\bibitem[{{Jin} {et~al.}(2023){Jin}, {Done}, {Ward}, {Panessa}, {Liu}, \& {Liu}}]{Jin2023MNRAS} {Jin}, C., {Done}, C., {Ward}, M., {et~al.} 2023, \mnras, 518, 6065

\bibitem[{{Jin} {et~al.}(2022){Jin}, {Wu}, \& {Feng}}]{Jin2022ApJ} {Jin}, J.-J., {Wu}, X.-B., \& {Feng}, X.-T. 2022, \apj, 926, 184

\bibitem[{{Kara} {et~al.}(2016){Kara}, {Alston}, {Fabian}, {Cackett}, {Uttley}, {Reynolds}, \& {Zoghbi}}]{Kara2016MNRAS} {Kara}, E., {Alston}, W.~N., {Fabian}, A.~C., {et~al.} 2016, \mnras, 462, 511

\bibitem[{{Kara} {et~al.}(2023){Kara}, {Barth}, {Cackett}, {Gelbord}, {Montano}, {Li}, {Santana}, {Horne}, {Alston}, {Buisson}, {Chelouche}, {Du}, {Fabian}, {Fian}, {Gallo}, {Goad}, {Grupe}, {Gonz{\'a}lez Buitrago}, {Hern{\'a}ndez Santisteban}, {Kaspi}, {Hu}, {Komossa}, {Kriss}, {Lewin}, {Lewis}, {Loewenstein}, {Lohfink}, {Masterson}, {McHardy}, {Mehdipour}, {Miller}, {Panagiotou}, {Parker}, {Pinto}, {Remillard}, {Reynolds}, {Rogantini}, {Wang}, {Wang}, \& {Wilkins}}]{Kara2023ApJ} {Kara}, E., {Barth}, A.~J., {Cackett}, E.~M., {et~al.} 2023, \apj, 947, 62

\bibitem[{{Kara} {et~al.}(2013){Kara}, {Fabian}, {Cackett}, {Miniutti}, \& {Uttley}}]{Kara2013MNRAS} {Kara}, E., {Fabian}, A.~C., {Cackett}, E.~M., {Miniutti}, G., \& {Uttley}, P.  2013, \mnras, 430, 1408

\bibitem[{{Kara} {et~al.}(2019){Kara}, {Steiner}, {Fabian}, {Cackett}, {Uttley}, {Remillard}, {Gendreau}, {Arzoumanian}, {Altamirano}, {Eikenberry}, {Enoto}, {Homan}, {Neilsen}, \& {Stevens}}]{Kara2019Natur} {Kara}, E., {Steiner}, J.~F., {Fabian}, A.~C., {et~al.} 2019, \nat, 565, 198

\bibitem[{{Kato} {et~al.}(2008){Kato}, {Fukue}, \& {Mineshige}}]{Kato2008book} Kato, S., Fukue, J., \& Mineshige, S. 1998, ed., Black-Hole Accretion Disks (Kyoto: Kyoto Univ. Press)

\bibitem[{{Katz}(1977)}]{Katz1977ApJ} {Katz}, J.~I. 1977, \apj, 215, 265

\bibitem[{{Kawanaka} \& {Mineshige}(2024)}]{Kawanaka2024} {Kawanaka}, N., \& {Mineshige}, S. 2024, PASJ, 76, 306

\bibitem[{{King} {et~al.}(2001){King}, {Davies}, {Ward}, {Fabbiano}, \& {Elvis}}]{King2001ApJ} {King}, A.~R., {Davies}, M.~B., {Ward}, M.~J., {Fabbiano}, G., \& {Elvis}, M.  2001, \apjl, 552, L109

\bibitem[{{Koratkar} \& {Blaes}(1999)}]{Koratkar1999PASP} {Koratkar}, A., \& {Blaes}, O. 1999, \pasp, 111, 1

\bibitem[{{Laha} {et~al.}(2022){Laha}, {Meyer}, {Roychowdhury}, {Becerra Gonzalez}, {Acosta-Pulido}, {Thapa}, {Ghosh}, {Behar}, {Gallo}, {Kriss}, {Panessa}, {Bianchi}, {La Franca}, {Scepi}, {Begelman}, {Longinotti}, {Lusso}, {Oates}, {Nicholl}, \& {Cenko}}]{Laha2022ApJ} {Laha}, S., {Meyer}, E., {Roychowdhury}, A., {et~al.} 2022, \apj, 931, 5

\bibitem[{{Laor} \& {Netzer}(1989)}]{Laor1989MNRAS} {Laor}, A., \& {Netzer}, H. 1989, \mnras, 238, 897

\bibitem[{{Lasota}(2001)}]{Lasota2001NewAR} {Lasota}, J.-P. 2001, \nar, 45, 449

\bibitem[{{Li} {et~al.}(2022){Li}, {Ho}, {Ricci}, {Trakhtenbrot}, {Arcavi}, {Kara}, \& {Hiramatsu}}]{Li2022paper1} {Li}, R., {Ho}, L.~C., {Ricci}, C., {et~al.} 2022, \apj, 933, 70

\bibitem[{{Li} {et~al.}(2024){Li}, {Ricci}, {Ho}, {Trakhtenbrot}, {Kara}, {Masterson}, \& {Arcavi}}]{Li2024paper2} {Li}, R., {Ricci}, C., {Ho}, L. C., et al. 2024, \apj, submitted

\bibitem[{{Li} {et~al.}(2018){Li}, {Songsheng}, {Qiu}, {Hu}, {Du}, {Lu}, {Huang}, {Bai}, {Bian}, {Yuan}, {Ho}, \& {Wang}}]{Li2018ApJ} {Li}, Y.-R., {Songsheng}, Y.-Y., {Qiu}, J., {et~al.} 2018, \apj, 869, 137

\bibitem[{{Liu} {et~al.}(2021){Liu}, {Luo}, {Brandt}, {Brotherton}, {Gallagher}, {Ni}, {Shemmer}, \& {Timlin}}]{Liu2021ApJ} {Liu}, H., {Luo}, B., {Brandt}, W.~N., {et~al.} 2021, \apj, 910, 103


\bibitem[{{Lodato} \& {Rossi}(2011)}]{Lodato2011MNRAS} {Lodato}, G., \& {Rossi}, E.~M. 2011, \mnras, 410, 359

\bibitem[{{Masterson} {et~al.}(2022){Masterson}, {Kara}, {Ricci}, {Garc{\'\i}a}, {Fabian}, {Pinto}, {Kosec}, {Remillard}, {Loewenstein}, {Trakhtenbrot}, \& {Arcavi}}]{Masterson2022} {Masterson}, M., {Kara}, E., {Ricci}, C., {et~al.} 2022, \apj, 934, 35

\bibitem[{{Mastroserio} {et~al.}(2020){Mastroserio}, {Ingram}, \& {van der Klis}}]{Mastroserio2020MNRAS} {Mastroserio}, G., {Ingram}, A., \& {van der Klis}, M. 2020, \mnras, 498, 4971

\bibitem[{{Middleton} {et~al.}(2009){Middleton}, {Done}, {Ward}, {Gierli{\'n}ski}, \& {Schurch}}]{Middleton2009MNRAS} {Middleton}, M., {Done}, C., {Ward}, M., {Gierli{\'n}ski}, M., \& {Schurch}, N.  2009, \mnras, 394, 250

\bibitem[{{Miller} {et~al.}(2015){Miller}, {Kaastra}, {Miller}, {Reynolds}, {Brown}, {Cenko}, {Drake}, {Gezari}, {Guillochon}, {Gultekin}, {Irwin}, {Levan}, {Maitra}, {Maksym}, {Mushotzky}, {O'Brien}, {Paerels}, {de Plaa}, {Ramirez-Ruiz}, {Strohmayer}, \& {Tanvir}}]{Miller2015Natur} {Miller}, J.~M., {Kaastra}, J.~S., {Miller}, M.~C., {et~al.} 2015, \nat, 526, 542

\bibitem[{{Mineshige} {et~al.}(2000){Mineshige}, {Kawaguchi}, {Takeuchi}, \& {Hayashida}}]{Mineshige2000PASJ} {Mineshige}, S., {Kawaguchi}, T., {Takeuchi}, M., \& {Hayashida}, K. 2000, \pasj, 52, 499

\bibitem[{{Miniutti} {et~al.}(2023){Miniutti}, {Giustini}, {Arcodia}, {Saxton}, {Chakraborty}, {Read}, \& {Kara}}]{Miniutti2023AA} {Miniutti}, G., {Giustini}, M., {Arcodia}, R., {et~al.} 2023, \aap, 674, L1

\bibitem[{{Mitsuda} {et~al.}(1984){Mitsuda}, {Inoue}, {Koyama}, {Makishima}, {Matsuoka}, {Ogawara}, {Shibazaki}, {Suzuki}, {Tanaka}, \& {Hirano}}]{Mitsuda1984PASJ} {Mitsuda}, K., {Inoue}, H., {Koyama}, K., {et~al.} 1984, \pasj, 36, 741

\bibitem[{{Nandra} {et~al.}(2007){Nandra}, {O'Neill}, {George}, \& {Reeves}}]{Nandra2007MNRAS} {Nandra}, K., {O'Neill}, P.~M., {George}, I.~M., \& {Reeves}, J.~N. 2007, \mnras, 382, 194

\bibitem[{{Narayan} \& {Yi}(1994)}]{Narayan1994ApJ} {Narayan}, R., \& {Yi}, I. 1994, \apjl, 428, L13

\bibitem[{{Nicholls} {et~al.}(2018){Nicholls}, {Brimacombe}, {Kiyota}, {Stone}, {Cruz}, {Trappett}, {Vallely}, {Stanek}, {Kochanek}, {Brown}, {Shields}, {Thompson}, {Shappee}, {Holoien}, {Prieto}, {Bersier}, {Dong}, {Bose}, {Chen}, {Stritzinger}, \& {Holmbo}}]{Nicholls2018ATel} {Nicholls}, B., {Brimacombe}, J., {Kiyota}, S., {et~al.} 2018, The Astronomer's Telegram, 11391

\bibitem[{{Noda} \& {Done}(2018)}]{Noda2018MNRAS} {Noda}, H., \& {Done}, C. 2018, \mnras, 480, 3898

\bibitem[{{Novikov} \& {Thorne}(1973)}]{Novikov1973blho} Novikov, I. D., \& Thorne, K. S. 1973, in Black Holes, ed. C. DeWitt, \& B. DeWitt (New York: Gordon and Breach), 343

\bibitem[{{Ohsuga}(2006)}]{Ohsuga2006ApJ} {Ohsuga}, K. 2006, \apj, 640, 923

\bibitem[{{Ohsuga} \& {Mineshige}(2011)}]{Ohsuga2011ApJ} {Ohsuga}, K., \& {Mineshige}, S. 2011, \apj, 736, 2

\bibitem[{{Ohsuga} {et~al.}(2005){Ohsuga}, {Mori}, {Nakamoto}, \& {Mineshige}}]{Ohsuga2005ApJ} {Ohsuga}, K., {Mori}, M., {Nakamoto}, T., \& {Mineshige}, S. 2005, \apj, 628, 368

\bibitem[{{Orosz} \& {Bailyn}(1997)}]{Orosz1997ApJ} {Orosz}, J.~A., \& {Bailyn}, C.~D. 1997, \apj, 477, 876

\bibitem[{{Pasham} {et~al.}(2014){Pasham}, {Strohmayer}, \& {Mushotzky}}]{Pasham2014Natur} {Pasham}, D.~R., {Strohmayer}, T.~E., \& {Mushotzky}, R.~F. 2014, \nat, 513, 74

\bibitem[{{Peng} \& {Chen}(2021)}]{Peng2021MNRAS} {Peng}, P., \& {Chen}, X. 2021, \mnras, 505, 1324

\bibitem[{{Penna} {et~al.}(2013){Penna}, {Narayan}, \&
  {S{\k{a}}dowski}}]{Penna2013MNRAS}
{Penna}, R.~F., {Narayan}, R., \& {S{\k{a}}dowski}, A. 2013, \mnras, 436, 3741

\bibitem[{{Petrucci} {et~al.}(2020){Petrucci}, {Gronkiewicz}, {Rozanska},
  {Belmont}, {Bianchi}, {Czerny}, {Matt}, {Malzac}, {Middei}, {De Rosa},
  {Ursini}, \& {Cappi}}]{Petrucci2020AA}
{Petrucci}, P.~O., {Gronkiewicz}, D., {Rozanska}, A., {et~al.} 2020, \aap, 634,
  A85

\bibitem[{{Pringle}(1981)}]{Pringle1981ARAA}
{Pringle}, J.~E. 1981, \araa, 19, 137

\bibitem[{{Proga} {et~al.}(2000){Proga}, {Stone}, \& {Kallman}}]{Proga2000ApJ} {Proga}, D., {Stone}, J.~M., \& {Kallman}, T.~R. 2000, \apj, 543, 686


\bibitem[Quataert et al.(1999)]{Quataert1999} Quataert, E., Di Matteo, T., Narayan, R., \& Ho, L. C. 1999, \apj, 525, L89

\bibitem[{{Rees}(1988)}]{Rees1988Natur} {Rees}, M.~J. 1988, \nat, 333, 523

\bibitem[{{Remillard} \& {McClintock}(2006)}]{Remillard2006ARAA} {Remillard}, R.~A., \& {McClintock}, J.~E. 2006, \araa, 44, 49

\bibitem[{{Ricci} {et~al.}(2020{\natexlab{a}}){Ricci}, {Kara}, {Loewenstein}, {Trakhtenbrot}, {Arcavi}, {Remillard}, {Fabian}, {Gendreau}, {Arzoumanian}, {Li}, {Ho}, {MacLeod}, {Cackett}, {Altamirano}, {Gand hi}, {Kosec}, {Pasham}, {Steiner}, \& {Chan}}]{Ricci2020ApJL} {Ricci}, C., {Kara}, E., {Loewenstein}, M., {et~al.} 2020, \apjl, 898, L1

\bibitem[{{Ricci} {et~al.}(2021{\natexlab{b}}){Ricci}, {Loewenstein}, {Kara}, {Remillard}, {Trakhtenbrot}, {Arcavi}, {Gendreau}, {Arzoumanian}, {Fabian}, {Li}, {Ho}, {MacLeod}, {Cackett}, {Altamirano}, {Gandhi}, {Kosec}, {Pasham}, {Steiner}, \& {Chan}}]{Ricci2021ApJS} {Ricci}, C., {Loewenstein}, M., {Kara}, E., {et~al.} 2021, \apjs, 255, 7

\bibitem[{{Ricci} \& {Trakhtenbrot}(2023)}]{Ricci2023NatAs}
{Ricci}, C., \& {Trakhtenbrot}, B. 2023, Nature Astronomy, 7, 1282

\bibitem[{{Roberts}(2007)}]{Roberts2007ApSS} {Roberts}, T.~P. 2007, \apss, 311, 203

\bibitem[{{Ruan} {et~al.}(2019){Ruan}, {Anderson}, {Eracleous}, {Green}, {Haggard}, {MacLeod}, {Runnoe}, \& {Sobolewska}}]{Ruan2019ApJ} {Ruan}, J.~J., {Anderson}, S.~F., {Eracleous}, M., {et~al.} 2019, \apj, 883, 76

\bibitem[{{Rybicki} \& {Lightman}(1986)}]{Rybicki1986book} Rybicki, G. B., \& Lightman, A. P. 1986, Radiative Processes in Astrophysics (Wiley-VCH)

\bibitem[{{S{\c a}dowski}(2009)}]{Sadowski2009ApJS} {S{\c a}dowski}, A. 2009, \apjs, 183, 171

\bibitem[{{Sadowski}(2011)}]{Sadowski2011} {S{\c a}dowski}, A. 2011, Ph.D. Thesis (Nicolaus Copernicus Astronomical Center, Polish Academy of Sciences)

\bibitem[{{S{\k{a}}dowski} {et~al.}(2014){S{\k{a}}dowski}, {Narayan},
  {McKinney}, \& {Tchekhovskoy}}]{Sadowski2014MNRAS}
{S{\k{a}}dowski}, A., {Narayan}, R., {McKinney}, J.~C., \& {Tchekhovskoy}, A.
  2014, \mnras, 439, 503

\bibitem[{{Salpeter}(1964)}]{Salpeter1964ApJ} {Salpeter}, E.~E. 1964, \apj, 140, 796

\bibitem[{{Saxton} {et~al.}(2017){Saxton}, {Read}, {Komossa}, {Lira}, {Alexander}, \& {Wieringa}}]{Saxton2017AA} {Saxton}, R.~D., {Read}, A.~M., {Komossa}, S., {et~al.} 2017, \aap, 598, A29

\bibitem[{{Shakura} \& {Sunyaev}(1973)}]{Shakura1973AA} {Shakura}, N.~I., \& {Sunyaev}, R.~A. 1973, \aap, 24, 337

\bibitem[{{Shappee} {et~al.}(2014){Shappee}, {Prieto}, {Grupe}, {Kochanek}, {Stanek}, {De Rosa}, {Mathur}, {Zu}, {Peterson}, {Pogge}, {Komossa}, {Im}, {Jencson}, {Holoien}, {Basu}, {Beacom}, {Szczygie{\l}}, {Brimacombe}, {Adams}, {Campillay}, {Choi}, {Contreras}, {Dietrich}, {Dubberley}, {Elphick}, {Foale}, {Giustini}, {Gonzalez}, {Hawkins}, {Howell}, {Hsiao}, {Koss}, {Leighly}, {Morrell}, {Mudd}, {Mullins}, {Nugent}, {Parrent}, {Phillips}, {Pojmanski}, {Rosing}, {Ross}, {Sand}, {Terndrup}, {Valenti}, {Walker}, \& {Yoon}}]{Shappee2014ApJ} {Shappee}, B.~J., {Prieto}, J.~L., {Grupe}, D., {et~al.} 2014, \apj, 788, 48

\bibitem[Shen \& Ho(2014)]{Shen2014} Shen, Y., \& Ho, L. C. 2014, Nature, 513, 210

\bibitem[{{Shen} {et~al.}(2011){Shen}, {Richards}, {Strauss}, {Hall}, {Schneider}, {Snedden}, {Bizyaev}, {Brewington}, {Malanushenko}, {Malanushenko}, {Oravetz}, {Pan}, \& {Simmons}}]{Shen2011ApJS} {Shen}, Y., {Richards}, G.~T., {Strauss}, M.~A., {et~al.} 2011, \apjs, 194, 45

\bibitem[Shields(1978)]{Shields1978} Shields, G. A. 1978, Nature, 272, 706

\bibitem[{{So\l tan}(1982)}]{Soltan1982MNRAS} {So\l tan}, A. 1982, \mnras, 200, 115

\bibitem[{{Starkey} {et~al.}(2017){Starkey}, {Horne}, {Fausnaugh}, {Peterson}, {Bentz}, {Kochanek}, {Denney}, {Edelson}, {Goad}, {De Rosa}, {Anderson}, {Ar{\'e}valo}, {Barth}, {Bazhaw}, {Borman}, {Boroson}, {Bottorff}, {Brandt}, {Breeveld}, {Cackett}, {Carini}, {Croxall}, {Crenshaw}, {Dalla Bont{\`a}}, {De Lorenzo-C{\'a}ceres}, {Dietrich}, {Efimova}, {Ely}, {Evans}, {Filippenko}, {Flatland}, {Gehrels}, {Geier}, {Gelbord}, {Gonzalez}, {Gorjian}, {Grier}, {Grupe}, {Hall}, {Hicks}, {Horenstein}, {Hutchison}, {Im}, {Jensen}, {Joner}, {Jones}, {Kaastra}, {Kaspi}, {Kelly}, {Kennea}, {Kim}, {Kim}, {Klimanov}, {Korista}, {Kriss}, {Lee}, {Leonard}, {Lira}, {MacInnis}, {Manne-Nicholas}, {Mathur}, {McHardy}, {Montouri}, {Musso}, {Nazarov}, {Norris}, {Nousek}, {Okhmat}, {Pancoast}, {Parks}, {Pei}, {Pogge}, {Pott}, {Rafter}, {Rix}, {Saylor}, {Schimoia}, {Schn{\"u}lle}, {Sergeev}, {Siegel}, {Spencer}, {Sung}, {Teems}, {Turner}, {Uttley}, {Vestergaard}, {Villforth}, {Weiss}, {Woo}, {Yan}, {Young}, {Zheng}, \& {Zu}}]{Starkey2017ApJ} {Starkey}, D., {Horne}, K., {Fausnaugh}, M.~M., {et~al.} 2017, \apj, 835, 65

\bibitem[{{Sutton} {et~al.}(2013){Sutton}, {Roberts}, \& {Middleton}}]{Sutton2013MNRAS} {Sutton}, A.~D., {Roberts}, T.~P., \& {Middleton}, M.~J. 2013, \mnras, 435, 1758

\bibitem[{{Svensson} \& {Zdziarski}(1994)}]{Svensson1994ApJ} {Svensson}, R., \& {Zdziarski}, A.~A. 1994, \apj, 436, 599

\bibitem[{{Takahashi} {et~al.}(2016){Takahashi}, {Ohsuga}, {Kawashima}, \& {Sekiguchi}}]{Takahashi2016ApJ} {Takahashi}, H.~R., {Ohsuga}, K., {Kawashima}, T., \& {Sekiguchi}, Y. 2016, \apj, 826, 23

\bibitem[{{Trakhtenbrot} {et~al.}(2019){Trakhtenbrot}, {Arcavi}, {MacLeod}, {Ricci}, {Kara}, {Graham}, {Stern}, {Harrison}, {Burke}, {Hiramatsu}, {Hosseinzadeh}, {Howell}, {Smartt}, {Rest}, {Prieto}, {Shappee}, {Holoien}, {Bersier}, {Filippenko}, {Brink}, {Zheng}, {Li}, {Remillard}, \& {Loewenstein}}]{Trakhtenbrot2019ApJ} {Trakhtenbrot}, B., {Arcavi}, I., {MacLeod}, C.~L., {et~al.} 2019, \apj, 883, 94

\bibitem[{{Tran} {et~al.}(2011){Tran}, {Lyke}, \& {Mader}}]{Tran2011ApJ} {Tran}, H.~D., {Lyke}, J.~E., \& {Mader}, J.~A. 2011, \apjl, 726, L21

\bibitem[{{Ulmer}(1999)}]{Ulmer1999ApJ} {Ulmer}, A. 1999, \apj, 514, 180

\bibitem[{{Uttley} {et~al.}(2014){Uttley}, {Cackett}, {Fabian}, {Kara}, \& {Wilkins}}]{Uttley2014ARAA} {Uttley}, P., {Cackett}, E.~M., {Fabian}, A.~C., {Kara}, E., \& {Wilkins}, D.~R. 2014, \aapr, 22, 72

\bibitem[{{Vanden Berk} {et~al.}(2001){Vanden Berk}, {Richards}, {Bauer}, {Strauss}, {Schneider}, {Heckman}, {York}, {Hall}, {Fan}, {Knapp}, {Anderson}, {Annis}, {Bahcall}, {Bernardi}, {Briggs}, {Brinkmann}, {Brunner}, {Burles}, {Carey}, {Castander}, {Connolly}, {Crocker}, {Csabai}, {Finkbeiner}, {Friedman}, {Frieman}, {Fukugita}, {Gunn}, {Hennessy}, {Ivezi{\'c}}, {Kent}, {Kunszt}, {Lamb}, {Leger}, {Long}, {Loveday}, {Lupton}, {Meiksin}, {Merelli}, {Munn}, {Newberg}, {Newcomb}, {Nichol}, {Owen}, {Pier}, {Pope}, {Rockosi}, {Schlegel}, {Siegmund}, {Smee}, {Snir}, {Stoughton}, {Stubbs}, {SubbaRao}, {Szalay}, {Szokoly}, {Tremonti}, {Uomoto}, {Waddell}, {Yanny}, \& {Zheng}}]{VandenBerk2001AJ} {Vanden Berk}, D.~E., {Richards}, G.~T., {Bauer}, A., {et~al.} 2001, \aj, 122, 549

\bibitem[{{Voges} {et~al.}(1999){Voges}, {Aschenbach}, {Boller}, {Br{\"a}uninger}, {Briel}, {Burkert}, {Dennerl}, {Englhauser}, {Gruber}, {Haberl}, {Hartner}, {Hasinger}, {K{\"u}rster}, {Pfeffermann}, {Pietsch}, {Predehl}, {Rosso}, {Schmitt}, {Tr{\"u}mper}, \& {Zimmermann}}]{Voges1999AA} {Voges}, W., {Aschenbach}, B., {Boller}, T., {et~al.} 1999, \aap, 349, 389

\bibitem[{{Wandel} \& {Petrosian}(1988)}]{Wandel1988ApJ} {Wandel}, A., \& {Petrosian}, V. 1988, \apjl, 329, L11

\bibitem[Wang et al.(2023)]{Wang2023} Wang, J.-M., Zhai, S., Li, Y.-R., et al. 2023, \apj, 954, 84

\bibitem[{{Wang} \& {Zhou}(1999)}]{Wang1999ApJ} {Wang}, J.-M., \& {Zhou}, Y.-Y. 1999, \apj, 516, 420

\bibitem[{{Watarai} \& {Mineshige}(2003)}]{Watarai2003ApJ} {Watarai}, K., \& {Mineshige}, S. 2003, \apj, 596, 421

\bibitem[{{Wilkins} {et~al.}(2017){Wilkins}, {Gallo}, {Silva}, {Costantini}, {Brandt}, \& {Kriss}}]{Wilkins2017MNRAS} {Wilkins}, D.~R., {Gallo}, L.~C., {Silva}, C.~V., {et~al.} 2017, \mnras, 471, 4436

\bibitem[{{Wu} {et~al.}(2015){Wu}, {Wang}, {Fan}, {Yi}, {Zuo}, {Bian}, {Jiang}, {McGreer}, {Wang}, {Yang}, {Yang}, {Thompson}, \& {Beletsky}}]{Wu2015Natur} {Wu}, X.-B., {Wang}, F., {Fan}, X., {et~al.} 2015, \nat, 518, 512

\bibitem[{{Yang} {et~al.}(2021){Yang}, {Wang}, {Fan}, {Barth}, {Hennawi}, {Nanni}, {Bian}, {Davies}, {Farina}, {Schindler}, {Ba{\~n}ados}, {Decarli}, {Eilers}, {Green}, {Guo}, {Jiang}, {Li}, {Venemans}, {Walter}, {Wu}, \& {Yue}}]{Yang2021ApJ} {Yang}, J., {Wang}, F., {Fan}, X., {et~al.} 2021, \apj, 923, 262


\bibitem[{{Yu} \& {Tremaine}(2002)}]{Yu2002MNRAS} {Yu}, Q., \& {Tremaine}, S. 2002, \mnras, 335, 965

\bibitem[{{Zauderer} {et~al.}(2011){Zauderer}, {Berger}, {Soderberg}, {Loeb}, {Narayan}, {Frail}, {Petitpas}, {Brunthaler}, {Chornock}, {Carpenter}, {Pooley}, {Mooley}, {Kulkarni}, {Margutti}, {Fox}, {Nakar}, {Patel}, {Volgenau}, {Culverhouse}, {Bietenholz}, {Rupen}, {Max-Moerbeck}, {Readhead}, {Richards}, {Shepherd}, {Storm}, \& {Hull}}]{Zauderer2011Natur} {Zauderer}, B.~A., {Berger}, E., {Soderberg}, A.~M., {et~al.} 2011, \nat, 476, 425

\bibitem[{{Zdziarski} {et~al.}(2022){Zdziarski}, {You}, \& {Szanecki}}]{Zdziarski2022ApJ} {Zdziarski}, A.~A., {You}, B., \& {Szanecki}, M. 2022, \apjl, 939, L2

\bibitem[{{Zhu} {et~al.}(2019){Zhu}, {Zhang}, {Jiang}, {Kataoka}, {Birnstiel}, {Dullemond}, {Andrews}, {Huang}, {P{\'e}rez}, {Carpenter}, {Bai}, {Wilner}, \& {Ricci}}]{Zhu2019ApJ} {Zhu}, Z., {Zhang}, S., {Jiang}, Y.-F., {et~al.} 2019, \apjl, 877, L18
\end{thebibliography}
\end{document}